# Covariant quantizations in plane and curved spaces

# J. L. M. Assirati & D. M. Gitman

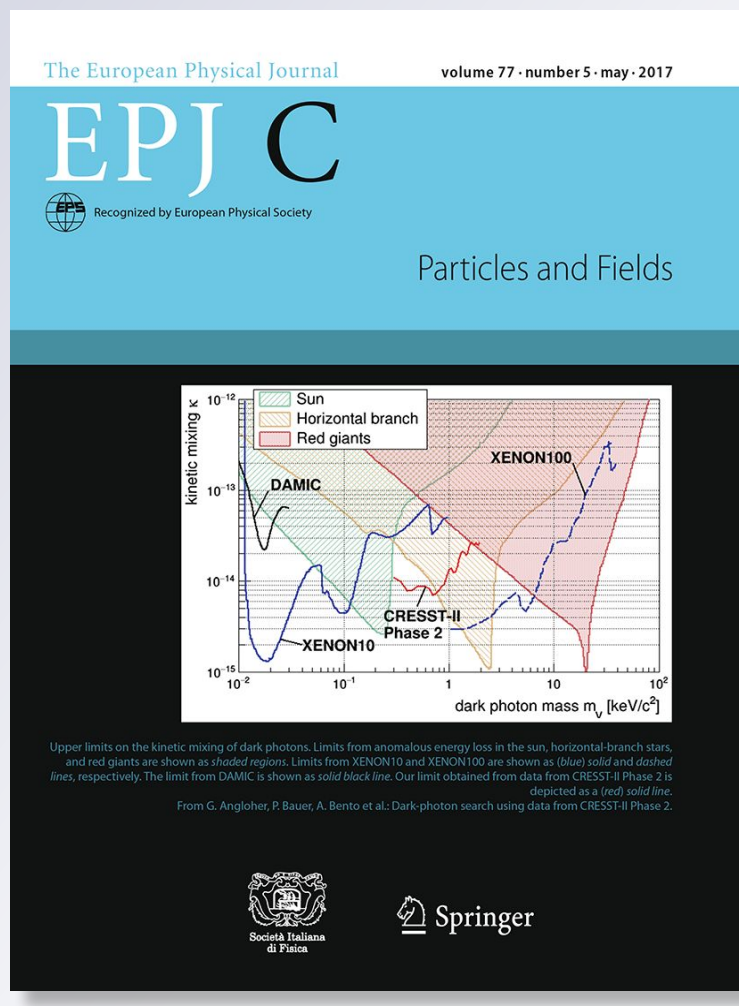

# Covariant quantizations in plane and curved spaces

J. L. M. Assirati[1,a], D. M. Gitman[2,3,4,b]

[1] Institute of Physics, University of São Paulo, São Paulo, Brazil
[2] Department of Physics, Tomsk State University, Tomsk, Russia
[3] P.N.Lebedev Physical Institute, Moscow, Russia
[4] Institute of Physics, University of São Paulo, São Paulo, Brazil

**Abstract** We present covariant quantization rules for nonsingular finite-dimensional classical theories with flat and curved configuration spaces. In the beginning, we construct a family of covariant quantizations in flat spaces and Cartesian coordinates. This family is parametrized by a function $\omega(\theta), \theta \in (1, 0)$, which describes an ambiguity of the quantization. We generalize this construction presenting covariant quantizations of theories with flat configuration spaces but already with arbitrary curvilinear coordinates. Then we construct a so-called minimal family of covariant quantizations for theories with curved configuration spaces. This family of quantizations is parametrized by the same function $\omega(\theta)$. Finally, we describe a more wide family of covariant quantizations in curved spaces. This family is already parametrized by two functions, the previous one $\omega(\theta)$ and by an additional function $\Theta(x, \xi)$. The above mentioned minimal family is a part at $\Theta = 1$ of the wide family of quantizations. We study constructed quantizations in detail, proving their consistency and covariance. As a physical application, we consider a quantization of a non-relativistic particle moving in a curved space, discussing the problem of a quantum potential. Applying the covariant quantizations in flat spaces to an old problem of constructing quantum Hamiltonian in polar coordinates, we directly obtain a correct result.

## Contents

[a] e-mail: assirati@if.usp.br
[b] e-mail: gitman@if.usp.br

# 1 Introduction

1.1 Quantization

In general, for a physicist, quantization means constructing a quantum theory (QT) of a given physical system. In the case when a classical Hamiltonian theory of the physical system does exist, quantization means constructing a QT of the physical system on the basis of such already existing classical theory. Here, one may speak of a quantum deformation of the classical theory. Such a quantization must obey a correspondence principle, which means that the QT has to reproduce predictions of the classical theory in a classical limit (big masses, macroscopic scales, smooth potentials, and so on). The quantization problem usually does not have an unique solution. The only criterion whether an adequate QT is constructed remains the coincidence of its predictions with the experiment. Even if a classical theory of a physical system does exist, it is difficult to formulate quantization rules in the general case. However, an experience in the quantization of simple systems as a free particle, a harmonic oscillator, and a non-relativistic particle in potential fields, was used to formulate a consistent scheme of operator quantization of classical Hamiltonian systems with finite degrees of freedom, without constraints, and with a flat phase space [1]. Such a scheme is called the operator canonical quantization. If points in the phase space of the classical theory are labeled by coordinates $x = (x^a)$ and momenta $p = (p_a)$, $a = 1, 2, \ldots, D$, then the operator canonical quantization is often formulated as some rules of constructing quantum operators $\hat{F}$ that correspond to classical functions $F(x, p)$ describing physical functions of the system.[1] All such operators $\hat{F}$ act in a Hilbert space $\mathfrak{H}$ of vectors that represent possible quantum states of the physical system under consideration. The above mentioned correspondence $F(x, p) \rightarrow \hat{F}$ is denoted as the $Q$-operation, $\hat{F} = Q(F)$, and is often called simply quantization. The quantization $Q(F)$ should obey some properties listed here [4–7]:

(i) The operation $Q(F)$ is linear and injective,

$$Q(\alpha F_1 + \beta F_2) = \alpha Q(F_1) + \beta Q(F_2),$$

$$F_1 \neq F_2 \Rightarrow Q(F_1) \neq Q(F_2), \quad \alpha, \beta \in \mathbb{C}. \tag{1}$$

(ii) If $F(x, p)$ is a real function then

$$(Q(F))^\dagger = Q(F), \quad F^* = F, \tag{2}$$

which implies that any quantization $Q(F)$ assigns a self-adjoint operator to a real function $F(x, p)$.
If $F(x, p)$ is a complex function and property, (1) holds then $(Q(F))^\dagger = Q(F^*)$.
Quantizations that satisfy condition (2) are called hermitian quantizations[2] in what follows.

(iii) A quantization $Q(F)$ that obeys the correspondence principle, must satisfy the following relations[3]

$$[Q(F), Q(G)] = i\hbar Q(\{F, G\}) + O(\hbar^2), \tag{3}$$

for any two physical functions $F$ and $G$, where $\{F, G\}$ are Poisson brackets,[4]

$$\{F, G\} = \frac{\partial F}{\partial x^a} \frac{\partial G}{\partial p_a} - \frac{\partial F}{\partial p_a} \frac{\partial G}{\partial x^a}. \tag{4}$$

It is convenient to formulate the correspondence principle in terms of operator symbols and their star products. We recall that a function $F_{(Q)}$ is called a symbol of the operator $\hat{F}$ with respect to a quantization $Q$ if $Q(F_{(Q)}) = \hat{F}$,

$F_{(Q)}$ is $Q$-symbol of $\hat{F}$ if $Q(F_{(Q)}) = \hat{F}$. (5)

Let $Q(F)$ and $Q'(F)$ be two different quantizations, and let $F_{(Q)}$ and $F_{(Q')}$ be $Q$-symbol and $Q'$-symbol of the same operator $\hat{F}$, then we have a trivial, in a sense, equality

$$Q'\left(F_{(Q')}\right) = Q\left(F_{(Q)}\right). \tag{6}$$

Let $F_{(Q)}$ and $G_{(Q)}$ be $Q$-symbols of operators $\hat{F}$ and $\hat{G}$. Then the star product $F_{(Q)} * G_{(Q)}$ is the $Q$-symbol of the operator $\hat{F}\hat{G}$, namely, $Q(F_{(Q)} * G_{(Q)}) = \hat{F}\hat{G}$, for detail see e.g. [7,9]. A quantization $Q(F)$ satisfies the correspondence principle if

---

[1] Since there are no constraints in the phase space, any real classical function in the phase space $F(x, p)$ is physical [2,3].

[2] We do not discuss here peculiarities connected to the difference between hermiticity and self-adjointness; see [8].

[3] If there are constraints in the classical Hamiltonian formulation of the system under consideration, then these commutation relations are modified as follows:

$[Q(F), Q(G)] = i\hbar Q(\{F, G\}|_{D(\Phi)}) + O(\hbar^2),$

where $\{F, G\}|_{D(\Phi)}$ are Dirac brackets and $\Phi(x, p) = 0$ is a complete set of second-class constraints, including gauge-fixing complementary conditions; see [3].

[4] The summation convention over repeated sub- and superscripts is assumed throughout, unless otherwise explicitly stated.

$$F_{(Q)} * G_{(Q)} = F_{(Q)} G_{(Q)} + O(\hbar),$$
$$(F_{(Q)} * G_{(Q)} - G_{(Q)} * F_{(Q)}) = i\hbar \{F_{(Q)}, G_{(Q)}\} + O(\hbar^2). \quad (7)$$

Quantizations that satisfy properties (i)–(iii), we call conditionally consistent quantizations.

If state vectors in the Hilbert space $\mathfrak{H}$ are chosen in the coordinate representation (as functions $\psi(x)$) then we have to find a coordinate realization or representation for all possible operators $\hat{F} = Q(F)$ that satisfy the above listed properties. First of all, we have to fix the right hand side of Eq. (3) (to fix in a way the term $O(\hbar^2)$) and then to find a quantization that satisfies these equations and the above listed properties.

Which difficulties can be met in the course of the realization of the above program? One of these difficulties is known as the ordering problem. It can easily be seen in the case when the phase space is flat and coordinates of the configuration space are Cartesian. In such a case, usually, for basic operators $\hat{x}^a$, $\hat{p}_a$, and for constant numbers $\hat{\alpha} = Q(\alpha)$, commutation relations (3) are chosen as

$$[\hat{\alpha}, \hat{x}^b] = [\hat{\alpha}, \hat{p}_b] = [\hat{x}^a, \hat{x}^b] = [\hat{p}_a, \hat{p}_b] = 0,$$
$$[\hat{x}^a, \hat{p}_b] = i\hbar \delta_{ab}, \quad a, b = 1, \ldots, D, \quad (8)$$

and their corresponding realization as

$$\hat{\alpha} = \alpha \hat{I}, \quad Q(x^a) = \hat{x}^a = x^a,$$
$$Q(p_a) = \hat{p}_a = -i\frac{\partial}{\partial x^a} = -i\partial_a. \quad (9)$$

Then one has to find a realization for all the operators $\hat{F} = Q(F)$. A naive rule (ordering rule) $\hat{F} = Q(F) = F(\hat{x}, \hat{p})$ formally obeys Eq. (3) but contains an ambiguity since the operators $\hat{x}$ and $\hat{p}$ do not commute. For example, one can consider, $xp$ or $px$ orderings, the Weyl ordering, and so on [7,10]. In spite of the fact that there exist some natural mathematical and physical restrictions on possible forms of quantum operators $\hat{F} = Q(F) = F(\hat{x}, \hat{p})$ (such as the operator boundedness from above, self-adjointness, the absence of anomalies, and so on), they do not remove completely the above mentioned ambiguity in the general case. In fact, such ambiguities reflect the above mentioned possibility to exist distinct quantum theories having one and the same classical limit. One cannot a priori advocate a given operator ordering, in principle, only an experiment as a final arbitrator may help in choosing an adequate quantum version of a given classical theory, and, thus, to provide arguments in favor of a certain operator ordering.

Quantization of a non-relativistic particle Hamiltonian

$$H(x, p) = \frac{1}{2m} g^{\mu\nu}(x) p_\mu p_\nu \quad (10)$$

in a curved space has a long story. A direct construction of a quantum operator that corresponds to classical function (10) in the general case meets an ordering problem, which creates ambiguities in the final result. These ambiguities affect the energy spectrum of the particle, which explains attempts to remove or to reduce these ambiguities applying different theoretical considerations. The first important work in this direction was done by DeWitt [11] who derived a covariant[5] quantum Hamiltonian (using a path-integral approach) in the following form:

$$\hat{H} = -\frac{\hbar^2}{2m} g^{-1/2} \partial_\mu g^{1/2} g^{\mu\nu} \partial_\nu + \frac{\hbar^2 \varkappa}{m} R(x), \quad \varkappa = \frac{1}{12}, \quad (11)$$

where the kinetic operator is proportional to the Laplace–Beltrami operator and an additional term proportional to the scalar curvature $R(x)$ appeared. In subsequent works that used the same path-integral approach the same [12] or slightly different forms for $\hat{H}$ were obtained, they differ by values of the dimensionless constant $\varkappa$, e.g. $\varkappa = \frac{1}{8}$ in Ref. [13] and $\varkappa = \frac{1}{6}$ in Refs. [14,15]. One has to mention attempts to quantize the particle motion in curved space canonically considering it as a motion on a hypersurface embedded in an Euclidean space [16–18]. To this end, they used a machinery elaborated for quantizing constraint systems [2,3]. In such a way, the authors of Refs. [16,17] derived the operator (11) with $\varkappa = \frac{1}{8}$, whereas according to the conclusion of the work [18] the constrained systems theory does not contribute to an unambiguous description of the quantum dynamics of non-relativistic particles in curved spaces.

Covariant canonical quantization of relativistic particles in flat and curved spaces meets additional difficulties due to the gauge nature of the corresponding actions and already discussed ambiguities related to ordering problems; see Refs. [19–23].

One ought to say that quantization in arbitrary (curvilinear) coordinates in flat spaces or in general coordinates in curved spaces reveals more problems, in particular, the problem of fixing the right hand side of Eq. (3), the problem of a realization of the commutation relations for the operators $\hat{x}$ and $\hat{p}$ and the above mentioned ordering problem. The quantization in arbitrary coordinate system in a configuration space attracted attention already in early days of quantum mechanics [10,24] and until present is regularly discussed in the literature; see e.g. Refs. [25–34].

To illustrate the problem, the case of a massive particle moving in a spherically symmetric field is usually considered. The corresponding classical and quantum Hamiltonians written in Cartesian coordinates are

$$H^{\text{C}}(\mathbf{r}, \mathbf{p}) = \frac{\mathbf{p}^2}{2m} + U(r) \Longrightarrow \hat{H}^{\text{C}} = -\frac{\hbar^2}{2m}\nabla^2 + U(r). \tag{12}$$

---

[5] We do not mention work in which noncovariant quantum Hamiltonians were derived.

However, taking into account the field symmetry, the use of polar coordinates is preferable. Being transformed to these coordinates $(x'^1, x'^2, x'^3) = (r, \theta, \varphi)$, the quantum Hamiltonian (12) takes the form

$$\hat{H}^{\text{P}} = -\frac{\hbar^2}{2m}\frac{1}{\sqrt{g'}}\partial_{\mu'}\sqrt{g'}g^{\mu'\nu'}\partial_{\nu'} + U(r), \tag{13}$$

where $g^{\mu'\nu'}$ and $g' = |\det g^{\mu'\nu'}|$ are the metric tensor and its determinant in polar coordinates. Its kinetic part is proportional to the Laplace–Beltrami operator. On the other side, we could start with the classical Hamiltonian, written from the beginning in polar coordinates,

$$\begin{aligned} H^{\text{P}}(r, \theta, \varphi) &= \frac{1}{2m}g^{\mu'\nu'}p_{\mu'}p_{\nu'} + U(r) \\ &= \frac{1}{2m}[p_r^2 + r^{-2}(p_\theta^2 + p_\varphi^2\sin^{-2}\theta)] + U(r). \end{aligned} \tag{14}$$

Then there appears the first question: how do we transform the classical function $H^{\text{P}}(r, \theta, \varphi)$, given by Eq. (14), into the corresponding quantum operator? If we make the naive replacement $r, \theta, \varphi \to r, \theta, \varphi$, and $p_r \to -i\hbar\partial_r$, $p_\theta \to -i\hbar\partial_\theta$, $p_\varphi \to -i\hbar\partial_\varphi$, in this, we obtain an operator, which obviously differs from $\hat{H}^{\text{P}}$ given by Eq. (13). Moreover, since the classical Hamiltonian $H(r, \theta, \varphi)$ is an essentially nontrivial function of noncommuting operators in polar coordinates and their momenta, an ordering problem appears real. Historically, Schrödinger [35,36] and Eckart [37] chose the Hamiltonian for the hydrogen atom in the Cartesian coordinates in the form (12) and then obtained (13) as done above. Then Dirac [38] in agreement with Pauli [39] derived Hamiltonian (13) quantizing classical function (14) under special assumptions. Afterwards, the same result was obtained in different ways in many publications; see e.g. [24].

Quantizations in curved spaces meet additional difficulties in the realization of the above program and impose additional restrictions on the ordering rules. Namely, the general covariance has to be maintained on the quantum level. And here one meets a serious problem with all the known orderings that were formulated in plane spaces and in Cartesian coordinates.

Our article represents an attempt to treat all the above mentioned problems from a unique point of view.[6] It is supposed that a Hamiltonian formulation of the classical theory of the physical system under consideration does exist and there are no constraints in the phase space. To start with, we construct a family of covariant quantizations in flat spaces and Cartesian coordinates. This family is parametrized by a function $\omega(\theta)$, $\theta \in (1, 0)$. Then we represent a minimal covariant generalization of the quantizations in flat spaces to curved spaces. This generalization is parametrized by the same function $\omega(\theta)$. Then we construct an extended family of covariant quantizations in curved space. This family is already parametrized by two functions, the previous one $\omega(\theta)$ and by an additional function $\Theta(x, \xi)$. The minimal generalization represents a part of the extended family of quantizations and corresponds to the extended quantizations with $\Theta = 1$. We study constructed quantizations in detail, proving their consistency, covariance and classical limit. As physical applications, we consider the above mentioned old problem of constructing a quantum Hamiltonian in polar coordinates and a quantization of a non-relativistic particle moving in a curved space.

1.2 Covariance of a quantization

Let a physical system in a reference frame $K_x$ is described by coordinates $x = (x^a, a = 1, \ldots, D)$ and its classical theory be given by a nonsingular Lagrangian[7] $L(x, \dot{x}, t)$. Then the corresponding momenta are $p = (p_a = \partial L/\partial \dot{x}^a)$. In a reference frame $K_{x'}$ with coordinates $x' = (x'^b, b = 1, \ldots, D)$, related to the coordinates $x$ by an arbitrary invertible time-independent coordinate transformation $x'^b = \varphi^b(x)$, the same system is described classically by the Lagrangian $L'(x', \dot{x}', t)$, which is related to the Lagrangian $L(x, \dot{x}, t)$ as a scalar,

$$L'(x', \dot{x}', t) = L(x, \dot{x}, t) = L\left(\varphi^{-1}(x'), \frac{\partial x}{\partial x'^a}(x')\dot{x}'^a\right). \tag{15}$$

The momenta $p'_b = \partial L'/\partial \dot{x}'^b$ are related to the momenta $p_a$ as $p'_b = p_\alpha \partial x^\alpha/\partial x'^b$.

Transformations

$$x'^b = \varphi^b(x), \quad p'_b = p_\alpha \frac{\partial x^\alpha}{\partial x'^b}(x) \tag{16}$$

are called point-like canonical transformations, they form a subgroup of all time-independent canonical transformations [25]. Classical functions $F'(x', p')$, which describe physical quantities in $K_{x'}$, are related to classical functions $F(x, p)$, which describe the same physical quantities in $K_x$, as scalars, $F'(x', p') = F(x, p)$.

Let, in an arbitrary reference frame $K_x$, a quantization $Q$ be constructed and a transformation law be given, which allows one to restore a quantization $Q'$ in any other (different from $K_x$) arbitrary reference frame $K_{x'}$. Let $\mathfrak{H}$ and $\mathfrak{H}'$ be Hilbert spaces corresponding to the quantizations $Q$ and $Q'$. Then if vectors $|\psi'\rangle \in \mathfrak{H}'$ and vectors $|\psi\rangle \in \mathfrak{H}$ that represent the same physical states are related by a unitary operator $U$, i.e., $|\psi'\rangle = U|\psi\rangle$ and quantizations $Q$ and $Q'$ of the same physical quantities $F$ and $F'$ are related as

$$Q'(F') = U Q(F) U^{\dagger}, \quad (17)$$

then the quantum theories generated by the quantizations $Q$ and $Q'$ are equivalent and we say that the quantization $Q$ is covariant under transformations (16).

---

[6] It should be noted that a part of the results presented below are a compilation of the contents of the Ph.D. dissertation [40], written by J. L. M. Assirati under D. Gitman's supervision.

[7] In particular, systems with nonsingular Lagrangians do not have constraints in Hamiltonian formulation; see [3].

It follows from Eq. (16) that a quantization $Q$ from a covariant set and a unitary operator $U$ from Eq. (17) are related by the conditions

$$\hat{x}'^a = U Q(\varphi^a(x)) U^{\dagger}, \quad \hat{p}'_b = U Q\left(p_\alpha \frac{\partial x^\alpha}{\partial x'^b}(x)\right) U^{\dagger}. \quad (18)$$

Moreover, Eq. (18) determines the operator $U$ uniquely up to a phase factor. Indeed, let us suppose that there exists another unitary operator $V : \mathfrak{H} \to \mathfrak{H}'$ which also satisfies Eq. (18). Multiplying the first equation of (18) by $UV^{\dagger}$ from the right and inserting an identity $V^{\dagger}V$, we obtain $\hat{x}'^a UV^{\dagger} = UV^{\dagger}VQ(\varphi^a(x))V^{\dagger} = UV^{\dagger}\hat{x}'^a$, and therefore $[UV^{\dagger}, \hat{x}'^a] = 0$. In the same manner, we can derive $[UV^{\dagger}, \hat{p}'_a] = 0$ from the second Eq. (18). But the Hilbert space $\mathfrak{H}'$ must be irreducible with respect to the canonical operators $\hat{x}'^a$ and $\hat{p}'_a$. Therefore, $UV^{\dagger} = \lambda \hat{I}$, or $U = \lambda V$. As both operators are unitary, $V = e^{i\phi}U$.

Let, the configuration space of a physical system is a plane $D$-dimensional pseudo-Euclidean space, which allows one to introduce a Cartesian coordinate system[8] $K_x^C$, where the metric tensor $g_{\mu\nu}$ takes the form,

$$g_{ab} = \mathrm{diag}\left(\underbrace{1, 1, \ldots, 1}_{r}, \underbrace{-1, -1, \ldots, -1}_{s}\right). \quad (19)$$

There exist an infinite number of Cartesian coordinates $x'$ (coordinate systems $K_{x'}^C$), and the corresponding coordinate transformations between them $x'^\alpha = M^\alpha{}_b x^b$ form orthogonal ($s = 0$) or pseudo-orthogonal ($s \neq 0$) group $O(r, s)$, $K_x^C \xrightarrow{O(r,s)} K_{x'}^C$. Matrix elements $M_b^a$ satisfy the well-known conditions $M^\alpha{}_a g_{\alpha\beta} M^\beta{}_b = g_{ab}$, which means invariance of the metric tensor (19) under the transformations from the group $O(r, s)$. In a coordinate system $K_{x'}^C$, the same system is described by the Lagrangian $L'(x', \dot{x}')$ which is related to the Lagrangian in $K_x^C$ as a scalar,

$$L'(x', \dot{x}') = L(x, \dot{x}) = L(M^{-1}x', M^{-1}\dot{x}'). \quad (20)$$

Momenta $p'_a = \partial L'/\partial \dot{x}'^a$ in $K_{x'}^C$ are related to momenta $p_b = \partial L/\partial \dot{x}^b$ in $K_x$ as $p'_a = p_b (M^{-1})^b{}_a$. The transformations

$$x'^\alpha = M^\alpha{}_b x^b, \quad p'_a = p_b (M^{-1})^b{}_a, \quad (21)$$

of the phase-space variables form a subgroup of the group of the point-like canonical transformations (16).

## 2 Covariant quantizations in flat spaces and Cartesian coordinates

### 2.1 Covariant $Q_\Omega$ quantizations

In this section, it is supposed that a classical theory, which we are going to quantize, is described by coordinates $x$ generated by a Cartesian coordinate system $K_x^C$ in a flat space. Here and in what follows, the upper script C means that the corresponding quantity belongs to a Cartesian coordinate system.

Let a function $F(x, p)$ describing a physical quantity that can be presented by the Fourier integral,

$$F(x, p) = \int \widetilde{F}(\eta, \xi) e^{i(\eta x + p\xi)} d\eta d\xi,$$
$$\widetilde{F}(\eta, \xi) = (2\pi)^{-2D} \int F(x, p) e^{-i(\eta x + p\xi)} dx dp, \quad (22)$$

where $\tau = (\tau_a)$, $s = (s^a)$, $\tau x = \tau_a x^a$, $ps = p_a s^a$, and $\widetilde{F}(\tau, s)$ is a double Fourier transform of $F(x, p)$ with respect to $x$ and $p$.

In this case, defining a quantization $Q^c(e^{i(\eta x + p\xi)})$, we, due to the linearity of $Q^c$, define the quantization $Q^c$ for all the functions $F(x, p)$ of the form (22),

$$Q^c(F) = \int \widetilde{F}(\eta, \xi) Q^c(e^{i(\eta x + p\xi)}) d\eta d\xi. \quad (23)$$

In the general case, there are an infinite number of different quantizations $Q^c(e^{i(\eta x + p\xi)})$. Let us consider a family of quantizations $Q_{\boldsymbol{\Omega}}^c$ that are parametrized by a weight function $\boldsymbol{\Omega}(\eta, \xi, \alpha, \beta)$ of real variables $\eta, \xi, \alpha$, and $\beta$ as follows:

$$Q_{\boldsymbol{\Omega}}^c(e^{i(\eta x + p\xi)}) = \int \boldsymbol{\Omega}(\eta, \xi, \alpha, \beta) e^{i(\alpha \hat{x} + \hat{p}\beta)} d\alpha d\beta. \quad (24)$$

Then

$$Q_{\boldsymbol{\Omega}}^c(F) = \int \widetilde{F}(\eta, \xi) \boldsymbol{\Omega}(\eta, \xi, \alpha, \beta) e^{i(\alpha \hat{x} + \hat{p}\beta)} d\alpha d\beta d\eta d\xi. \quad (25)$$

However, the correspondence principle imposes restrictions on possible weight functions $\boldsymbol{\Omega}(\eta, \xi, \alpha, \beta)$. It must have the form, see Sect. 2.1.1,

$$\boldsymbol{\Omega}(\eta, \xi, \alpha, \beta) = \boldsymbol{\Omega}(\eta, \xi) \delta(\alpha - \eta) \delta(\beta - \xi). \quad (26)$$

Additional restrictions follow from covariance considerations; see Sect. 2.1.2. Namely, functions $\Omega(\eta, \xi)$ must depend on their arguments $\eta$ and $\xi$ via the scalar product $\alpha = \eta \xi = \eta_a \xi^a$ only, namely, $\boldsymbol{\Omega}(\eta, \xi) = \Omega(k)$, $k = \hbar \eta \xi$,

---

[8] Here and in what follows, we denote a Cartesian coordinate system by an superscript C.

and satisfy the initial condition $\Omega(0) = 1$. Thus, any quantization

$$Q^c_\Omega(F) = \int \widetilde{F}(\eta, \xi) \Omega(\hbar\eta\xi) e^{i(\eta\hat{x}+\hat{p}\xi)} d\eta d\xi, \quad \Omega(0) = 1, \tag{27}$$

satisfies the correspondence principle and is covariant with respect to the group $O(r, s)$.

### 2.1.1 Restrictions form the correspondence principle

I. Let us consider quantization (25). One can easily verify that conditions (we call them Heisenberg conditions)

$$\begin{aligned}
\left[\hat{x}^a, Q^c_\Omega(F)\right] &= i\hbar Q^c_\Omega\left(\frac{\partial F}{\partial p_a}\right), \\
\left[\hat{p}_a, Q^c_\Omega(F)\right] &= -i\hbar Q^c_\Omega\left(\frac{\partial F}{\partial x^a}\right),
\end{aligned} \tag{28}$$

provide the classical limit of the Heisenberg equations with the Hamiltonian $\hat{H} = Q^c_\Omega(H)$,

$$\dot{\hat{x}}^a = Q^c_\Omega\left(\frac{\partial H}{\partial p_a}\right), \quad \dot{\hat{p}}_a = -Q^c_\Omega\left(\frac{\partial H}{\partial x^a}\right). \tag{29}$$

Using the fact that the Fourier transform (22) satisfies the relation

$$\frac{\partial \widetilde{\widetilde{F}}}{\partial p_a}(\eta, \xi) = i\xi^a \widetilde{\widetilde{F}}(\eta, \xi), \tag{30}$$

we can write the first equation of (29) as follows:

$$\begin{aligned}
\left[\hat{x}^a, Q^c_\Omega(F)\right] = -\hbar \int \xi^a \widetilde{\widetilde{F}}(\eta, \xi) \boldsymbol{\Omega}(\eta, \xi, \alpha, \beta) \\
\times e^{i(\alpha\hat{x}+\hat{p}\beta)} d\alpha d\beta d\eta d\xi.
\end{aligned} \tag{31}$$

On the other side, it follows from the first equation of (28) that

$$\begin{aligned}
\left[\hat{x}^a, Q^c_\Omega(F)\right] = -\hbar \int \beta^a \widetilde{\widetilde{F}}(\eta, \xi) \boldsymbol{\Omega}(\eta, \xi, \alpha, \beta) \\
\times e^{i(\alpha\hat{x}+\hat{p}\beta)} d\alpha d\beta d\eta d\xi.
\end{aligned} \tag{32}$$

Since $F$ is, in a sense, an arbitrary function, we obtain $(\beta^a - \xi^a) \boldsymbol{\Omega}(\eta, \xi, \alpha, \beta) = 0$. In the same manner, we obtain $(\alpha_a - \eta_a) \boldsymbol{\Omega}(\eta, \xi, \alpha, \beta) = 0$. These equations imply that $\boldsymbol{\Omega}(\eta, \xi, \alpha, \beta) = \boldsymbol{\Omega}(\eta, \xi)\delta(\beta - \xi)\delta(\alpha - \eta)$, justifying Eq. (26).

Usually the following initial conditions for the weight function $\boldsymbol{\Omega}(\eta, \xi)$ are assumed

$$\boldsymbol{\Omega}(\eta, 0) = \boldsymbol{\Omega}(0, \xi) = 1. \tag{33}$$

Their fulfillment guarantees the so-called von Neumann properties of a quantization [1],

$$Q^c_\Omega(F(x)) = F(\hat{x}), \quad Q^c_\Omega(G(p)) = G(\hat{p}), \tag{34}$$

and, in particular, conditions (9). To be more precise, the operators $F(\hat{x})$ and $G(\hat{p})$ in Eq. (34) are understood as follows:[9]

$$\begin{aligned}
F(\hat{x}) &= \int \tilde{F}(\eta) e^{i\eta\hat{x}} d\eta, \quad \tilde{F}(\eta) = (2\pi)^{-D} \int F(x) e^{-i\eta x} dx, \\
G(\hat{p}) &= \int \tilde{F}(\xi) e^{i\xi\hat{p}} d\eta, \quad \tilde{G}(\xi) = (2\pi)^{-D} \int G(p) e^{-i\xi p} dx,
\end{aligned} \tag{35}$$

where partial Fourier transforms $\tilde{F}(\eta)$ and $\tilde{G}(\xi)$ are denoted by only one tilde above.

It should be noted that here exist more general conditions than von Neumann ones (34), which have to be fulfilled for any quantization $Q^c_\Omega$. Indeed, let for a given index $a$ the classical Hamiltonian $H(x, p)$ does not depend on both canonical variables $x^a$ and $p_a$, i.e., $\partial H/\partial x^a = \partial H/\partial p_a = 0$. Then it is natural to suppose that the quantization $Q^c_\Omega$ must satisfy the following generalized conditions (we call them the generalized von Neumann conditions):

$$\begin{aligned}
Q^c_\Omega\left(F(x^a) H(x, p)\right) &= F(\hat{x}^a) Q^c_\Omega(H) = Q^c_\Omega(H) F(\hat{x}^a), \\
Q^c_\Omega\left(G(p_a) H(x, p)\right) &= G(\hat{p}_a) Q^c_\Omega(H) = Q^c_\Omega(H) G(\hat{p}_a).
\end{aligned} \tag{36}$$

It is easily to see that the latter conditions, together with the property $Q^c_\Omega(1) = \hat{I}$, imply Eq. (34).

However, the initial conditions (33) on weight functions $\boldsymbol{\Omega}(\eta, \xi)$ are already not sufficient to provide Eq. (36). The adequate generalization of conditions (33) reads

$$\begin{aligned}
\frac{\partial}{\partial \xi^a}\left[\boldsymbol{\Omega}(\eta, \xi)|_{\eta_a=0}\right] &= \frac{\partial}{\partial \eta_a}\left[\boldsymbol{\Omega}(\eta, \xi)|_{\xi^a=0}\right] = 0, \\
\boldsymbol{\Omega}(0, 0) &= 1.
\end{aligned} \tag{37}$$

### 2.1.2 Restrictions from the covariance

Let us consider a function $E(x, p) = e^{i(\eta x + p\xi)}$. In terms of transformed by (21) phase-space variables $x'$ and $p'$, this function reads $E(x, p) = E'(x', p') = e^{i(\eta' x' + p'\xi')}$, where $\eta'_\alpha = \eta_\mu (M^{-1})^\mu{}_\alpha$ and $\xi'^\alpha = M^\alpha{}_\mu \xi^\mu$. According to Eqs. (24) and (26),

$$\begin{aligned}
Q^{c'}_\Omega(E') &= \boldsymbol{\Omega}(\eta', \xi') e^{i(\eta'\hat{x}'+\hat{p}'\xi')} \\
&= \boldsymbol{\Omega}(\eta', \xi') U e^{i(\eta'M\hat{x}+\hat{p}M^{-1}\xi')} U^\dagger \\
&= U \left[\boldsymbol{\Omega}(\eta M^{-1}, M\xi) e^{i(\eta\hat{x}+\hat{p}\xi)}\right] U^\dagger.
\end{aligned}$$

On the other side, see definition (17), quantizations $Q^c_\Omega$ and $Q^{c'}_\Omega$ are equivalent if $\boldsymbol{\Omega}(\eta M^{-1}, M\xi) = \boldsymbol{\Omega}(\eta, \xi)$. Because $\xi$, $\eta$ and $M \in O(r, s)$ are arbitrary, the function $\boldsymbol{\Omega}$ must depend on the invariants $I_1(\xi) = g_{ab}\xi^a\xi^b$, $I_2(\eta) = g^{ab}\eta_a\eta_b$ and $I_3(\eta, \xi) = \eta_a\xi^a = \eta\xi$ only,

$$\mathbf{\Omega}(\eta, \xi) = f(I_1(\xi), I_2(\eta), I_3(\eta, \xi)). \tag{38}$$

---

[9] When $F$ and $G$ are polynomials this definition is reduced to a simple substitution $x \to \hat{x}$ and $p \to \hat{p}$ in $F(x)$ and $G(p)$, respectively.

For functions of the form (38), two first conditions (37) imply

$$\forall \xi^a, \eta_a : \xi^a \frac{\partial f}{\partial I_1} = \eta_a \frac{\partial f}{\partial I_2} = 0 \Longrightarrow \frac{\partial f}{\partial I_1} = \frac{\partial f}{\partial I_2} = 0. \tag{39}$$

Thus, for $Q^c_{\mathbf{\Omega}}$ to be covariant with respect to the group transformations $O(r, s)$, $\mathbf{\Omega}(\eta, \xi)$ must be a function of the invariant combination $I_3(\eta, \xi) = \eta_a\xi^a = \eta\xi$ only,

$$\mathbf{\Omega}(\eta, \xi) = \Omega(k), \quad k = \hbar\eta\xi. \tag{40}$$

One can easily verify that initial condition (37) for the function $\mathbf{\Omega}(\eta, \xi)$ holds when the initial condition $\Omega(0) = 1$ holds for the function $\Omega(k)$.

Thus, the final statement of Sect. 2.1 is proved.

### 2.2 Basic quantizations

Let a weight function $\Omega(k)$ have the form

$$\Omega(k) = \Omega_{(\theta)}(k) = e^{i\theta k} = e^{i\hbar\theta\eta\xi}.$$

It defines a covariant quantization $Q^c_{\Omega_{(\theta)}} \equiv Q^c_{(\theta)}$,

$$Q^c_{(\theta)}(F) = \int \widetilde{\widetilde{F}}(\eta, \xi) e^{i\hbar\theta\eta\xi} e^{i(\eta\hat{x}+\hat{p}\xi)} d\eta d\xi. \tag{41}$$

In what follows, we call (41) the basic quantizations.

The basic quantization $Q^c_{(\theta)}$ can be written in a more symmetric form. To this end, we use the relation

$$e^{i(\eta\hat{x}+\hat{p}\xi)} = e^{-i\hbar\theta\eta\xi} e^{i\left(\frac{1}{2}+\theta\right)\hat{p}\xi} e^{i\eta\hat{x}} e^{i\left(\frac{1}{2}-\theta\right)\hat{p}\xi}, \tag{42}$$

which can be obtained from the Baker–Campbell–Hausdorff formula $e^{\hat{A}+\hat{B}} = e^{-\frac{1}{2}[\hat{A},\hat{B}]} e^{\hat{A}} e^{\hat{B}}$. Thus,

$$Q^c_{(\theta)}(F) = \int \widetilde{\widetilde{F}}(\eta, \xi) e^{i\left(\frac{1}{2}+\theta\right)\hat{p}\xi} e^{i\eta\hat{x}} e^{i\left(\frac{1}{2}-\theta\right)\hat{p}\xi} d\eta d\xi. \tag{43}$$

The basic quantizations $Q^c_{(\theta)}(F)$ are covariant for any $\theta$.

However, the basic quantizations are hermitian only for $\theta = 0$. Indeed, using the formal hermiticity of the operators $\hat{x}$ and $\hat{p}$ and the property $(\widetilde{\widetilde{F}}(\eta, \xi))^* = \widetilde{\widetilde{F}}^*(-\eta, -\xi)$ of the double Fourier transforms, we obtain the relation

$$[Q^c_{(\theta)}(F)]^\dagger = \int \widetilde{\widetilde{F}}^*(-\eta, -\xi) e^{-i\hbar\theta\eta\xi} e^{-i(\eta\hat{x}+\hat{p}\xi)} d\eta d\xi$$
$$= \int \widetilde{\widetilde{F}}^*(\eta, \xi) e^{-i\hbar\theta\eta\xi} e^{i(\eta\hat{x}+\hat{p}\xi)} d\eta d\xi = Q^c_{(-\theta)}(F^*), \tag{44}$$

which proves the above assertion.

The importance of the basic quantizations is stressed by the following assertion: Any covariant with respect to a group $O(r, s)$ quantization $Q^c_{\Omega}(F)$ can be presented as an integral over basic quantizations.

To demonstrate this fact, we introduce the Fourier transform $\omega(\theta)$, $\theta \in \mathbb{R}$ of a weight function $\Omega(k)$,

$$\Omega(k) = \int_{-\infty}^{+\infty} \omega(\theta) e^{ik\theta} d\theta,$$
$$\omega(\theta) = \frac{1}{2\pi} \int_{-\infty}^{+\infty} \Omega(k) e^{-i\theta k} dk. \tag{45}$$

Weight functions that define covariant quantizations satisfy initial condition (41). As a consequence, Fourier transforms of such weight functions are normalized,

$$\Omega(0) = 1 \Longrightarrow \int_{-\infty}^{+\infty} \omega(\theta) d\theta = 1. \tag{46}$$

In other words, Fourier transforms of weight functions that define covariant quantizations can be treated as complex measures $\omega(\theta)$ of the total weight 1 on the real line.

Substituting $\Omega(k)$ from Eq. (45) into Eq. (27), we obtain

$$Q^c_{\Omega}(F) = \int_{-\infty}^{+\infty} \int \widetilde{\widetilde{F}}(\eta, \xi) e^{i\hbar\theta\alpha} e^{i(\eta\hat{x}+\hat{p}\xi)} d\eta d\xi \omega(\theta) d\theta$$
$$= \int_{-\infty}^{+\infty} Q^c_{(\theta)}(F) \omega(\theta) d\theta = Q^c_{[\omega]}(F). \tag{47}$$

Thus, any covariant quantization with a weight function $\Omega(k)$, can be represented either by Eq. (27) or by Eqs. (47) and (46) via the Fourier transform of the weight function $\Omega(k)$. In the latter case, we denote the corresponding quantization as $Q^c_{[\omega]}(F)$.

If $\omega(\theta)$ is the Fourier transform of a weight function $\Omega(k)$, then there is the following correspondence between the introduced notation, $Q^c_{[\omega]}(F) = Q^c_{\Omega}(F)$.

It follows from Eq. (47) that any quantization $Q^c_{[\omega]}(F)$ with $\omega^*(-\theta) = \omega(\theta)$ is hermitian,

$$\left.\begin{array}{l} Q^c_{[\omega]}(F) = \int_{-\infty}^{+\infty} Q^c_{(\theta)}(F) \omega(\theta) d\theta \\ \omega^*(-\theta) = \omega(\theta) \end{array}\right\} \Longrightarrow [Q^c_{[\omega]}(F)]^\dagger$$
$$= Q^c_{[\omega]}(F^*). \tag{48}$$

It follows from Eq. (47) that

$$Q^c_{[\omega]}(F)\big|_{\omega(\theta)=\delta(\theta-\theta')} = Q^c_{(\theta')}(F). \tag{49}$$

### 2.3 Quantizations of homogeneous polynomials

Consider functions $F(x, p)$ of the phase-space variables that have the following form:

$$F(x, p) = T^{a_1 \cdots a_n}(x) p_{a_1} \cdots p_{a_n}, \tag{50}$$

where $x$-dependent coefficients $T^{a_1 \cdots a_n}(x)$ are symmetric under any change of the indices $a_1 \cdots a_n$. We call such functions homogeneous polynomials in what follows. One can see that the Fourier transform $\widetilde{\widetilde{F}}(\eta, \xi)$ of a homogeneous polynomial (50) can be written as

$$\widetilde{\widetilde{F}}(\xi, \eta) = i^n \tilde{T}^{a_1 \cdots a_n}(\eta) \frac{\partial^n \delta(\xi^{a_1} \xi^{a_2} \cdots \xi^{a_n})}{\partial \xi^{a_1} \cdots \partial \xi^{a_n}}. \quad (51)$$

Substituting the latter representation for $\widetilde{\widetilde{F}}(\xi, \eta)$ into Eq. (43), we obtain the basic quantization of $F(x, p)$ in the following form:

$$\begin{aligned} Q^c_{(\theta)}(F) &= i^n \int \frac{\partial^n \delta(\xi^{a_1} \xi^{a_2} \cdots \xi^{a_n})}{\partial \xi^{a_1} \cdots \partial \xi^{a_n}} e^{i\left(\frac{1}{2}+\theta\right)\hat{p}\xi} \\ &\quad \times \left( \int \tilde{T}^{a_1 \cdots a_n}(\eta) e^{i\eta\hat{x}} \mathrm{d}\eta \right) e^{i\left(\frac{1}{2}-\theta\right)\hat{p}\xi} \mathrm{d}\xi \\ &= (-i)^n \frac{\partial^n}{\partial \xi^{a_1} \cdots \partial \xi^{a_n}} \left[ e^{i\left(\frac{1}{2}+\theta\right)\hat{p}\xi} T^{a_1 \cdots a_n}(\hat{x}) e^{i\left(\frac{1}{2}-\theta\right)\hat{p}\xi} \right]_{\xi=0}. \end{aligned} \quad (52)$$

Using a generalization of the Leibnitz formula (see Appendix A.2), we can write

$$\begin{aligned} &\frac{\partial^n}{\partial \xi^{a_1} \cdots \partial \xi^{a_n}} \left[ e^{i\left(\frac{1}{2}+\theta\right)\hat{p}\xi} T^{a_1 \cdots a_n}(\hat{x}) e^{i\left(\frac{1}{2}-\theta\right)\hat{p}\xi} \right] \\ &= \sum_{k=0}^{n} \binom{n}{k} \left[ \frac{\partial^n \exp i\left(\frac{1}{2}+\theta\right)\hat{p}\xi}{\partial \xi^{a_1} \cdots \partial \xi^{a_k}} \right] T^{a_1 \cdots a_n}(\hat{x}) \\ &\quad \times \left[ \frac{\partial^n \exp i\left(\frac{1}{2}-\theta\right)\hat{p}\xi}{\partial \xi^{a_{k+1}} \cdots \partial \xi^{a_n}} \right]. \end{aligned} \quad (53)$$

Then it follows from (52) that the basic quantization $Q^c_{(\theta)}$ of a homogeneous polynomial has the form

$$Q^c_{(\theta)}(F) = \sum_{k=0}^{n} C^n_k(\theta) \hat{p}_{a_1} \cdots \hat{p}_{a_k} T^{a_1 \cdots a_n}(\hat{x}) \hat{p}_{a_{k+1}} \cdots \hat{p}_{a_n}, \quad (54)$$

where

$$C^n_k(\theta) = \binom{n}{k} \left(\frac{1}{2}+\theta\right)^k \left(\frac{1}{2}-\theta\right)^{n-k}, \\ C^n_k(-\theta) = C^n_{n-k}(\theta). \quad (55)$$

According to Eq. (47) the quantization $Q^c_{[\omega]}$ of the homogeneous polynomial reads

$$Q^c_{[\omega]}(F) = \sum_{k=0}^{n} \omega^n_k \hat{p}_{a_1} \cdots \hat{p}_{a_k} T^{a_1 \cdots a_n}(\hat{x}) \hat{p}_{a_{k+1}} \cdots \hat{p}_{a_n}, \quad (56)$$

where

$$\omega^n_k = \int_{-\infty}^{+\infty} C^n_k(\theta) \omega(\theta) \mathrm{d}\theta. \quad (57)$$

The sum $\sum_{k=0}^{n} C^n_k(\theta)$ is reduced to a unit according to the binomial theorem,

$$\sum_{k=0}^{n} C^n_k(\theta) = \left[ \left(\frac{1}{2}+\theta\right) + \left(\frac{1}{2}-\theta\right) \right]^n = 1. \quad (58)$$

Together with the normalization (46) this implies that

$$\sum_{k=0}^{n} \omega^n_k = \int_{-\infty}^{+\infty} \sum_{k=0}^{n} C^n_k(\theta) \omega(\theta) \mathrm{d}\theta = 1. \quad (59)$$

For a hermitian quantization $Q^c_{[\omega]}$ (see Eq. (48)) the coefficients $\omega^n_k$ have the properties

$$\begin{aligned} \left(\omega^n_k\right)^* &= \int_{-\infty}^{+\infty} C^n_k(\theta) \omega(-\theta) \mathrm{d}\theta \\ &= \int_{-\infty}^{+\infty} C^n_k(-\theta) \omega(\theta) \mathrm{d}\theta = \omega^n_{n-k}. \end{aligned} \quad (60)$$

### 2.4 Factorization property of covariant quantizations

In the previous sections, we have considered Heisenberg conditions (28), the generalized von Neumann conditions (36), and conditions that provide the covariance and self-adjointness of a quantization. However, it is possible to require some additional properties for a quantization and derive corresponding formal conditions that provide these properties. Let us consider some of them.

Let a function $F(x, p)$ be represented as a product of two functions $G(x^1, p_1, \ldots, x^k, p_k)$ and $H(x^{k+1}, p_{k+1}, \ldots x^D, p_D)$,

$$F(x, p) = G(x^1, p_1, \ldots, x^k, p_k) H(x^{k+1}, p_{k+1}, \ldots x^D, p_D). \quad (61)$$

It is easily to verify that any quantization $Q^c_{[\omega]}(F)$ given by the rules (47) produces, in this case, commuting operators $Q^c_{[\omega]}(G)$ and $Q^c_{[\omega]}(H)$. Then we ask the question: which quantizations satisfies the following property:

$$Q^c(F) = Q^c(G) Q^c(H), \quad (62)$$

which we call the factorization property in what follows.

Let us derive the corresponding formal conditions on functions $\omega(\theta)$ that provide the latter property.

First, taking the structure of the functions $G$, $H$, and $F$ into account, we find that their Fourier transforms have the forms

$$\begin{aligned} \widetilde{\widetilde{G}}(\eta, \xi) &= \bar{G}(\eta_1, \xi^1, \ldots, \eta_k, \xi^k) \delta(\eta_{k+1}) \\ &\quad \times \delta(\xi^{k+1}) \ldots \delta(\eta_D) \delta(\xi^D), \\ \widetilde{\widetilde{H}}(\eta, \xi) &= \bar{H}(\eta_{k+1}, \xi^{k+1}, \ldots, \eta_D, \xi^D) \delta(\eta_1) \\ &\quad \times \delta(\xi^1) \ldots \delta(\eta_k) \delta(\xi^k), \\ \widetilde{\widetilde{F}}(\eta, \xi) &= \bar{G}(\eta_1, \xi^1, \ldots, \eta_k, \xi^k) \\ &\quad \times \bar{H}(\eta_{k+1}, \xi^{k+1}, \ldots, \eta_D, \xi^D). \end{aligned} \quad (63)$$

Then we consider the basic quantization $Q^c_{(\theta)}$ of functions (61). It is easily to verify that it satisfies the property

$$Q^{\mathrm{c}}_{(\theta)}(F) = \int \bar{G} e^{i\hbar\theta \sum_{a=1}^{k} \eta_a \xi^a} e^{i \sum_{a=1}^{k}(\eta_a \hat{x}^a + \hat{p}_a \xi^a)}$$

$$\times \prod_{a=1}^{k} d\eta_a d\xi^a \int \bar{H} e^{i\hbar\theta \sum_{a=k+1}^{D} \eta_a \xi^a} e^{i \sum_{a=k+1}^{D}(\eta_a \hat{x}^a + \hat{p}_a \xi^a)}$$

$$\times \prod_{a=k+1}^{D} d\eta_a d\xi^a = Q^{\mathrm{c}}_{(\theta)}(G) Q^{\mathrm{c}}_{(\theta)}(H). \tag{64}$$

Thus, any basic quantization $Q^{\mathrm{c}}_{(\theta)}$ satisfies the factorization property.

Moreover, one can see that among quantizations $Q^{\mathrm{c}}_{\Omega}$, which can be represented by Eq. (47) as $Q^{\mathrm{c}}_{[\omega]}(F)$, only basic quantization $Q^{\mathrm{c}}_{(\theta)}$ satisfies the factorization property. Indeed, due to property (62) the quantization $Q^{\mathrm{c}}_{[\omega]}(F)$ of functions $F$ given by Eq. (61) can be represented as

$$Q^{\mathrm{c}}_{[\omega]}(F) = \int Q^{\mathrm{c}}_{(\theta)}(G) Q^{\mathrm{c}}_{(\theta)}(H) \omega(\theta) d\theta$$
$$= \int Q^{\mathrm{c}}_{(\theta_1)}(G) Q^{\mathrm{c}}_{(\theta_2)}(H) \omega(\theta_1) \delta(\theta_2 - \theta_1) d\theta_1 d\theta_2.$$

On the other side,

$$Q^{\mathrm{c}}_{[\omega]}(G) Q^{\mathrm{c}}_{[\omega]}(H) = \int Q^{\mathrm{c}}_{(\theta_1)}(G) Q^{\mathrm{c}}_{(\theta_2)}(H) \omega(\theta_1) \omega(\theta_2) d\theta_1 d\theta_2$$

Equation (54) implies that functions $G$ and $H$ can be chosen in such a way that the operator $Q^{\mathrm{c}}_{(\theta_1)}(G) Q^{\mathrm{c}}_{(\theta_2)}(H)$ turns out to be a polynomial (with operator coefficients) in $\theta_1$ and $\theta_2$ of any given degree. Thus, condition (62) can be satisfied only if $\omega(\theta_1)\omega(\theta_2) = \omega(\theta_1)\delta(\theta_2 - \theta_1)$. Then, since the relation $f(\theta_1)\delta(\theta_2-\theta_1) = f(\theta_2)\delta(\theta_2-\theta_1)$ holds true for any function $f(\theta)$, we obtain

$$f(\theta_1)\omega(\theta_1)\omega(\theta_2) = f(\theta_2)\omega(\theta_1)\omega(\theta_2).$$

Integrating this relation in $\theta_2$ and taking into account Eq. (46), we obtain

$$f(\theta_1)\omega(\theta_1) = k\omega(\theta_1), \quad k = \int_{-\infty}^{\infty} f(\theta)\omega(\theta)d\theta.$$

The latter equality holds for any $f(\theta)$ only if $\omega(\theta) = \delta(\theta - \lambda)$. Then $k = f(\lambda)$ and $Q^{\mathrm{c}}_{[\omega]}(F) = Q^{\mathrm{c}}_{(\lambda)}(F)$, which proves the statement.

## 2.5 Commonly discussed quantizations

I. The basic quantizations $Q^{\mathrm{c}}_{\left(\pm\frac{1}{2}\right)}(F)$ with $\theta = \pm\frac{1}{2}$ are, in fact, the well-known $px$ and $xp$ quantizations, in other words they produce $px$ and $xp$ ordered operators; see e.g. [7]. Indeed, using Eq. (43), in particular, Eq. (54), one can see that the quantization $Q^{\mathrm{c}}_{\left(\frac{1}{2}\right)}(F) = Q^{\mathrm{px}}(F)$ acting on classical functions $F(x, p)$ substitutes all $x$ by the operators $\hat{x}$, and all $p$ by the operators $\hat{p}$, placing at the same time all $\hat{p}$ on the left of all $\hat{x}$. The quantization $Q^{\mathrm{c}}_{\left(-\frac{1}{2}\right)}(F) = Q^{\mathrm{xp}}(F)$ acts similarly, placing all $\hat{x}$ on the left of all $\hat{p}$.

Using Eq. (43), we obtain

$$Q^{\mathrm{c}}_{\left(\frac{1}{2}\right)}(F) = Q^{\mathrm{px}}(F) = \int \widetilde{F}(\eta, \xi) e^{i\hat{p}\xi} e^{i\eta\hat{x}} d\eta d\xi,$$
$$Q^{\mathrm{c}}_{\left(-\frac{1}{2}\right)}(F) = Q^{\mathrm{xp}}(F) = \int \widetilde{F}(\eta, \xi) e^{i\eta\hat{x}} e^{i\hat{p}\xi} d\eta d\xi. \tag{65}$$

These quantizations are covariant and satisfy the factorization property. However, using Eq. (44), one can see that $(Q^{\mathrm{xp}}(F))^{\dagger} = Q^{\mathrm{px}}(F)$, which means that the quantizations $Q^{\mathrm{xp}}$ and $Q^{\mathrm{px}}$ are not hermitian in the general case.

II. The basic quantization with $\theta = 0 \iff \Omega = 1 \iff \omega(\theta) = \delta(\theta)$ is, in fact, the Weyl quantization [41], which is here denoted by $Q^{\mathrm{w}}(F)$,

$$Q^{\mathrm{w}}(F) = Q^{\mathrm{c}}_{(0)}(F) = Q^{\mathrm{c}}_1(F) = \int \widetilde{F}(\eta, \xi) e^{i(\eta\hat{x} + \xi\hat{p})} d\eta d\xi. \tag{66}$$

As follows from the above consideration, the Weyl quantization is covariant, satisfies the factorization property, is hermitian and produces symmetric in $\hat{x}$ and $\hat{p}$ operators. As follows from Eqs. (43) and (54) the Weyl quantization of homogeneous polynomials reads

$$Q^{\mathrm{w}}\left(T^{a_1 \cdots a_n}(x) p_{a_1} \cdots p_{a_n}\right)$$
$$= \frac{1}{2^n} \sum_{k=0}^{n} \binom{n}{k} \hat{p}_{a_1} \cdots \hat{p}_{a_k} T^{a_1 \cdots a_n}(\hat{x}) \hat{p}_{a_{k+1}} \cdots \hat{p}_{a_n}. \tag{67}$$

It should be noted that the Weyl quantization (the Weyl ordering) plays an important role in quantum mechanical applications. There is a convenient formula for the Weyl quantization,

$$Q^{\mathrm{w}}(F) = e^{\hat{x}\frac{\partial}{\partial t} + \hat{p}\frac{\partial}{\partial s}} F(t, s)\Big|_{t=s=0}. \tag{68}$$

III. In the article [10], Born and Jordan introduce a specific ordering, which is now called the Born–Jordan ordering or the Born–Jordan quantization. In terms of the introduced above constructions, the many-dimensional generalization of the Born–Jordan quantization $Q^{\mathrm{BJ}}(F)$ can be written as

$$Q^{\mathrm{BJ}}(F) = \int_{-\frac{1}{2}}^{+\frac{1}{2}} Q^{\mathrm{c}}_{(\theta)}(F) d\theta = \int_{-\infty}^{+\infty} Q^{\mathrm{c}}_{(\theta)}(F) \omega(\theta) d\theta$$
$$\implies \omega(\theta) = \begin{cases} 1, & |\theta| < \frac{1}{2} \\ 0, & |\theta| \geq \frac{1}{2} \end{cases} \implies \Omega(k)$$

$$= \int_{-\infty}^{+\infty} \omega(\theta) e^{ik\theta} d\theta = \frac{2}{k} \sin\left(\frac{k}{2}\right)$$

$$\implies Q^{\text{BJ}}(F) = \int \widetilde{\widetilde{F}}(\eta,\xi) \frac{2}{\hbar\eta\xi} \sin\left(\frac{\hbar\eta\xi}{2}\right)$$
$$\times e^{i(\eta\hat{x}+\hat{p}\xi)} d\eta d\xi. \tag{69}$$

The Born–Jordan quantization is covariant and hermitian but does not satisfy the factorization property. However, for two functions of the form $F = F(x)$ and $G = G(p)$ the following exact result holds:

$$[Q^{\text{BJ}}(F), Q^{\text{BJ}}(G)] = i\hbar Q^{\text{BJ}}(\{F,G\})$$
$$\equiv i\hbar Q^{\text{BJ}}\left(\frac{\partial F}{\partial x^a}\frac{\partial G}{\partial p_a}\right). \tag{70}$$

To verify Eq. (23), we use a relation, which holds for the double Fourier transform of the product $FG$,

$$\left(\frac{\partial F}{\partial x^a}\frac{\partial G}{\partial p_a}\right)^{\approx}(\eta,\xi) = -\eta_a \xi^a (FG)^{\approx}(\eta,\xi). \tag{71}$$

Here and in what follows, we often use the following notation for the double Fourier transform $\widetilde{\widetilde{(A)}}(\eta,\xi)$ of a function $A(x,p)$:

$$\widetilde{\widetilde{(A)}}(\eta,\xi) = (A)^{\approx}(\eta,\xi)$$

Then, taking into account the definition of the basic quantization, we can write

$$Q^{\text{c}}_{(\theta)}\left(\frac{\partial F}{\partial x^a}\frac{\partial G}{\partial p_a}\right) = -\int (\widetilde{\widetilde{FG}})\eta_a \xi^a e^{i\hbar\theta\eta\xi} e^{i(\eta\hat{x}+\hat{p}\xi)} d\eta d\xi$$
$$= -\frac{1}{i\hbar}\frac{\partial}{\partial\theta}Q^{\text{c}}_{(\theta)}(FG). \tag{72}$$

Using Eq. (72) in (70), we obtain

$$Q^{\text{BJ}}\left(\frac{\partial F}{\partial x^a}\frac{\partial G}{\partial p_a}\right) = \frac{1}{i\hbar}\left[Q^{\text{xp}}(FG) - Q^{\text{px}}(FG)\right]$$
$$= \frac{1}{i\hbar}\left[F(\hat{x}), G(\hat{p})\right], \tag{73}$$

which proves the validity of Eq. (70).

The Born–Jordan quantization of homogeneous polynomials is given by Eq. (56) where

$$\omega_k^n = \binom{n}{k}\int_{-\frac{1}{2}}^{+\frac{1}{2}}\left(\frac{1}{2}+\theta\right)^k\left(\frac{1}{2}-\theta\right)^{n-k} d\theta = \frac{1}{n+1}.$$

Thus,

$$Q^{\text{BJ}}\left(T^{a_1\cdots a_n}(x) p_{a_1}\cdots p_{a_n}\right)$$
$$= \frac{1}{n+1}\sum_{k=0}^n \hat{p}_{a_1}\cdots \hat{p}_{a_k} T^{a_1\cdots a_n}(\hat{x})\hat{p}_{a_{k+1}}\cdots \hat{p}_{a_n}. \tag{74}$$

### 2.6 Operator symbols and classical limit

For the Weyl ordering, we have $\Omega = 1$, such that the corresponding quantization is given by Eq. (66). Comparing it with the general formula (27), we see that

$$Q^{\text{c}}_\Omega(F) = Q^{\text{w}}(G), \quad \widetilde{\widetilde{G}}(\eta,\xi) = \Omega(\hbar\eta\xi)\widetilde{\widetilde{F}}(\eta,\xi). \tag{75}$$

One can see from definition (22) that

$$\Omega(-\hbar\Pi) F(x,p) = \int \Omega(\hbar\eta\xi)\widetilde{\widetilde{F}}(\eta,\xi) e^{i(\eta x+p\xi)} d\eta d\xi,$$

where

$$\Pi \equiv \sum_{\mu=1}^D \frac{\partial^2}{\partial x^\mu \partial p_\mu}. \tag{76}$$

Thus, we have

$$\Omega(\hbar\eta\xi)\widetilde{\widetilde{F}}(\eta,\xi) = [\Omega(-\hbar\Pi) F]^{\approx} \implies Q^{\text{c}}_\Omega(F)$$
$$= Q^{\text{w}}(\Omega(-\hbar\Pi) F). \tag{77}$$

We note that if $F(x,p)$ is a polynomial in $x$ or in $p$, there exists a number $n < \infty$ such that $\Delta^n F = 0$, and $\Omega(-\hbar\Pi) F$ always exist for such functions $F(x,p)$.

The operator $\Omega(-\hbar\Pi)$ has an inverse one. Indeed, it follows from the relation $\widetilde{\widetilde{F}}(\eta,\xi) = [\Omega(\hbar\eta\xi)]^{-1}\widetilde{\widetilde{G}}(\eta,\xi)$ that $F = [\Omega(-\hbar\Pi)]^{-1} G$.

Now we turn to operator symbols. Their notion in the general case was defined in Sect. 1.1. Here we slightly change the notation introduced there, namely, $F_{(Q_\Omega)}(x,p) \implies F_\Omega(x,p)$. A function $F_\Omega(x,p)$ will be called an $\Omega$-symbol of an operator $\hat{F}$ if $\hat{F} = Q_\Omega(F_\Omega)$. Then the star product $F_\Omega * G_\Omega$ is the $\Omega$-symbol of the operator $\hat{F}\hat{G}$, namely,

$$Q_\Omega(F_\Omega * G_\Omega) = \hat{F}\hat{G} = Q_\Omega(F_\Omega) Q_\Omega(G_\Omega). \tag{78}$$

Similarly, we define w-symbols (Weyl-symbols) $F^{\text{w}}$ by the relation $\hat{F} = Q^{\text{w}}(F^{\text{w}})$, and the star product $(F^{\text{w}} * G^{\text{w}})$ with respect to the Weyl quantization as follows $\hat{F}\hat{G} = Q^{\text{w}}(F^{\text{w}} * G^{\text{w}})$.

The star product of w-symbols can be expressed via the corresponding w-symbols as follows:

$$(F^{\text{w}} * G^{\text{w}})(x,p) = \exp\left[\frac{i\hbar}{2}\left(\frac{\partial^2}{\partial x^1 \partial p_2} - \frac{\partial^2}{\partial x^2 \partial p_1}\right)\right]$$
$$\times F^{\text{w}}(x^1, p_1) G^{\text{w}}(x^2, p_2)\Big|_{\substack{x^1=x^2=x \\ p_1=p_2=p}}$$
$$= F^{\text{w}}(x,p) G^{\text{w}}(x,p) + \frac{i\hbar}{2}\{F^{\text{w}}, G^{\text{w}}\}(x,p) + O(\hbar^2). \tag{79}$$

Expressing the quantization $Q^{\text{c}}_\Omega(F)$ in terms of the Weyl quantization $Q^{\text{w}}(F)$ with account taken (66) and (78), we obtain

$$Q^{\mathrm{w}}(\Omega(-\hbar\Pi)(F_\Omega * G_\Omega))$$
$$= Q^{\mathrm{w}}(\Omega(-\hbar\Pi)F_\Omega)Q^{\mathrm{w}}(\Omega(-\hbar\Pi)G_\Omega)$$
$$= Q^{\mathrm{w}}(\Omega(-\hbar\Pi)F^{\mathrm{w}} * \Omega(-\hbar\Pi)G^{\mathrm{w}}). \quad (80)$$

The equality (80) of operators implies the equality of the corresponding symbols,

$$(F_\Omega * G_\Omega) = \Omega^{-1}(-\hbar\Pi)(\Omega(-\hbar\Pi)F^{\mathrm{w}} * \Omega(-\hbar\Pi)G^{\mathrm{w}}).$$

According to initial condition (27), we have

$$\Omega(-\hbar\Pi) = 1 - a\hbar\Pi + O(\hbar^2),$$
$$\Omega^{-1}(-\hbar\Pi) = 1 + a\hbar\Delta + O(\hbar^2),$$

where $a = \mathrm{d}\Omega(k)/\mathrm{d}k|_{k=0}$.

Then, using expansion (79), we obtain

$$(F_\Omega * G_\Omega) = F_\Omega G_\Omega + \frac{i\hbar}{2}\{F_\Omega, G_\Omega\}$$
$$+ \left.\frac{\mathrm{d}\Omega(k)}{\mathrm{d}k}\right|_{k=0} \hbar \sum_{\mu=1}^{D} \left(\frac{\partial F_\Omega}{\partial x^\mu}\frac{\partial G_\Omega}{\partial p_\mu} + \frac{\partial G_\Omega}{\partial x^\mu}\frac{\partial F_\Omega}{\partial p_\mu}\right)$$
$$+ O(\hbar^2), \quad (81)$$

which implies the relation

$$(F_\Omega * G_\Omega) - (G_\Omega * F_\Omega) = i\hbar\{F_\Omega, G_\Omega\} + O(\hbar^2). \quad (82)$$

The two last equations are just conditions (7) that justify the fulfillment of the correspondence principle for the quantization $Q_\Omega^{\mathrm{c}}(F)$.

## 3 Covariant quantizations in flat spaces in general coordinates

Our aim here is to generalize quantizations $Q_{(\theta)}^{\mathrm{c}}(F)$ (43), and $Q_\Omega^{\mathrm{c}}(F)$ (27) or $Q_{[\omega]}^{\mathrm{c}}(F)$ (47), formulated in flat spaces in Cartesian coordinate systems and covariant with respect to the group $O(r, s)$ transformations, to quantizations formulated also in flat spaces but already in arbitrary coordinate systems and covariant under arbitrary coordinate transformations.

### 3.1 Alternative forms of covariant quantizations in flat spaces in Cartesian coordinates

To construct the above mentioned generalizations of quantizations in flat spaces in Cartesian coordinates, we first are going to avoid using expansions of coordinate parts of classical functions in Fourier integrals. The reason is simple, not all the coordinate systems allow such expansions. Thus, we consider now a partial Fourier transform $\tilde{F}(x, \xi)$ of a function $F(x, p)$ with respect to the momenta $p$ only,[10]

$$\tilde{F}(x, \xi) = (2\pi\hbar)^{-D} \int F(x, p)\mathrm{e}^{-\frac{i}{\hbar}p\xi}\mathrm{d}p,$$
$$F(x, p) = \int \tilde{F}(x, \xi)\mathrm{e}^{\frac{i}{\hbar}p\xi}\mathrm{d}\xi. \quad (83)$$

To express the basic quantization $Q_{(\theta)}^{\mathrm{c}}(F)$ in terms of the partial transform $\tilde{F}(x, \xi)$, we can start with Eq. (43), written as

$$Q_{(\theta)}^{\mathrm{c}}(F) = \int \mathrm{e}^{i\left(\frac{1}{2}+\theta\right)\hat{p}\xi} \left(\int \tilde{\tilde{F}}(\eta, \xi)\mathrm{e}^{i\eta\hat{x}}\mathrm{d}\eta\right) \mathrm{e}^{i\left(\frac{1}{2}-\theta\right)\hat{p}\xi}\mathrm{d}\xi. \quad (84)$$

Then we have express the integral in brackets containing $\tilde{\tilde{F}}(\eta, \xi)$ in terms of $\tilde{F}(x, \xi)$. To this end, we write

$$\int \tilde{\tilde{F}}(\eta, \xi)\mathrm{e}^{i\eta x}\mathrm{d}\eta = (2\pi)^{-D} \int F(x, p)\mathrm{e}^{-ip\xi}\mathrm{d}p$$
$$= \hbar^D \tilde{F}(x, \hbar\xi). \quad (85)$$

In what follows, we will use a supposition that there exist eigenvectors[11] $|x\rangle$ of the operators $\hat{x}^\mu$ with the following properties:

$$\hat{x}^\mu |x\rangle = x^\mu |x\rangle, \quad \langle x|y\rangle = \delta(x-y), \quad \int |x\rangle\langle x|\mathrm{d}x = \hat{I},$$
$$\langle x|\hat{p}_\mu = -i\hbar\frac{\partial}{\partial x^\mu}\langle x|,$$
$$f(\hat{x}) = \int f(x)|x\rangle\langle x|\mathrm{d}x \quad \forall f(x). \quad (86)$$

This allows us to rewrite Eq. (85) as

$$\int \tilde{\tilde{F}}(\eta, \xi)\mathrm{e}^{i\eta\hat{x}}\mathrm{d}\eta = \hbar^D \tilde{F}(\hat{x}, \hbar\xi)$$
$$= \hbar^D \int \tilde{F}(x, \hbar\xi)|x\rangle\langle x|\mathrm{d}x. \quad (87)$$

Substituting spectral decomposition (87) into representation (84), we obtain

$$Q_{(\theta)}^{\mathrm{c}}(F) = \int \tilde{F}(x, \xi)\mathrm{e}^{\frac{i}{\hbar}\left(\frac{1}{2}+\theta\right)\hat{p}\xi}|x\rangle\langle x|\mathrm{e}^{\frac{i}{\hbar}\left(\frac{1}{2}-\theta\right)\hat{p}\xi}\mathrm{d}x\mathrm{d}\xi. \quad (88)$$

Using in (88) coordinate representation (86) of the momentum operators and relations

---

[10] To minimize occurrences of $\hbar$ in the following formulas, we introduce this constant in the complex exponential.

[11] Of course they are generalized vectors which do not belong to the Hilbert space.

$$\langle x | e^{\frac{i}{\hbar}\left(\frac{1}{2}-\theta\right)\hat{p}\xi} = e^{\left(\frac{1}{2}-\theta\right)\xi^\mu \frac{\partial}{\partial x'^\mu}} \langle x | = \left\langle x + \left(\frac{1}{2}-\theta\right)\xi \right|,$$

$$e^{\frac{i}{\hbar}\left(\frac{1}{2}+\theta\right)\hat{p}\xi} |x\rangle = \left(\langle x | e^{-\frac{i}{\hbar}\left(\frac{1}{2}+\theta\right)\hat{p}\xi}\right)^\dagger = \left| x - \left(\frac{1}{2}+\theta\right)\xi \right\rangle,$$
(89)

we finally obtain

$$Q^{\text{c}}_{(\theta)}(F) = \int \tilde{F}(x,\xi) \left| x - \left(\frac{1}{2}+\theta\right)\xi \right\rangle$$
$$\times \left\langle x + \left(\frac{1}{2}-\theta\right)\xi \right| dx d\xi. \quad (90)$$

Representation (90) for the basic quantization $Q^{\text{c}}_{(\theta)}(F)$ does not contain a Fourier transform of the function $F(x, p)$ with respect to the coordinates $x$, it is expressed via the partial Fourier transform of the function $F(x, p)$ with respect to the momenta only. Equation (90) can be interpreted as a spectral decomposition of the operator $Q^{\text{c}}_{(\theta)}(F)$ into the operators

$$\left| x - \left(\frac{1}{2}+\theta\right)\xi \right\rangle \left\langle x + \left(\frac{1}{2}-\theta\right)\xi \right|, \quad (91)$$

which are operators of a projection with a subsequent rotation.

Using Eq. (90) in Eq. (47), we construct the following representation for the quantization $Q^{\text{c}}_{[\omega]}(F)$:

$$Q^{\text{c}}_{[\omega]}(F) = \int \tilde{F}(x,\xi) \hat{D}(x,\xi,\omega) dx d\xi, \quad (92)$$

where

$$\hat{D}(x,\xi,\omega) = \int \left| x - \left(\frac{1}{2}+\theta\right)\xi \right\rangle \left\langle x + \left(\frac{1}{2}-\theta\right)\xi \right| \omega(\theta) d\theta. \quad (93)$$

In the case, when $F(x, p)$ is a polynomial in $p$, the partial Fourier transform $\tilde{F}(x, \xi)$ can be written as

$$\tilde{F}(x,\xi) = (2\pi\hbar)^{-D} F(x, i\hbar\partial_\xi) \int e^{-\frac{i}{\hbar}p\xi} dp$$
$$= F(x, i\hbar\partial_\xi) \delta(\xi), \quad (94)$$

such that

$$Q^{\text{c}}_{[\omega]}(F) = \int F(x, -i\hbar\partial_\xi) \hat{D}(x,\xi,\omega) \Big|_{\xi=0} dx. \quad (95)$$

### 3.2 Spectral representations of covariant quantizations in flat spaces in arbitrary coordinate systems

#### 3.2.1 General construction

We start with a covariant consistent (hermitian in particular) quantization $Q^{\text{c}}_{[\omega]}(F)$ which is formulated in a Cartesian coordinate system $K^{\text{C}}_x$ with coordinates $x$ and produces a quantum theory in a Hilbert space $\mathfrak{H}$. Let us try to construct a quantization $Q'_{[\omega]}(F')$ in the same flat space but in an arbitrary coordinate system $K_{x'}$ with coordinates $x' = \varphi(x)$. Quantization $Q'_{[\omega]}(F')$ must produce an equivalent quantum theory in a Hilbert space $\mathfrak{H}'$, in particular, $Q'_{[\omega]}(F')$ must also be a consistent quantization. We suppose that both quantum theories are related by a unitary operator $U : \mathfrak{H} \xrightarrow{U} \mathfrak{H}'$ in the spirit of Sect. 1.2. Equation (18) define this unitary operator, which depends, in the general case, on $\omega$. Thus, the quantization $Q'_{[\omega]}$ in $K_{x'}$ is related to the quantization $Q_{[\omega]}$ in $K^{\text{C}}_x$ as

$$Q'_{[\omega]}(F') = U Q_{[\omega]}(F) U^\dagger, \quad (96)$$

which coincides by the construction with quantization (47) if coordinates $x'$ are Cartesian.

One can demonstrate that the quantization $Q'_{[\omega]}$ can be formulated in terms of intrinsic elements inherent to the coordinate system $K_{x'}$, and, thus, is independent on the quantization $Q^{\text{c}}_{[\omega]}(F)$ formulated in the Cartesian coordinate system $K^{\text{C}}_x$.

Using Eq. (34), we obtain

$$Q^{\text{c}}_{[\omega]}\left(\varphi^\mu(x)\right) = \varphi^\mu(\hat{x}) \implies \hat{x}'^\mu = U\varphi^\mu(\hat{x}) U^\dagger. \quad (97)$$

On the other side, using Eqs. (54), (55), and (57), we obtain

$$Q^{\text{c}}_{[\omega]}\left(p_\alpha \frac{\partial x^\alpha}{\partial x'^\mu}\right) = \omega_{(-)} \widehat{\frac{\partial x^\alpha}{\partial x'^\mu}} \hat{p}_\alpha + \omega_{(+)} \hat{p}_\alpha \widehat{\frac{\partial x^\alpha}{\partial x'^\mu}},$$

$$\omega_{(-)} = \int_{-\infty}^{+\infty} \left(\frac{1}{2}-\theta\right) \omega(\theta) d\theta = \frac{1}{2} + i\varpi,$$

$$\omega_{(+)} = \int_{-\infty}^{+\infty} \left(\frac{1}{2}+\theta\right) \omega(\theta) d\theta = \frac{1}{2} - i\varpi,$$

$$\varpi = i \int_{-\infty}^{+\infty} \theta \omega(\theta) d\theta = \frac{d\Omega}{dk}\bigg|_{k=0},$$

$$\widehat{\frac{\partial x^\alpha}{\partial x'^\mu}} \equiv \frac{\partial x^\alpha}{\partial x'^\mu}\left(\varphi(\hat{x})\right). \quad (98)$$

Therefore,

$$Q^{\text{c}}_{[\omega]}\left(p_\alpha \frac{\partial x^\alpha}{\partial x'^\mu}\right) = \frac{1}{2}\left(\widehat{\frac{\partial x^\alpha}{\partial x'^\mu}} \hat{p}_\alpha + \hat{p}_\alpha \widehat{\frac{\partial x^\alpha}{\partial x'^\mu}}\right)$$
$$- i\varpi \left[\hat{p}_\alpha, \widehat{\frac{\partial x^\alpha}{\partial x'^\mu}}\right]. \quad (99)$$

The commutator in the right hand side of Eq. (99) can be found from Heisenberg conditions (28),

$$\left[\hat{p}_\alpha, \widehat{\frac{\partial x^\alpha}{\partial x'^\mu}}\right] = \left[\hat{p}_\alpha, Q^{\text{c}}_{[\omega]}\left(\frac{\partial x^\alpha}{\partial x'^\mu}(\varphi(\hat{x}))\right)\right]$$
$$= -i\hbar Q^{\text{c}}_{[\omega]}\left(\frac{\partial}{\partial x^\alpha}\left(\frac{\partial x^\alpha}{\partial x'^\mu}\right)\right). \quad (100)$$

On the other side, taking into account the relation

$$\frac{\partial}{\partial x^\alpha}\left(\frac{\partial x^\alpha}{\partial x'^\mu}\right) = \frac{\partial^2 x^\alpha}{\partial x'^\mu \partial x'^\nu} \frac{\partial x'^\nu}{\partial x^\alpha} = \Gamma'^\alpha_{\alpha\mu} \equiv \Gamma'_\mu,$$

where $\Gamma'^\nu_{\alpha\mu}$ are Christoffel symbols, we obtain

$$\left[\hat{p}_\alpha, \widehat{\frac{\partial x^\alpha}{\partial x'^\mu}}\right] = -i\hbar \Gamma'_\mu(\varphi(\hat{x})), \quad (101)$$

such that finally

$$Q^c_{[\varpi]}\left(p_\alpha \frac{\partial x^\alpha}{\partial x'^\mu}\right) = \frac{1}{2}\left(\widehat{\frac{\partial x^\alpha}{\partial x'^\mu}}\hat{p}_\alpha + \hat{p}_\alpha \widehat{\frac{\partial x^\alpha}{\partial x'^\mu}}\right) - \hbar\varpi \Gamma'_\mu(\varphi(\hat{x})). \quad (102)$$

Obviously the left hand side of Eq. (102) can be considered as a self-adjoint operator if $\varpi$ is real, which, in turn, provides the hermiticity of the quantization $Q'_{[\varpi]}$. At the same time, real $\varpi$ provides the hermiticity of the quantization $Q^c_{[\varpi]}$. In what follows, we suppose that quantizations $Q^c_{[\varpi]}$ are hermitian.

It follows from Eq. (101) that

$$Q^c_{[\varpi]}\left(p_\alpha \frac{\partial x^\alpha}{\partial x'^\mu}\right) = \widehat{\frac{\partial x^\alpha}{\partial x'^\mu}}\hat{p}_\alpha - \hbar\left(\frac{i}{2} + \varpi\right)\Gamma'_\mu(\varphi(\hat{x})). \quad (103)$$

Then, taking into account Eq. (97), we obtain

$$U\widehat{\frac{\partial x^\alpha}{\partial x'^\mu}}U^\dagger = U\frac{\partial x^\alpha}{\partial x'^\mu}(\varphi(\hat{x}))U^\dagger = \frac{\partial x^\alpha}{\partial x'^\mu}(\hat{x}'),$$
$$U\Gamma'_\mu(\varphi(\hat{x}))U^\dagger = \Gamma'_\mu(\hat{x}'). \quad (104)$$

Thus, Eq. (18) takes the form

$$\hat{p}'_\mu = \frac{\partial x^\alpha}{\partial x'^\mu}(\hat{x}')U\hat{p}_\alpha U^\dagger - \hbar\left(\frac{i}{2} + \varpi\right)\Gamma'_\mu(\hat{x}'). \quad (105)$$

Equations (97) and (105) determine the unitary transformation $U$ up to a global phase.

Let $|x\rangle$ are eigenvectors defined by Eq. (86). They are generalized states related to the Hilbert space $\mathfrak{H}$. Then the generalized states

$$|x'\rangle' = |\varphi(x)\rangle' = U|x\rangle \quad (106)$$

are related to the Hilbert space $\mathfrak{H}'$. Taking into account Eq. (97), one can see that $|x'\rangle'$ are eigenvectors of the operators $\hat{x}'^\mu$. Indeed,

$$\hat{x}'^\mu|\varphi(x)\rangle' = U\varphi^\mu(\hat{x})U^\dagger|\varphi(x)\rangle' = U\varphi^\mu(\hat{x})|x\rangle$$
$$= \varphi^\mu(x)U|x\rangle = \varphi^\mu(x)|\varphi(x)\rangle' \implies \hat{x}'^\mu|x'\rangle' = x'^\mu|x'\rangle'. \quad (107)$$

These vectors satisfy the orthonormalization and completeness relations

$$\langle x'|'y'\rangle' = \langle x|U^\dagger U|y\rangle = \langle x|y\rangle = \delta(x-y)$$
$$= \frac{\delta(x'-y')}{|\det(\partial x/\partial x')|} \quad (108)$$

$$\implies \langle x'|'y'\rangle' = \frac{\delta(x'-y')}{\sqrt{g'(x')}},$$
$$\int |x'\rangle'\langle x'|'\sqrt{g'(x')}dx' = \hat{I}. \quad (109)$$

Here $\langle x'|'$ are bra-vectors associated to the ket-vectors $|x'\rangle'$, $g' = |\det g'_{\mu\nu}(x')|$, and we have used the well-known property $\delta(f(x')) = \delta(x'-y')|\partial f/\partial x'|^{-1}$ of the $\delta$-function, where $f(x') = \varphi^{-1}(x') - \varphi^{-1}(y')$ (the equation $f(x') = 0$ has only one root since $\varphi(x)$ is a bijection).

Using Eq. (105) and the relations

$$\langle x'|'\frac{\partial x^\alpha}{\partial x'^\mu}(\hat{x}')U\hat{p}_\alpha U^\dagger = \frac{\partial x^\alpha}{\partial x'^\mu}(x')\langle x'|'U\hat{p}_\alpha U^\dagger$$
$$= \frac{\partial x^\alpha}{\partial x'^\mu}(x')\langle x|\hat{p}_\alpha U^\dagger$$
$$= -i\hbar\frac{\partial x^\alpha}{\partial x'^\mu}(x')\frac{\partial}{\partial x^\alpha}\langle x|U^\dagger$$
$$= -i\hbar\frac{\partial x^\alpha}{\partial x'^\mu}(x')\frac{\partial}{\partial x^\alpha}\langle \varphi(x)|'$$
$$= -i\hbar\frac{\partial}{\partial x'^\mu}\langle x'|',$$

we derive a coordinate representation of the momentum operator $\hat{p}'_\mu$,

$$\langle x'|'\hat{p}'_\mu = -i\hbar\left[\frac{\partial}{\partial x'^\mu} + \left(\frac{1}{2} - i\varpi\right)\Gamma'_\mu(x')\right]\langle x'|'. \quad (110)$$

The state $|\psi\rangle' = U|\psi\rangle$ in $\mathfrak{H}'$ corresponds to any given state $|\psi\rangle \in \mathfrak{H}$. Using Eq. (106), we see that wave functions are scalars under the coordinate transformations $x' = \varphi(x)$,

$$\psi'(x') = \langle x'|'\psi\rangle' = \langle x|U^\dagger U|\psi\rangle = \langle x|\psi\rangle = \psi(x). \quad (111)$$

Now we can construct a quantization $Q'_{[\varpi]}$ in an arbitrary reference frame $K_{x'}$ according to definition (96), using representation (92) for the quantization $Q^c_{[\varpi]}$ in a Cartesian reference frame $K^C_x$,

$$Q'_{[\varpi]}(F') = UQ^c_{[\varpi]}(F)U^\dagger$$
$$= \int \tilde{F}(x,\xi)U\hat{D}(x,\xi,\omega)U^\dagger dxd\xi. \quad (112)$$

In the integral in the right side of Eq. (112), we perform a variable change,

$$x' = \varphi(x), \quad \xi' = \frac{\partial x'}{\partial x}\xi,$$
$$dxd\xi = |\det(\partial x/\partial x')|^2 dx'd\xi' = g'(x')dx'd\xi'. \quad (113)$$

We note that $g = |\det(g_{\mu\nu})| = 1$ since $K^C_x$ is a Cartesian reference frame. Next, we need a transformation law for $\tilde{F}(x,\xi)$. Using the relations

$$F(x,p) = F'(x',p'), \quad p'_\mu = p_\nu \frac{\partial x^\nu}{\partial x'^\mu}$$

already discussed in Sect. 1.2, and definition (83) of $\tilde{F}(x, \xi)$, we obtain

$$\tilde{F}(x, \xi) = \frac{1}{(2\pi\hbar)^D} \int F'\left(x', p\frac{\partial x}{\partial x'}\right) e^{-\frac{i}{\hbar}p\xi} dp$$
$$= \frac{1}{(2\pi\hbar)^D} \int F'(x', p') e^{-\frac{i}{\hbar}p'\frac{\partial x'}{\partial x}\xi}$$
$$\times \left|\det(\partial x'/\partial x)\right| dp' = \frac{\tilde{F}'(x', \xi')}{\sqrt{g'(x')}},$$
$$g'(x') = \left|\det(\partial x/\partial x')\right|. \quad (114)$$

Next, we have to calculate the operator $U\hat{D}(x, \xi, \omega) U^\dagger$, which we denote by $\hat{D}'(x', \xi', \omega)$. Using Eqs. (93) and (106), we derive for it the following representation:

$$\hat{D}'(x', \xi', \omega) = U\hat{D}(x, \xi, \omega) U^\dagger$$
$$= \int U \left|x - \left(\frac{1}{2} + \theta\right)\xi\right\rangle \left\langle x + \left(\frac{1}{2} - \theta\right)\xi\right| U^\dagger \omega(\theta) d\theta$$
$$= \int \left|\varphi\left(x - \left(\frac{1}{2} + \theta\right)\xi\right)\right\rangle'$$
$$\times \left\langle\varphi\left(x + \left(\frac{1}{2} - \theta\right)\xi\right)\right|' \omega(\theta) d\theta. \quad (115)$$

Using the definition of the exponential function in the general coordinate system $K_{x'}$, its transformation law and its form in the Cartesian reference frame $K_x^C$, see Eqs. (269), and (272) in Appendix A.3, we obtain

$$\left|\varphi\left(x - \left(\frac{1}{2} + \theta\right)\xi\right)\right\rangle' \left\langle\varphi\left(x + \left(\frac{1}{2} - \theta\right)\xi\right)\right|'$$
$$= \left|\gamma'\left(x', -\left(\frac{1}{2} + \theta\right)\xi'\right)\right\rangle' \left\langle\gamma'\left(x', \left(\frac{1}{2} - \theta\right)\xi'\right)\right|'. \quad (116)$$

Thus, the quantization $Q'_{[\omega]}(F')$ has the following form:

$$Q'_{[\omega]}(F') = \int \tilde{F}'(x', \xi') \hat{D}'(x', \xi', \omega) \sqrt{g'(x')} dx'd\xi',$$
$$\hat{D}'(x', \xi', \omega) = \int \left|\gamma'\left(x', -\left(\frac{1}{2} + \theta\right)\xi'\right)\right\rangle'$$
$$\times \left\langle\gamma'\left(x', \left(\frac{1}{2} - \theta\right)\xi'\right)\right|' \omega(\theta) d\theta. \quad (117)$$

It follows from (117) that the quantization $Q'_{[\omega]}(F')$, defined originally by Eq. (96), is, in fact, formulated in terms of the intrinsic characteristics ($g'_{\mu\nu}$ and $\Gamma'^\mu_{\alpha\beta}$) of the coordinate system $K_{x'}$, and does not depend on the Cartesian reference frame $K_x^C$ at all.

Now we are going to formulate once again the quantizations $Q_{[\omega]}(F)$ in arbitrary coordinate systems $K_x$ with coordinates $x$ (in contrast to the previous consideration now by $x$ we denote already arbitrary coordinates) without any reference to a quantization in a Cartesian reference frame.

In the coordinate system $K_x$, the metric tensor is $g_{\mu\nu}(x)$, and Christoffel symbols $\Gamma^\mu_{\alpha\beta}(x)$ are expressed via $g_{\mu\nu}(x)$ according to Eq. (252), the exponential function $\gamma(x, \xi)$ is defined via $\Gamma^\mu_{\alpha\beta}(x)$; see Appendix A.3. Canonical variables $x$ and $p$ are represented in quantum theory by operators $\hat{x}^\mu$ and $\hat{p}_\mu$ that act in the corresponding Hilbert space $\mathfrak{H}$. We define a complete set $|x\rangle_\varpi$ in $\mathfrak{H}$ of generalized eigenvectors of the operators $\hat{x}^\mu$, which satisfy the following relations:

$$\hat{x}^\mu |x\rangle_\varpi = x^\mu |x\rangle_\varpi, \quad {}_\varpi\langle x|y\rangle_\varpi = g^{-1/2}(x)\delta(x-y),$$
$${}_\varpi\langle x|\hat{p}_\mu = -i\hbar\left[\partial_\mu + \left(\frac{1}{2} - i\varpi\right)\Gamma_\mu(x)\right]{}_\varpi\langle x|. \quad (118)$$

The vectors $|x\rangle_\varpi$ are related to the vectors $|x\rangle_0$ as

$$|x\rangle_\varpi = e^{-i\varpi \ln\sqrt{g(x)}} |x\rangle_0, \quad (119)$$

which can be checked (taking into account Eq. (257)) by the following calculations:

$$(i/\hbar) e^{i\varpi \ln\sqrt{g(x)}}{}_0\langle x|\hat{p}_\mu = e^{i\varpi \ln\sqrt{g(x)}}\left[\partial_\mu + \frac{1}{2}\Gamma_\mu(x)\right]{}_0\langle x|$$
$$= \left[e^{i\varpi \ln\sqrt{g(x)}}\partial_\mu e^{-i\varpi \ln\sqrt{g(x)}} + \frac{1}{2}\Gamma_\mu(x)\right] e^{i\varpi \ln\sqrt{g(x)}}{}_0\langle x|$$
$$= \left[\partial_\mu + \left(\frac{1}{2} - i\varpi\right)\Gamma_\mu(x)\right] e^{i\varpi \ln\sqrt{g(x)}}{}_0\langle x|. \quad (120)$$

In terms of vectors $|x\rangle_\varpi$, we construct an operator $\hat{D}(x, \xi, \omega)$ by analogy with (115), and we define the quantization $Q_{[\omega]}(F)$ in $K_x$ by the equations

$$Q_{[\omega]}(F) = \int \tilde{F}(x, \xi) \hat{D}(x, \xi, \omega) \sqrt{g(x)} dx d\xi,$$
$$\hat{D}(x, \xi, \omega) = \int \left|\gamma\left(x, -\left(\frac{1}{2} + \theta\right)\xi\right)\right\rangle_\varpi$$
$$\times {}_\varpi\left\langle\gamma\left(x, \left(\frac{1}{2} - \theta\right)\xi\right)\right| \omega(\theta) d\theta. \quad (121)$$

If $F(x, p)$ are homogeneous polynomials in $p$, we use Eq. (121) to obtain an analog of representation (95),

$$Q_{[\omega]}(F) = \int F(x, -i\hbar\partial_\xi) \hat{D}(x, \xi, \omega)\Big|_{\xi=0} \sqrt{g(x)} dx. \quad (122)$$

### 3.2.2 Covariance and correspondence principle

I. One can demonstrate that, in an arbitrary coordinate system $K_{x''}$ (different from $K_x$ and $K_{x'}$), the quantization $Q''_{[\omega]}(F'') = VQ_{[\omega]}(F)V^\dagger$, where the unitary operator $V$ relates Hilbert spaces $\mathfrak{H}$ and $\mathfrak{H}''$, and the quantization $Q'_{[\omega]}(F')$ are related as $Q''_{[\omega]}(F'') = WQ'_{[\omega]}(F')W^\dagger$ with a unitary operator $W = VU^\dagger$. Since coordinate systems $K_{x'}$ and $K_{x''}$ are arbitrary and there exists a unitary equivalence between the corresponding quantum theories, the quantization $Q_{[\omega]}(F)$ given by Eq. (121) is covariant according to the definition formulated in Sect. 1.2.

II. In an arbitrary coordinate system $K_{x'}$ the quantization $Q'_{[\omega]}(F')$ given by Eq. (96) or Eq. (117) satisfies the correspondence principle. To prove this statement, we can write, see Eq. (5),

$$\begin{aligned}Q'_{[\omega]}\left(F'_{[\omega]} * G'_{[\omega]}\right) &= Q'_{[\omega]}\left(F'_{[\omega]}\right) Q'_{[\omega]}\left(G'_{[\omega]}\right) \\ &= U^\dagger Q_{[\omega]}\left(F_{[\omega]}\right) Q_{[\omega]}\left(G_{[\omega]}\right) U \\ &= U^\dagger Q_{[\omega]}\left(F_{[\omega]} * G_{[\omega]}\right) U \\ &= Q'_{[\omega]}\left(\left(F_{[\omega]} * G_{[\omega]}\right)'\right),\end{aligned} \quad (123)$$

where $F_{[\omega]}$ is $[\omega]$-symbol of the operator $\hat{F}$, i.e., $\hat{F} = Q_{[\omega]}\left(F_{[\omega]}\right)$ and so on. It follows from Eq. (123) that

$$\begin{aligned}F'_{[\omega]} * G'_{[\omega]} &= \left(F_{[\omega]} * G_{[\omega]}\right)' = \left[F_{[\omega]} G_{[\omega]} + O(\hbar)\right]' \\ &= F'_{[\omega]} G'_{[\omega]} + O(\hbar), \\ F'_{[\omega]} * G'_{[\omega]} &- G'_{[\omega]} * F'_{[\omega]} \\ &= \left(F_{[\omega]} * G_{[\omega]}\right)' - \left(G_{[\omega]} * F_{[\omega]}\right)' \\ &= \left[i\hbar\left\{F_{[\omega]}, G_{[\omega]}\right\} + O(\hbar^2)\right]' \\ &= i\hbar\left\{F'_{[\omega]}, G'_{[\omega]}\right\}' + O(\hbar^2),\end{aligned} \quad (124)$$

where $\{,\}$ and $\{,\}'$ are Poisson brackets in coordinates $(x, p)$ and $(x', p')$, respectively. The fact that the equality $\{F, G\} = \{F', G'\}'$ holds true for any functions $F$ and $G$ was also used, since the phase-space variables $(x, p)$ and $(x', p')$ are related by a time-independent point canonical transformation.

Equation (124) proves that the quantization $Q_{[\omega]}(F)$ satisfies the correspondence principle.

### 3.3 Coordinate representation

Let a covariant quantization $Q_{[\omega]}(F)$ be formulated in arbitrary coordinate systems $K_x$ with coordinates $x$. This quantization defines a quantum theory in a Hilbert space $\mathfrak{H}$. Canonical variables $x$ and $p$ are represented in the quantum theory by operators $\hat{x}^\mu$ and $\hat{p}_\mu$. And let $|x\rangle_\varpi$ be a complete set in $\mathfrak{H}$ of generalized eigenvectors of the operators $\hat{x}^\mu$ with properties (118). These eigenvectors define a coordinate representation of state vectors $|\psi\rangle$ (wave functions $\psi(x)$) in the Hilbert space $\mathfrak{H}$ as follows:

$$|\psi\rangle \Longleftrightarrow {}_\varpi\langle x|\psi\rangle = \psi(x). \quad (125)$$

One can easily verify, by the analogy with the derivation of Eq. (111) that the wave functions $\psi(x)$ are scalars under the general coordinate transformation $x' = \varphi(x)$.

If we define a coordinate representation $\hat{A}_x$ of an operator $\hat{A}$ as $\hat{A}_x \psi(x) = \langle x|\hat{A}|\psi\rangle$, then we can derive a transformation law for the operators $\left[Q_{[\omega]}(F)\right]_x$ as

$$\begin{aligned}\left[Q'_{[\omega]}(F')\right]_{x'} \psi'(x') &= {}_\varpi\langle x'|Q'_{[\omega]}(F')|\psi\rangle' \\ &= {}_\varpi\langle x|U^\dagger Q'_{[\omega]}(F') U|\psi\rangle = {}_\varpi\langle x|Q_{[\omega]}(F)|\psi\rangle \\ &= \left[Q_{[\omega]}(F)\right]_x \psi(x).\end{aligned} \quad (126)$$

Namely in this sense, we interpret operators $\left[Q_{[\omega]}(F)\right]_x$ as scalars.

This fact can be used in calculating the coordinate representation for the quantization $Q_{[\omega]}(F)$ of homogeneous polynomials (50). Namely, we calculate first $Q^C_{[\omega]}(F)$ in a Cartesian coordinate systems $K^C_x$ with coordinates $x$, where $\Gamma^\mu_{\alpha\beta} = 0$ and $(\hat{p}_\mu)_x = -i\hbar\partial_\mu$. Using the latter operators in Eq. (56), we obtain

$$\begin{aligned}&\left[Q^C_{[\omega]}\left(T^{\mu_1\cdots\mu_n}(x) p_{\mu_1}\cdots p_{\mu_n}\right)\right]_x \\ &= (-i\hbar)^n \sum_{k=0}^n \omega_k^n \frac{\partial}{\partial x^{\mu_1}} \cdots \frac{\partial}{\partial x^{\mu_k}} T^{\mu_1\cdots\mu_n} \\ &\quad \times (x) \frac{\partial}{\partial x^{\mu_{k+1}}} \cdots \frac{\partial}{\partial x^{\mu_n}}.\end{aligned} \quad (127)$$

In an arbitrary coordinate system $K_x$ with coordinates $x$, the corresponding scalar operator can be obtained from Eq. (127) by substituting ordinary derivatives by covariant ones (253), $\partial_\mu \to \nabla_\mu$. Thus,

$$\begin{aligned}&\left[Q_{[\omega]}\left(T^{\mu_1\cdots\mu_n}(x) p_{\mu_1}\cdots p_{\mu_n}\right)\right]_x \\ &= (-i\hbar)^n \sum_{k=0}^n \omega_k^n \nabla_{\mu_1}\cdots\nabla_{\mu_k} T^{\mu_1\cdots\mu_n}(x) \nabla_{\mu_{k+1}}\cdots\nabla_{\mu_n}.\end{aligned} \quad (128)$$

Now, using the obtained results, we present quantizations of some frequently encountered functions in arbitrary curvilinear coordinates:

I. Let $F(x, p) = F(x)$.

In this case $\tilde{F}(x, \xi) = F(x)\delta(\xi)$. It follows from the second equation (121), with account taken of the property $\gamma(x, 0) = x$, that $\hat{D}(x, 0, \omega) = |x\rangle\langle x|$. Using these facts in the first equation of (121), we obtain

$$Q_{[\omega]}(F(x)) = \int F(x)|x\rangle\langle x|\sqrt{g(x)} dx = F(\hat{x}). \quad (129)$$

In the coordinate representation, this quantization reads

$$\left[Q_{[\omega]}(F(x))\right]_x = F(x). \quad (130)$$

II. Let $F(x, p) = T^\alpha(x) p_\alpha$.
In this case, using Eq. (98) in Eq. (128) we obtain

$$\begin{aligned}\left[Q_{[\omega]}\left(T^\alpha(x) p_\alpha\right)\right]_x &= -i\hbar \left[\omega_{(-)} T^\alpha \nabla_\alpha + \omega_{(+)} \nabla_\alpha T^\alpha\right] \\ &= -i\hbar \left[\omega_{(-)} T^\alpha \partial_\alpha + \omega_{(+)} (\partial_\alpha + \Gamma_\alpha) T^\alpha\right] \\ &= -i\hbar \left[\frac{1}{2} \left(T^\alpha \partial_\alpha + \partial_\alpha T^\alpha + \Gamma_\alpha T^\alpha\right) \right. \\ & \quad \left. - i\varpi \left(\partial_\alpha T^\alpha + \Gamma_\alpha T^\alpha\right)\right]. \end{aligned} \quad (131)$$

III. A coordinate representation for the momentum operator follows from (131) when $T^\alpha = \delta^\alpha_\mu$,

$$\left[Q_{[\omega]}(p_\mu)\right]_x = (\hat{p}_\mu)_x = -i\hbar \left[\partial_\mu + \left(\frac{1}{2} - i\varpi\right) \Gamma_\mu\right], \quad (132)$$

which is consistent with Eq. (118).
Using the substitution

$$\frac{\partial}{\partial x^\mu} \to \frac{i}{\hbar} \hat{p}_\mu - \left(\frac{1}{2} - i\varpi\right) \Gamma_\mu(\hat{x}), \quad x^\mu \to \hat{x}^\mu, \quad (133)$$

we always can pass from the coordinate representation to a representation in terms of abstract operators $\hat{x}$ and $\hat{p}_\alpha$. Doing this in Eq. (131), we obtain

$$\begin{aligned}Q_{[\omega]}\left(T^\alpha(x) p_\alpha\right) &= \frac{1}{2}\left[T^\alpha(\hat{x}) \hat{p}_\alpha + \hat{p}_\alpha T^\alpha(\hat{x})\right] \\ &\quad - \hbar\varpi \partial_\alpha T^\alpha(\hat{x}). \end{aligned} \quad (134)$$

IV. Let $F(x, p) = g^{\mu\nu}(x) p_\mu p_\nu = p^2$. In fact, we deal here with a free particle Hamiltonian written in arbitrary coordinates.
We can use Eq. (128) to find a coordinate representation,

$$\begin{aligned}\left[Q_{[\omega]}\left(g^{\mu\nu}(x) p_\mu p_\nu\right)\right]_x &= -\hbar^2 \left(\omega_0^2 g^{\mu_1\mu_2} \nabla_{\mu_1} \nabla_{\mu_2} \right. \\ &\quad \left. + \omega_1^2 \nabla_{\mu_1} g^{\mu_1\mu_2} \nabla_{\mu_2} + \omega_2^2 \nabla_{\mu_1} \nabla_{\mu_2} g^{\mu_1\mu_2}\right) \\ &= -\hbar^2 \left(\omega_0^2 + \omega_1^2 + \omega_2^2\right) g^{-1/2} \partial_\mu g^{1/2} g^{\mu\nu} \partial_\nu.\end{aligned}$$

Since the covariant derivative of the metric tensor is zero, we obtain

$$\begin{aligned}g^{\mu\nu} \nabla_\mu \nabla_\nu &= \nabla_\mu \nabla_\nu g^{\mu\nu} \\ &= \nabla_\mu g^{\mu\nu} \nabla_\nu = g^{-1/2} \partial_\mu g^{1/2} g^{\mu\nu} \partial_\nu,\end{aligned} \quad (135)$$

where, in the rhs of Eq. (135), we recognize the Laplace–Beltrami operator. Then, taking into account property (59) of the coefficients $\omega_k^n$, we finally obtain the following result:

$$\left[Q_{[\omega]}\left(g^{\mu\nu}(x) p_\mu p_\nu\right)\right]_x = -\hbar^2 g^{-1/2} \partial_\mu g^{1/2} g^{\mu\nu} \partial_\nu, \quad (136)$$

which is independent of the function $\omega(\theta)$. Thus, the covariant quantization of the free particle Hamiltonian in the coordinate representation is unique and, therefore, is also unique in the abstract representation.
Using the relation $\Gamma_\mu = \partial_\mu \ln(\sqrt{g})$, see (252), Eq. (132) can be rewritten as

$$(\hat{p}_\mu)_x = -i\hbar g^{-1/4} e^{i\varpi g^{1/2}} \partial_\mu g^{1/4} e^{-i\varpi g^{1/2}}, \quad (137)$$

which implies the inverse relation

$$\partial_\mu = \frac{i}{\hbar} g^{1/4} e^{-i\varpi g^{1/2}} (\hat{p}_\mu)_x g^{-1/4} e^{i\varpi g^{1/2}}. \quad (138)$$

Substituting this relation into (136) together with the trivial substitution $x^\mu \to \hat{x}$, we obtain a quantization of the function $g^{\mu\nu}(x) p_\mu p_\nu$ in terms of abstract operators $\hat{x}$ and $\hat{p}_\alpha$,

$$\begin{aligned}Q_{[\omega]}&\left(g^{\mu\nu}(x) p_\mu p_\nu\right) \\ &= e^{-i\varpi g^{1/2}} g^{-1/4} \hat{p}_\mu g^{1/2} g^{\mu\nu} \hat{p}_\nu g^{-1/4} e^{i\varpi g^{1/2}}. \end{aligned} \quad (139)$$

V. Let $F(x, p) = T^{\mu\nu} p_\mu p_\nu$ with $T^{\mu\nu} = \frac{1}{2}\left(\delta^\mu_\alpha \delta^\nu_\beta + \delta^\mu_\beta \delta^\nu_\alpha\right)$, i.e., $F = p_\alpha p_\beta$.
Let us consider cases when functions $\omega(\theta)$ are real. In particular, the Weyl and Born–Jordan orderings belong to such cases. For real $\omega(\theta)$, coefficients $\omega_k^n$ (57) satisfy the relation $\omega_k^n = \omega_{n-k}^n$. Thus,

$$\begin{aligned}\left[Q_{[\omega]}\left(p_\alpha p_\beta\right)\right]_x &= (-i\hbar)^2 \left(\omega_0^2 T^{\mu\nu} \nabla_\mu \nabla_\nu \right. \\ &\quad \left. + \omega_1^2 \nabla_\mu T^{\mu\nu} \nabla_\nu + \omega_2^2 \nabla_\mu \nabla_\nu T^{\mu\nu}\right) \\ &= (-i\hbar)^2 \left[\omega_0^2 \left(T^{\mu\nu} \nabla_\mu \nabla_\nu + \nabla_\mu \nabla_\nu T^{\mu\nu}\right) \right. \\ &\quad \left. + \omega_1^2 \nabla_\mu T^{\mu\nu} \nabla_\nu\right], \quad 2\omega_0^2 + \omega_1^2 = 1.\end{aligned} \quad (140)$$

Operators that contain covariant derivatives are calculated separately here:

$$\begin{aligned}T^{\mu\nu} \nabla_\mu \nabla_\nu &= \partial_\alpha \partial_\beta - \Gamma^\rho_{\alpha\beta} \partial_\rho, \\ \nabla_\mu T^{\mu\nu} \nabla_\nu &= \partial_\alpha \partial_\beta + \frac{1}{2}\left(\Gamma_\alpha \partial_\beta + \Gamma_\beta \partial_\alpha\right), \\ \nabla_\mu \nabla_\nu T^{\mu\nu} &= \partial_\alpha \partial_\beta + \partial_\rho \Gamma^\rho_{\alpha\beta} + \Gamma_\rho \Gamma^\rho_{\alpha\beta} + \Gamma_\alpha \Gamma_\beta \\ &\quad + \partial_\beta \Gamma_\alpha + \left(\Gamma_\alpha \partial_\beta + \Gamma_\beta \partial_\alpha\right). \end{aligned} \quad (141)$$

Here we have taken into account that $\partial_\alpha \Gamma_\beta = \partial_\beta \Gamma_\alpha$.

Substituting (141) into (140), we obtain

$$[Q_{[\omega]}(p_\alpha p_\beta)]_x = -\hbar^2 \Big[\partial_\alpha \partial_\beta + \frac{1}{2}(\Gamma_\alpha \partial_\beta + \Gamma_\beta \partial_\alpha)$$
$$+ \omega_0^2 \Big(\partial_\beta \Gamma_\alpha + \partial_\rho \Gamma^\rho_{\alpha\beta} + \Gamma_\rho \Gamma^\rho_{\alpha\beta} + \Gamma_\alpha \Gamma_\beta\Big)\Big]. \quad (142)$$

To obtain a representation for $Q_{[\omega]}(p_\alpha p_\beta)$ in terms of abstract operators $\hat{x}$ and $\hat{p}_\alpha$, we have to use substitutions (133) into (142), taking into account that $\varpi = 0$ since $\omega(\theta)$ is real in the case under consideration. Thus,

$$Q_{[\omega]}(p_\alpha p_\beta) = \hat{p}_\alpha \hat{p}_\beta - \hbar^2 \Big[-\frac{1}{2}\partial_\beta \Gamma_\alpha - \frac{1}{4}\Gamma_\alpha \Gamma_\beta$$
$$+ \omega_0^2 \Big(\partial_\beta \Gamma_\alpha + \partial_\rho \Gamma^\rho_{\alpha\beta} + \Gamma_\rho \Gamma^\rho_{\alpha\beta} + \Gamma_\alpha \Gamma_\beta\Big)\Big]. \quad (143)$$

For the Weyl quantization, $\omega_0^2 = \frac{1}{4}$, and the latter expression takes a simple form:

$$Q^w(p_\alpha p_\beta) = \hat{p}_\alpha \hat{p}_\beta - \frac{\hbar^2}{4}\Big(R_{\alpha\beta} + \Gamma^\rho_{\alpha\sigma}\Gamma^\sigma_{\beta\rho}\Big)$$
$$= \hat{p}_\alpha \hat{p}_\beta - \frac{\hbar^2}{4}\Gamma^\rho_{\alpha\sigma}\Gamma^\sigma_{\beta\rho}, \quad (144)$$

where the Ricci tensor $R_{\alpha\beta} = \partial_\rho \Gamma^\rho_{\alpha\beta} - \partial_\beta \Gamma_\alpha + \Gamma_\rho \Gamma^\rho_{\alpha\beta} - \Gamma^\rho_{\alpha\sigma}\Gamma^\sigma_{\beta\rho}$ was set to zero since the configuration space is a plane in the case under consideration.

## 4 Covariant quantizations in curved spaces

### 4.1 A minimal generalization of quantizations in flat spaces to curved spaces

In this section, we construct a class of covariant quantizations already in curved spaces. For simplicity we consider Riemannian spaces which can be covered by a unique coordinate systems $K_x$ that generate coordinate $x$ with metric tensors $g_{\mu\nu}(x)$. First, in this section, we construct a class of quantizations $\mathbf{Q}_{[\omega]}(F)$, which is, in fact, a direct generalization of the quantizations $Q_{[\omega]}(F)$ to the curved space case. Such a generalization can be called minimal and is parametrized by the same function $\omega(\theta)$. We prove its consistency and covariance under general coordinate transformations. Then in Sect. 4, we construct an extended class $\mathbf{Q}_{[\omega,\Theta]}(F)$ of quantizations in a curved space, which is already parametrized by the function $\omega(\theta)$ and a new function $\Theta(x,\xi)$. This class is a generalization of the quantizations $\mathbf{Q}_{[\omega]}(F)$ and incudes these quantizations as a particular case at $\Theta = 1$. We present a rather nontrivial prove of the consistency and covariance of this class of quantizations under general coordinate transformations.

We note that consideration and definitions of Sect. 1.2 are fully applicable to the case under consideration. We use also definitions (118) and (121) literally in the case of a curved space with corresponding connections $\Gamma^\mu_{\alpha\beta}(x)$.

One can easily verify the construction (121) is consistently defined for arbitrary metric even for any curved space metric. In particular, the important ingredient part of this construction, is the exponential function, which is defined in Appendix A.3 in arbitrary Riemannian space. Thus, we may suppose that a covariant quantization in any curved space is given by Eq. (121) with the metric tensor and all the ingredient constructions corresponding to such a space. To justify this supposition, we have to verify the consistency and the covariance of such a quantization. For definiteness, we write this quantization explicitly,

$$\mathbf{Q}_{[\omega]}(F) = \int \tilde{F}(x,\xi)\hat{D}(x,\xi,\omega)\sqrt{g(x)}\mathrm{d}x\mathrm{d}\xi, \quad (145)$$

where $\hat{D}(x,\xi,\omega)$ is given by Eq. (121). In fact, as in the plane space case, Eq. (145) represents a family of operators parametrized by a normalized function $\omega(\theta)$ (see Eq. (46)), introduced in Sect. 2.2.

Similar to the plane space case, one can verify that quantization (145) is hermitian if $\omega(\theta) = \omega^*(-\theta)$.

In the same manner as in the plane space case (see Sect. 2.1.1), one can verify that quantization (145) satisfies the properties

$$\mathbf{Q}_{[\omega]}(1) = \hat{I}, \quad \mathbf{Q}_{[\omega]}(F(x)) = F(\hat{x}). \quad (146)$$

Let us consider the quantization $\mathbf{Q}_{[\omega]}(F(x,p))$, where $F(x,p) = T^{\mu_1\cdots\mu_n}(x)p_{\mu_1}\cdots p_{\mu_n}$ is homogeneous polynomial. We assume that the quantization (145) satisfies Eq. (122). Therefore, we have to calculate derivatives in $\xi$ of the quantity $\hat{D}(x,\xi,\omega)$,

$$T^{\mu_1\cdots\mu_n}\frac{\partial}{\partial \xi^{\mu_1}}\cdots\frac{\partial}{\partial \xi^{\mu_n}}\Big|\gamma\Big(x,-\Big(\frac{1}{2}+\theta\Big)\xi\Big)\Big\rangle$$
$$\times \Big\langle\gamma\Big(x,\Big(\frac{1}{2}-\theta\Big)\xi\Big)\Big|$$
$$= T^{\mu_1\cdots\mu_n}\sum_{k=0}^{n}\binom{n}{k}$$
$$\times \Big[\frac{\partial}{\partial \xi^{\mu_1}}\cdots\frac{\partial}{\partial \xi^{\mu_k}}\Big|\gamma\Big(x,-\Big(\frac{1}{2}+\theta\Big)\xi\Big)\Big\rangle\Big]$$
$$\times \Big[\frac{\partial}{\partial \xi^{\mu_{k+1}}}\cdots\frac{\partial}{\partial \xi^{\mu_n}}\Big\langle\gamma\Big(x,\Big(\frac{1}{2}-\theta\Big)\xi\Big)\Big|\Big]. \quad (147)$$

Using Eqs. (247) and (276) from Appendix A.3, we obtain

$$\frac{\partial}{\partial \xi^{\mu_1}}\cdots\frac{\partial}{\partial \xi^{\mu_k}}\Big|\gamma\Big(x,-\Big(\frac{1}{2}+\theta\Big)\xi\Big)\Big\rangle\Big|_{\xi=0}$$
$$= (-1)^k\Big(\frac{1}{2}+\theta\Big)^k \nabla_{(\mu_1}\ldots\nabla_{\mu_k)}|x\rangle,$$

$$\frac{\partial}{\partial \xi^{\mu_{k+1}}} \cdots \frac{\partial}{\partial \xi^{\mu_n}} \left\langle \gamma \left(x, \left(\frac{1}{2} - \theta\right) \xi\right) \right\rangle \bigg|_{\xi=0}$$
$$= \left(\frac{1}{2} - \theta\right)^{n-k} \nabla_{(\mu_{k+1}} \ldots \nabla_{\mu_n)} \langle x |. \qquad (148)$$

Then, taking into account the properties of coefficients (55), and the total symmetry of $T^{\mu_1 \cdots \mu_n}$ in the upper indices, we represent the quantity (147) as follows:

$$T^{\mu_1 \cdots \mu_n} \sum_{k=0}^{n} (-1)^k C_k^n(\theta) \left(\nabla_{\mu_1} \ldots \nabla_{\mu_k} |x\rangle\right)$$
$$\times \left(\nabla_{\mu_{k+1}} \ldots \nabla_{\mu_n} \langle x|\right). \qquad (149)$$

Using definition (57) of coefficients $\omega_k^n$, we find

$$T^{\mu_1 \cdots \mu_n} \frac{\partial}{\partial \xi^{\mu_1}} \cdots \frac{\partial}{\partial \xi^{\mu_n}} \hat{D}(x, \xi, \omega) \bigg|_{\xi=0}$$
$$= T^{\mu_1 \cdots \mu_n} \sum_{k=0}^{n} (-1)^k \omega_k^n \left(\nabla_{\mu_1} \ldots \nabla_{\mu_k} |x\rangle\right)$$
$$\times \left(\nabla_{\mu_{k+1}} \ldots \nabla_{\mu_n} \langle x|\right). \qquad (150)$$

The above results allow us to write

$$\mathbf{Q}_{[\omega]} \left(T^{\mu_1 \cdots \mu_n}(x) p_{\mu_1} \cdots p_{\mu_n}\right)$$
$$= \int F(x, -i\hbar \partial_\xi) \hat{D}(x, \xi, \omega) \bigg|_{\xi=0} \sqrt{g(x)} \mathrm{d}x$$
$$= (-i\hbar)^n \sum_{k=0}^{n} (-1)^k \omega_k^n \int T^{\mu_1 \cdots \mu_n}(x) \left(\nabla_{\mu_1} \ldots \nabla_{\mu_k} |x\rangle\right)$$
$$\times \left(\nabla_{\mu_{k+1}} \ldots \nabla_{\mu_n} \langle x|\right) \sqrt{g(x)} \mathrm{d}x. \qquad (151)$$

In the coordinate representation, we have

$$\left[\mathbf{Q}_{[\omega]} \left(T^{\mu_1 \cdots \mu_n}(x) p_{\mu_1} \cdots p_{\mu_n}\right)\right]_y$$
$$= \langle y| \mathbf{Q}_{[\omega]} \left(T^{\mu_1 \cdots \mu_n}(x) p_{\mu_1} \cdots p_{\mu_n}\right) |\psi\rangle$$
$$= (-i\hbar)^n \sum_{k=0}^{n} (-1)^k \omega_k^n \int I(x, y) \sqrt{g(x)} \mathrm{d}x,$$
$$I(x, y) = T^{\mu_1 \cdots \mu_n}(x) \left[\nabla_{\mu_1} \ldots \nabla_{\mu_k} \frac{\delta(x-y)}{\sqrt{g(x)}}\right]$$
$$\times \nabla_{\mu_{k+1}} \ldots \nabla_{\mu_n} \psi(x). \qquad (152)$$

The quantity $I(x, y)$ can be written as

$$I(x, y) = \nabla_{\mu_1} \left[\left(\nabla_{\mu_2} \ldots \nabla_{\mu_k} \frac{\delta(x-y)}{\sqrt{g(x)}}\right) T^{\mu_1 \cdots \mu_n}(x) \right.$$
$$\left. \times \nabla_{\mu_{k+1}} \ldots \nabla_{\mu_n} \psi(x)\right]$$
$$- \left(\nabla_{\mu_2} \ldots \nabla_{\mu_k} \frac{\delta(x-y)}{\sqrt{g(x)}}\right) \nabla_{\mu_1} T^{\mu_1 \cdots \mu_n}(x)$$
$$\times \nabla_{\mu_{k+1}} \ldots \nabla_{\mu_n} \psi(x). \qquad (153)$$

The first term in the right hand side of (153) is a covariant divergence which vanishes at infinity due to the presence of a delta function. Thus, the integral in the right hand side of

Eq. (152) is zero. Similarly, we can transform the second term in the right hand side of (153) and so on. Finally, we obtain

$$\left[\mathbf{Q}_{[\omega]} \left(T^{\mu_1 \cdots \mu_n}(x) p_{\mu_1} \cdots p_{\mu_n}\right)\right]_x$$
$$= (-i\hbar)^n \sum_{k=0}^{n} \omega_k^n \nabla_{\mu_1} \ldots \nabla_{\mu_k} T^{\mu_1 \cdots \mu_n}(x) \nabla_{\mu_{k+1}} \ldots \nabla_{\mu_n}. \qquad (154)$$

A coordinate representation for the momentum operator follows from (154) when $T^\alpha = \delta_b^\alpha$,

$$\left[\mathbf{Q}_{[\omega]}(p_a)\right]_x = \left(\hat{p}_a\right)_x = -i\hbar \left[\frac{\partial}{\partial x^a} + \left(\frac{1}{2} - i\varpi\right) \Gamma_a\right]. \qquad (155)$$

This result is similar to the one (132) in the case of a flat space.

Comparing (155) with Eq. (118), which, as was mentioned above, holds true in any curved space, we obtain

$$\mathbf{Q}_{[\omega]}(p_a) = \hat{p}_a, \qquad (156)$$

where $\hat{p}_a$ is the momentum operator in the abstract representation.

### 4.1.1 Verification of the covariance

Let us consider a coordinate change $x' = \varphi(x)$. Due to the property $\mathbf{Q}_{[\omega]}(F(x)) = F(\hat{x})$, first equation of (18) remains equal to Eq. (97),

$$\hat{x}'^\mu = U \varphi^\mu (\hat{x}) U^\dagger. \qquad (157)$$

Equation (134) follows only from Eq. (154) and definition (118), which still holds true in curved spaces. Taking into account the commutation relations between $\hat{x}^\mu$ and $\hat{p}_\mu$, it can be written as

$$\mathbf{Q}_{[\omega]}(T^\alpha p_\alpha) = T^\alpha(\hat{x}) \hat{p}_\alpha + \frac{1}{2}\left[\hat{p}_\alpha, T^\alpha(\hat{x})\right] - \hbar\varpi \frac{\partial T^\alpha}{\partial x^\alpha}(\hat{x})$$
$$= T^\alpha(\hat{x}) \hat{p}_\alpha - i\hbar \left(\frac{1}{2} - i\varpi\right) \frac{\partial T^\alpha}{\partial x^\alpha}(\hat{x}). \qquad (158)$$

Setting $T^\alpha = \frac{\partial x^\alpha}{\partial x'^\mu}(\varphi(x))$, we obtain

$$\mathbf{Q}_{[\omega]}\left(p_\alpha \frac{\partial x^\alpha}{\partial x'^\mu}\right) = \frac{\partial x^\alpha}{\partial x'^\mu}(\varphi(\hat{x})) \hat{p}_\alpha - i\hbar \left(\frac{1}{2} - i\varpi\right)$$
$$\times \frac{\partial^2 x^\alpha}{\partial x'^\mu \partial x'^\nu}(\varphi(\hat{x})) \frac{\partial x'^\nu}{\partial x^\alpha}(\hat{x}). \qquad (159)$$

Thus, taking into account (97), the second equation (18) takes the form

$$\hat{p}'_\mu = \frac{\partial x^\alpha}{\partial x'^\mu}(\hat{x}') U \hat{p}_\alpha U^\dagger - i\hbar \left(\frac{1}{2} - i\varpi\right) \frac{\partial^2 x^\alpha}{\partial x'^\mu \partial x'^\nu}(\hat{x}')$$
$$\times \frac{\partial x'^\nu}{\partial x^\alpha}\left(\varphi^{-1}(\hat{x}')\right). \qquad (160)$$

Let us suppose that vectors $|x\rangle \in \mathfrak{H}$ satisfy conditions (118), which are necessary for the definition of the quantity $\hat{D}(x, \xi, \omega)$ by (121), and the consider vectors $|x'\rangle' = U|x\rangle \in \mathfrak{H}'$. Following the same method as used in Eqs. (107) and (108), one can demonstrate that Eq. (97) and the unitarity of $U$ guarantee that Eq. (118) holds true for $|x'\rangle'$. Indeed, one can write

$$\begin{aligned}\langle x'|' \frac{\partial x^\alpha}{\partial x'^\mu}(\hat{x}') U \hat{p}_\alpha U^\dagger &= \frac{\partial x^\alpha}{\partial x'^\mu}(x') \langle x'|' U \hat{p}_\alpha U^\dagger \\ &= \frac{\partial x^\alpha}{\partial x'^\mu}(x') \langle x| \hat{p}_\alpha U^\dagger \\ &= -i\hbar \frac{\partial x^\alpha}{\partial x'^\mu}(x') \left[\frac{\partial}{\partial x^\alpha} + \left(\frac{1}{2} - i\varpi\right) \Gamma_\alpha(x)\right] \langle x| U^\dagger \\ &= -i\hbar \frac{\partial x^\alpha}{\partial x'^\mu}(x') \left[\frac{\partial}{\partial x^\alpha} + \left(\frac{1}{2} - i\varpi\right) \Gamma_\alpha(x)\right] \langle \varphi(x)|' \\ &= -i\hbar \left[\frac{\partial}{\partial x'^\mu} + \left(\frac{1}{2} - i\varpi\right) \frac{\partial x^\alpha}{\partial x'^\mu}(x') \Gamma_\alpha(\varphi^{-1}(x'))\right] \langle x'|'.\end{aligned}$$
(161)

Using (160), we may write

$$\langle x'|' \hat{p}'_\mu = -i\hbar \left[\frac{\partial}{\partial x'^\mu} + \left(\frac{1}{2} - i\varpi\right) \Gamma'_\mu(x')\right] \langle x'|',$$

$$\Gamma'_\mu(x') = \frac{\partial x^\alpha}{\partial x'^\mu}(x') \Gamma_\alpha(\varphi^{-1}(x')) + \frac{\partial^2 x^\alpha}{\partial x'^\mu \partial x'^\nu}(x') \frac{\partial x'^\nu}{\partial x^\alpha}(\varphi^{-1}(x')),$$

where $\Gamma'_\mu(x') = \Gamma'^\rho_{\rho\mu}$ is the contraction of the Levi-Civita connections in the reference frame $K_{x'}$.

Thus, vectors $|x'\rangle' = U|x\rangle \in \mathfrak{H}'$ satisfy Eq. (118), which define these vectors up to a global phase and, therefore, define the operator $\hat{D}'(x', \xi', \omega)$ in $\mathfrak{H}'$ as follows:

$$\hat{D}'(x', \xi', \omega) = \int \left|\gamma'\left(x', -\left(\frac{1}{2} + \theta\right)\xi'\right)\right\rangle' \times \left\langle\gamma'\left(x', \left(\frac{1}{2} - \theta\right)\xi'\right)\right|' \omega(\theta)\mathrm{d}\theta.$$

According to property (269) of the exponential function (see Appendix A.3), we have

$$U|\gamma(x, \xi)\rangle = |\varphi(\gamma(x, \xi))\rangle' = \left|\gamma'\left(\varphi(x), \frac{\partial x'}{\partial x}\xi\right)\right\rangle'$$

$$\Longrightarrow U\hat{D}(x, \xi, \omega)U^\dagger = \hat{D}'\left(\varphi(x), \frac{\partial x'}{\partial x}\xi, \omega\right).$$

Using definition (145), the variable changes $x' = \varphi(x)$ and $\xi' = \frac{\partial x'}{\partial x}\xi$, and the transformation law for the Fourier transform (114), we can calculate the operator $U\mathbf{Q}_{[\omega]}(F)U^\dagger$ to be

$$U\mathbf{Q}_{[\omega]}(F)U^\dagger = \int \tilde{F}(x, \xi) \hat{D}'\left(\varphi(x), \frac{\partial x'}{\partial x}\xi, \omega\right) \sqrt{g(x)}\mathrm{d}x\mathrm{d}\xi$$

$$= \int \tilde{F}(x, \xi) \hat{D}'\left(\varphi(x), \frac{\partial x'}{\partial x}\xi, \omega\right) \sqrt{g(x)}\mathrm{d}x\mathrm{d}\xi$$

$$= \int \tilde{F}'(x', \xi') \hat{D}'(x', \xi', \omega) \sqrt{g'(x')}\mathrm{d}x'\mathrm{d}\xi'$$

$$= \mathbf{Q}'_{[\omega]}(F'). \quad (162)$$

The latter formula demonstrates the covariance of the quantization $\mathbf{Q}_{[\omega]}(F)$.

### 4.1.2 Matrix elements and symbols of operators

Results of this section will be used in proving the validity of the correspondence principle for general class of quantizations in curved space constructed in Sect. 5. Since the quantizations $\mathbf{Q}_{[\omega]}(F)$ represent a part of this general class, the correspondence principle will be proved for the quantization $\mathbf{Q}_{[\omega]}(F)$ as well.

First we calculate matrix elements of operators $\mathbf{Q}_{[\omega]}(F)$ with respect to the basis $|x\rangle_\varpi$ which is defined by Eq. (118). To this end, we have to calculate matrix elements of the operator $\hat{D}(x, \xi, \omega)$ given by Eq. (121),

$$\varpi\langle y|\hat{D}(x, \xi, \omega)|z\rangle_\varpi = \int [g(y) g(z)]^{-1/2}$$
$$\times \delta^D\left(\gamma\left(x, -\left(\theta + \frac{1}{2}\right)\xi\right) - y\right)$$
$$\times \delta^D\left(\gamma\left(x, -\left(\theta - \frac{1}{2}\right)\xi\right) - z\right)\omega(\theta)\mathrm{d}\theta. \quad (163)$$

The product of delta functions under the integral sign in the right hand side of (163) can be written as

$$\delta^D\left(\gamma\left(x, -\left(\theta + \frac{1}{2}\right)\xi\right) - y\right)$$
$$\times \delta^D\left(\gamma\left(x, -\left(\theta - \frac{1}{2}\right)\xi\right) - z\right)$$
$$= \delta^{2D}(\Phi_\theta(x, \xi) - (y, z)), \quad (164)$$

where $(y, z) \in \mathbb{R}^{2D}$ and the function $\Phi_\theta : \mathbb{R}^{2D} \to \mathbb{R}^{2D}$ reads

$$\Phi_\theta(x, \xi) = \left(\gamma\left(x, -\left(\theta + \frac{1}{2}\right)\xi\right), \gamma\left(x, -\left(\theta - \frac{1}{2}\right)\xi\right)\right). \quad (165)$$

Using composition law (267), we find

$$\gamma\left(x, -\left(\theta \pm \frac{1}{2}\right)\xi\right) = \gamma\left(\gamma(x, -\theta\xi), \frac{\pm\beta(x, -\theta\xi)}{2\theta}\right). \quad (166)$$

Function (165) can be expressed via the geodesic flux $T_{-\theta}(x, \xi)$ (265) as follows:

**Fig. 1** The middle point and tangent vector

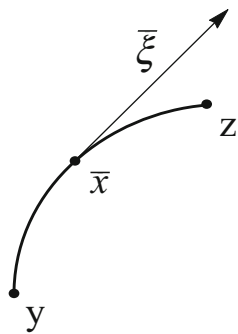

$$\Phi_\theta(x,\xi) = \Phi(T_{-\theta}(x,\xi)),$$
$$\Phi = \Phi_0 = \left(\gamma\left(x, \frac{-1}{2}\xi\right), \gamma\left(x, \frac{1}{2}\xi\right)\right). \quad (167)$$

This allows one to represent matrix elements (163) of the operator $\hat{D}(x,\xi,\omega)$ in the form

$$\langle y|_\varpi \hat{D}(x,\xi,\omega) |z\rangle_\varpi$$
$$= \frac{1}{\sqrt{g(y)g(z)}} \int \delta^{2D}(\Phi(T_{-\theta}(x,\xi)) - (y,z))\,\omega(\theta)\mathrm{d}\theta. \quad (168)$$

Let us use a well-known property of the delta function,

$$\delta(f(x)) = \sum_k \left|\det\left(\partial f/\partial x|_{x=r_k}\right)\right|^{-1} \delta(x - r_k),$$

where $r_k$ are roots of the equation $f(x) = 0$, in the case under consideration, we need to find roots of the equation $\Phi(T_{-\theta}(x,\xi)) = (y,z)$. We note that the latter equation is equivalent to the set

$$\Phi(\bar{x},\bar{\xi}) = (y,z), \quad T_{-\theta}(x,\xi) = (\bar{x},\bar{\xi}). \quad (169)$$

Due to the property (266), the second equation of this set has immediate solution $(x,\xi) = T_\theta(\bar{x},\bar{\xi})$ (Fig. 1). Solutions of the first equation are in one-to-one correspondence with geodesic lines which relate points $y$ and $z$. Indeed, let $K(\tau)$ be a geodesic line with $K(0) = y$ and $K(1) = z$. Therefore $\gamma(y,\zeta) = z$ with $\zeta = \dot{K}(0)$, and the length of the vector $\zeta$ is equal to the length (255) of the line $K$ i.e., $g_{\mu\nu}(y)\zeta^\mu\zeta^\nu = S[K]$. If we choose $\bar{x} = \gamma(y,\zeta/2)$ and $\bar{\xi} = 2\beta(y,\zeta/2)$, then, according to the composition law (267), we obtain

$$\left.\begin{array}{l}\gamma(\bar{x}, \frac{1}{2}\bar{\xi}) = \gamma(\gamma(y,\zeta/2), \beta(y,\zeta/2)) = \gamma(y,\zeta) = z\\ \gamma(\bar{x}, \frac{-1}{2}\bar{\xi}) = \gamma(\gamma(y,\zeta/2), -\beta(y,\zeta/2)) = \gamma(y,0) = y\end{array}\right\}$$
$$\implies \Phi(\bar{x},\bar{\xi}) = (y,z).$$

Thus, $\bar{x}$ is the middle point of the geodesic line $K$, and $\bar{\xi}$ is a tangent vector to $K$ in the point $\bar{x}$ in the $z$-direction (see Fig. 2 in Appendix A.3).

According to Eq. (261), $g_{\mu\nu}(\bar{x})\bar{\xi}^\mu\bar{\xi}^\nu = S[K]$. On the other side, if $\Phi(\bar{x},\bar{\xi}) = (y,z)$, we define $\zeta = -2\beta(\bar{x}, \frac{-1}{2}\bar{\xi})$. For the geodesic line $K(\tau) = \gamma(y,\tau\zeta)$, we have $K(0) = \gamma(y,0) = y$ and according to the composition law (267),

$$K(1) = \gamma\left(\gamma\left(\bar{x}, \frac{-1}{2}\bar{\xi}\right), -2\beta\left(\bar{x}, \frac{-1}{2}\bar{\xi}\right)\right)$$
$$= \gamma\left(\bar{x}, \frac{1}{2}\bar{\xi}\right) = z,$$

which means that the geodesic line $K(\tau)$ relates points $y$ and $z$. Thus, if there exist $n$ geodesic lines $K_{(1)}, \ldots, K_{(n)}$ (with middle points $\bar{x}_{(1)}, \ldots, \bar{x}_{(n)}$ and tangent vectors $\bar{\xi}_{(1)}, \ldots, \bar{\xi}_{(n)}$) that relate points $y$ and $z$, then there exist $n$ solutions $T_\theta(\bar{x}_{(1)}, \bar{\xi}_{(1)}), \ldots, T_\theta(\bar{x}_{(n)}, \bar{\xi}_{(n)})$ of the equation $\Phi(T_{-\theta}(x,\xi)) = (y,z)$.

Then we need to calculate a Jacobian of the composed function $\Phi(T_{-\theta}(x,\xi)) - (y,z)$. Since $(y,z)$ is constant, the determinant reads

$$\det\left(\frac{\partial \Phi(T_{-\theta}(x,\xi))}{\partial(x,\xi)}\right) = \det\left(\frac{\partial \Phi}{\partial(x,\xi)}(T_{-\theta}(x,\xi))\right)$$
$$\times \det\left(\frac{\partial T_{-\theta}}{\partial(x,\xi)}\right) = \det\left(\frac{\partial \Phi}{\partial(x,\xi)}(T_{-\theta}(x,\xi))\right)$$
$$\times \frac{g(x)}{g(\gamma(x,-\theta\xi))}.$$

In these calculations, we have used a Jacobian of the geodesic flux found in Appendix A.5. Thus, taking into account the relations

$$T_{-\theta}(T_\theta(\bar{x}_{(n)}, \bar{\xi}_{(n)})) = (\bar{x}_{(n)}, \bar{\xi}_{(n)}),$$
$$\left.\frac{g(x)}{g(\gamma(x,-\theta\xi))}\right|_{(x,\xi)=T_\theta(\bar{x}_{(k)}, \bar{\xi}_{(k)})} = \frac{g(\gamma(\bar{x}_{(k)}, \theta\bar{\xi}_{(k)}))}{g(\bar{x}_{(k)})},$$

we obtain

$$\frac{\delta^{2D}(\Phi(T_{-\theta}(x,\xi)) - (y,z))}{\sqrt{g(y)g(z)}}$$
$$= \sum_{k=1}^n \frac{\delta^{2D}((x,\xi) - T_\theta(\bar{x}_{(k)}, \bar{\xi}_{(k)}))}{\mathscr{M}(\bar{x}_{(k)}, \bar{\xi}_{(k)})\, g(\gamma(\bar{x}_{(k)}, \theta\bar{\xi}_{(k)}))}, \quad (170)$$

where a function $\mathscr{M}(x,\xi)$ is introduced,

$$\mathscr{M}(x,\xi) = g^{-1}(x)\sqrt{g\left(\gamma\left(x, \frac{-1}{2}\xi\right)\right) g\left(\gamma\left(x, \frac{1}{2}\xi\right)\right)}$$
$$\times \left|\det\left(\frac{\partial \Phi}{\partial(x,\xi)}(x,\xi)\right)\right|. \quad (171)$$

We recall that

$$g(y) = g\left(\gamma\left(\bar{x}_{(k)}, \frac{-1}{2}\bar{\xi}_{(k)}\right)\right),$$
$$g(z) = g\left(\gamma\left(\bar{x}_{(k)}, \frac{1}{2}\bar{\xi}_{(k)}\right)\right) \quad \forall k.$$

Using (170) in Eq. (168), we obtain

$$_\varpi \langle y| \hat{D}(x,\xi,\omega) |z\rangle_\varpi = \sum_{k=1}^{n} \frac{1}{\mathscr{M}(\bar{x}_{(k)}, \bar{\bar\xi}_{(k)})}$$
$$\times \int \frac{\delta^{2D}\left((x,\xi) - T_\theta\left(\bar{x}_{(k)}, \bar{\bar\xi}_{(k)}\right)\right)}{g\left(\gamma\left(\bar{x}_{(k)}, \theta\bar{\bar\xi}_{(k)}\right)\right)} \omega(\theta)d\theta. \quad (172)$$

According to definition (145), we calculate matrix elements of $\mathbf{Q}_{[\omega]}(F)$ taking into account Eq. (172) and the relation

$$\sqrt{g(x)}\bigg|_{(x,\xi)=T_{\hbar\theta}(\bar{x}_{(k)},\bar{\bar\xi}_{(k)})} = \sqrt{g\left(\gamma\left(\bar{x}_{(k)}, \hbar\theta\bar{\bar\xi}_{(k)}\right)\right)}.$$

Finally, we obtain

$$_\varpi \langle y| \mathbf{Q}_{[\omega]}(F) |z\rangle_\varpi = \sum_{k=1}^{n} \frac{1}{\mathscr{M}(\bar{x}_{(k)}, \bar{\bar\xi}_{(k)})}$$
$$\times \int \frac{\tilde{F}\left(T_\theta\left(\bar{x}_{(k)}, \bar{\bar\xi}_{(k)}\right)\right)}{\sqrt{g\left(\gamma\left(\bar{x}_{(k)}, \theta\bar{\bar\xi}_{(k)}\right)\right)}} \omega(\theta)d\theta. \quad (173)$$

### 4.1.3 Polynomial functions

The general formula (173), which is valid for any function $F(x, p)$, turns out to be simple enough if $F(x, p)$ is a polynomial in $p$. In this case,

$$\tilde{F}(x,\xi) = F\left(x, i\hbar\frac{\partial}{\partial\xi}\right)\delta\left(\xi^{a_1}\xi^{a_2}\cdots\xi^{a_n}\right)$$

is a distribution with the support $\xi = 0$. Then

$$\tilde{F}\left(T_\theta\left(\bar{x}_{(k)}, \bar{\bar\xi}_{(k)}\right)\right) = \tilde{F}\left(\gamma\left(\bar{x}_{(k)}, \theta\bar{\bar\xi}_{(k)}\right), \frac{1}{\theta}\beta\left(\bar{x}_{(k)}, \theta\bar{\bar\xi}_{(k)}\right)\right)$$

is always zero unless the equation

$$\frac{1}{\theta}\beta\left(\bar{x}_{(k)}, \theta\bar{\bar\xi}_{(k)}\right) = 0 \Leftrightarrow \bar{\bar\xi}_{(k)} = 0$$

is possible. Between all the geodesic lines that relate $y$ and $z$, the only minimal geodesic line $C_{yz}$ satisfies this property (see Sect. A.4) when $y \to z$. Thus, if $F$ is a polynomial, the sum in (173) is reduced to one unique term, which corresponds to the minimal geodesic line, and we obtain

$$_\varpi \langle y| \mathbf{Q}_{[\omega]}(F) |z\rangle_\varpi$$
$$= \frac{1}{\mathscr{M}(\bar{x}, \bar{\bar\xi})} \int \frac{\tilde{F}\left(T_\theta\left(\bar{x}, \bar{\bar\xi}\right)\right)}{\sqrt{g\left(\gamma\left(\bar{x}, \theta\bar{\bar\xi}\right)\right)}} \omega(\theta)d\theta, \quad (174)$$

where $\bar{x} = \bar{x}(y, z)$ and $\bar{\bar\xi} = \bar{\bar\xi}(y, z)$ are a middle point and a tangent vector of the minimal geodesic line $C_{yz}$.

Let us consider only functions that are polynomials in $p$. Taking into account that the Fourier transformation $F \to \tilde{F}$ acts on the variable $p$ only, Eq. (174) can be written as

$$_\varpi \langle y| \mathbf{Q}_{[\omega]}(F) |z\rangle_\varpi$$
$$= \frac{1}{\mathscr{M}(\bar{x}, \bar{\bar\xi})} \int \left(g^{-1/2}F\right)^{\sim}\left(T_\theta\left(\bar{x}, \bar{\bar\xi}\right)\right) \omega(\theta)d\theta. \quad (175)$$

Here, and in what follows, we often use the following notation for the partial Fourier transform $\tilde{A}(x, \xi)$ of a function $A(x, p)$:

$$\tilde{A}(x,\xi) = (A)^\sim(x,\xi).$$

If we set $\omega(\theta) = \delta(\theta)$ in $\mathbf{Q}_{[\omega]}(F)$, we obtain a minimal generalization of the Weyl quantization to the curved space case,

$$\mathbf{Q}_{[\omega]}(F)\big|_{\omega(\theta)=\delta(\theta)} = \mathbf{Q}^{\mathrm{w}}(F). \quad (176)$$

In this case $\varpi = 0$, according to Eq. (98). Then, taking into account that for the geodesic flux the relation $T_0(\bar{x}, \bar{\bar\xi}) = (\bar{x}, \bar{\bar\xi})$ holds true, the matrix elements of the minimal generalization of the Weyl quantization to the curved space case has the form

$$_0\langle y| \mathbf{Q}^{\mathrm{w}}(F) |z\rangle_0 = \frac{1}{\mathscr{M}(\bar{x}, \bar{\bar\xi})} \left(g^{-1/2}F\right)^{\sim}(\bar{x}, \bar{\bar\xi})$$
$$= \frac{\tilde{F}(\bar{x}, \bar{\bar\xi})}{\mathscr{M}(\bar{x}, \bar{\bar\xi})\sqrt{g(\bar{x})}}. \quad (177)$$

According to Sect. 2.6, a function $F_\Omega(x, p)$ is the $\Omega$-symbol of the operator $\hat{F}$ if $\hat{F} = Q_\Omega(F_\Omega)$, see Eq. (5). Following this notational logic, the function $F_{[\omega]}(x, p)$ is $[\omega]$-symbol of the operator is $\hat{F}$, if $\hat{F} = \mathbf{Q}_{[\omega]}(F_{[\omega]})$. Equations (175) and (177) relate matrix elements of an operator $\hat{F}$ with its $[\omega]$-symbols and its w-symbols $F^{\mathrm{w}}$ as

$$_\varpi \langle y| \hat{F} |z\rangle_\varpi$$
$$= \frac{1}{\mathscr{M}(\bar{x}, \bar{\bar\xi})} \int \left(g^{-1/2}F_{[\omega]}\right)^{\sim} T_\theta(\bar{x}, \bar{\bar\xi}) \omega(\theta)d\theta$$
$$_0\langle y| \hat{F} |z\rangle_0 = \frac{\widetilde{F^{\mathrm{w}}}(\bar{x}, \bar{\bar\xi})}{\mathscr{M}(\bar{x}, \bar{\bar\xi})\sqrt{g(\bar{x})}}, \quad (178)$$

where $\bar{x} = \bar{x}(y, z)$ and $\bar{\bar\xi} = \bar{\bar\xi}(y, z)$ are a middle point and a tangent vector of the minimal geodesic line that connects $y$ and $z$. It follows from Eq. (119) that

$$_\varpi \langle y| \hat{F} |z\rangle_\varpi = e^{i\varpi(\ln\sqrt{g(y)} - \ln\sqrt{g(z)})} {}_0\langle y| \hat{F} |z\rangle_0. \quad (179)$$

Taking into account that $y = \gamma\left(\bar{x}, \frac{-1}{2}\bar{\bar\xi}\right)$ and $z = \gamma\left(\bar{x}, \frac{1}{2}\bar{\bar\xi}\right)$, we obtain a relation between $[\omega]$-symbols and w-symbols,

$$\widetilde{F^{\mathrm{w}}}(x,\xi) = \Lambda_\varpi(x,\xi)\sqrt{g(x)}$$
$$\times \int \left(g^{-1/2}F_{[\omega]}\right)^{\sim} T_\theta(x,\xi)\omega(\theta)d\theta, \quad (180)$$

where

$$\Lambda_\varpi(x, \xi) = \exp\left\{i\varpi\left[\ln\sqrt{g\left(\gamma\left(x, \frac{1}{2}\xi\right)\right)}\right.\right.$$
$$\left.\left. - \ln\sqrt{g\left(\gamma\left(x, \frac{-1}{2}\xi\right)\right)}\right]\right\}$$
$$= \exp\left[i\varpi\xi^\mu \Gamma_\mu(x) + O\left(\xi^3\right)\right] = 1 + i\varpi\xi^\mu \Gamma_\mu(x)$$
$$- \left[\varpi\xi^\mu \Gamma_\mu(x)\right]^2 + O\left(\xi^3\right), \quad (181)$$

and the expansion of $\ln\sqrt{g\left(\gamma\left(x, \pm\frac{1}{2}\xi\right)\right)}$ was derived with the help of Eq. (278).

Below, we decompose the latter quantity (181) in a power series with respect to the Planck constant $\hbar$. To this end, we consider a partial Fourier transform $\tilde{A}(x, \xi)$ (given by Eq. (83)) of a function $A(x, p)$. Using Eq. (264), we calculate the derivative

$$\partial_\theta \tilde{A}(T_\theta(x, \xi)) = \partial_\theta \tilde{A}(x_\theta, \xi_\theta) = \partial_\mu \tilde{A}(x_\theta, \xi_\theta) \xi_\theta^\mu$$
$$- \frac{\partial \tilde{A}}{\partial \xi^\mu}(x_\theta, \xi_\theta) \Gamma^\mu_{\alpha\beta}(x_\theta) \xi_\theta^\alpha \xi_\theta^\beta. \quad (182)$$

According to Eq. (83), the multiplication of $\tilde{A}$ by $\xi^\mu$ is equivalent to the action of the derivative $-i\hbar\frac{\partial}{\partial p_\mu}$. Taking this into account, we can represent the term $\partial_\mu \tilde{A}(x_\theta, \xi_\theta)\xi_\theta^\mu$ as follows:

$$\partial_\mu \tilde{A}(x_\theta, \xi_\theta) \xi_\theta^\mu = \left(\partial_\mu A\right)^\sim (x_\theta, \xi_\theta) \xi_\theta^\mu$$
$$= -i\hbar \left(\frac{\partial^2 A}{\partial x^\mu \partial p_\mu}\right)^\sim (x_\theta, \xi_\theta).$$

Similarly, the action of the derivative $\frac{\partial}{\partial \xi^\mu}$ on $\tilde{A}$ corresponds to a multiplication $\tilde{A}$ by $-\frac{i}{\hbar}p_\mu$. Therefore, the next term can be written as

$$\frac{\partial \tilde{A}}{\partial \xi^\mu}(x_\theta, \xi_\theta) \Gamma^\mu_{\alpha\beta}(x_\theta) \xi_\theta^\alpha \xi_\theta^\beta$$
$$= -\frac{i}{\hbar}\left(p_\mu A\right)^\sim (x_\theta, \xi_\theta) \Gamma^\mu_{\alpha\beta}(x_\theta) \xi_\theta^\alpha \xi_\theta^\beta$$
$$= -\frac{i}{\hbar}\left[\Gamma^\mu_{\alpha\beta} p_\mu A\right]^\sim (x_\theta, \xi_\theta) \xi_\theta^\alpha \xi_\theta^\beta$$
$$= i\hbar \left[\Gamma^\mu_{\alpha\beta}\frac{\partial^2 (p_\mu A)}{\partial p_\alpha \partial p_\beta}\right]^\sim (x_\theta, \xi_\theta).$$

Thus, the derivative (182) can be rewritten as

$$\partial_\theta \tilde{A}(T_\theta(x, \xi)) = \left[-i\hbar \mathbf{\Pi}' A\right]^\sim (T_\theta(x, \xi)), \quad (183)$$

where the operator $\Pi'$ is defined as follows:

$$\mathbf{\Pi}' A(x, p) = \left(\frac{\partial^2}{\partial x^\mu \partial p_\mu} + 2\Gamma_\alpha \frac{\partial}{\partial p_\alpha}\right.$$
$$\left. + p_\mu \Gamma^\mu_{\alpha\beta} \frac{\partial^2}{\partial p_\alpha \partial p_\beta}\right) A(x, p). \quad (184)$$

Since the derivative in Eq. (183) is a function of $T_\theta(x, \xi)$, we obtain

$$\frac{\mathrm{d}^n}{\mathrm{d}\theta^n}\tilde{A}(T_\theta(x, \xi)) = \left[\left(-i\hbar\mathbf{\Pi}'\right)^n A\right]^\sim (T_\theta(x, \xi)),$$
$$\tilde{A}(T_\theta(x, \xi)) = \left[\sum_{n=0}^\infty \frac{\theta^n}{n!}\left(-i\hbar\mathbf{\Pi}'\right)^n A\right]^\sim (x, \xi). \quad (185)$$

Now we study the $\mathbf{Q}_{(w)}$-symbols (180). To this end, using the relation $\partial_\mu g^{-1/2} = -g^{-1/2}\Gamma_\mu(x)$, which follows from Eq. (252), we write

$$\mathbf{\Pi}'\left(g^{-1/2} F(x, p)\right) = \left[\left(\partial_a g^{-1/2} + 2g^{-1/2}\Gamma_\alpha\right.\right.$$
$$\left.\left. + g^{-1/2} p_\mu \Gamma^\mu_{\alpha\beta}\frac{\partial}{\partial p_\beta}\right)\frac{\partial}{\partial p_\alpha}\right] F(x, p)$$
$$= g^{-1/2}\mathbf{\Pi} F(x, p) \Longrightarrow \left(\mathbf{\Pi}'\right)^n\left(g^{-1/2} F(x, p)\right)$$
$$= g^{-1/2}\mathbf{\Pi}^n F(x, p),$$

where

$$\mathbf{\Pi} = \frac{\partial^2}{\partial x^\mu \partial p_\mu} + \Gamma_\alpha\frac{\partial}{\partial p_\alpha} + p_\mu \Gamma^\mu_{\alpha\beta}\frac{\partial^2}{\partial p_\alpha \partial p_\beta}. \quad (186)$$

With the help of Eq. (185), we derive the expansion

$$\left(g^{-1/2} F\right)^\sim T_\theta(x, \xi) = g^{-1/2}\left[\sum_{k=0}^\infty \frac{(i\theta)^n}{n!}(-\hbar\mathbf{\Pi})^n F\right]^\sim (x, \xi),$$

which allows one to obtain a power series with respect to the Planck constant $\hbar$ for the w-symbols (180),

$$\widetilde{F^{\mathrm{w}}}(x, \xi) = \Lambda_\varpi(x, \xi)\left[\sum_{k=0}^\infty \frac{\omega^n}{n!}(-\hbar\mathbf{\Pi})^n F_{[\omega]}\right]^\sim (x, \xi)$$
$$= \left[\Lambda_\varpi\left(x, -i\hbar\frac{\partial}{\partial p}\right)\sum_{k=0}^\infty \frac{\omega^n}{n!}(-\hbar\mathbf{\Pi})^n F_{[\omega]}\right]^\sim (x, \xi),$$
$$(187)$$

where

$$\omega^n = \int (i\theta)^n \omega(\theta)\mathrm{d}\theta = \frac{\mathrm{d}^n \Omega}{\mathrm{d}k^n}(0); \quad (188)$$

see Eq. (45). Due to the fact that the partial Fourier transform is invertible, Eq. (187) implies

$$F^{\mathrm{w}}(x, p) = \Lambda_\varpi\left(x, -i\hbar\frac{\partial}{\partial p}\right)\sum_{n=0}^\infty \frac{\omega^n}{n!}(-\hbar\mathbf{\Pi})^n F_{[\omega]}(x, p). \quad (189)$$

We note that the exponent (181) is an even function because an analytical extension of $\Lambda_\varpi$ to imaginary argument $\xi$ is real. Thus, Eq. (189) can be written as follows:

$$F^{\mathrm{w}}(x, p) = \Lambda_\varpi\left(x, -i\hbar\frac{\partial}{\partial p}\right)\Omega\left(-\hbar\mathbf{\Pi}\right) F_{[\omega]}(x, p). \quad (190)$$

The normalization condition $\omega^0 = 1$ implies that $\Omega(0) = 1$, similarly to the plane space case. The operator $\boldsymbol{\Pi}$ is a generalization of the operator (76) to the plane space.

### 4.2 An extended family of covariant quantizations in curved space

#### 4.2.1 Definition of the general quantization

It follows from Eq. (175) and from the fact that $\mathscr{M}(x, \xi) = 1$ in the plane space case, that in such a case

$$\varpi \langle y | \mathbf{Q}_{[\omega]}(F) | z \rangle_\varpi \big|_{\text{plane space}} = \varpi \langle y | Q_{[\omega]}(F) | z \rangle_\varpi$$
$$= \int (g^{-1/2} F)^\sim T_\theta(\bar{x}, \bar{\xi}) \omega(\theta) d\theta. \quad (191)$$

Thus, we can suppose that a possible generalization of the quantization $\mathbf{Q}_{[\omega]}$ (and of the quantization $Q_{[\omega]}$ as well) to a curved space case is a correspondence $F(x, p) \to \hat{F}$ which has the same property (191) of its matrix elements. As a candidate for such a correspondence, we offer the one $\mathbf{Q}_{[\omega,\Theta]}(F)$, which is already parametrized not only by the previously introduced function $\omega(\theta)$, but by a new function $\Theta(x, \xi)$. We define this correspondence by its matrix elements as

$$\varpi \langle y | \mathbf{Q}_{[\omega,\Theta]}(F)(F) | z \rangle_\varpi$$
$$= \frac{\Theta(\bar{x}, \bar{\xi})}{\mathscr{M}(\bar{x}, \bar{\xi})} \int \left(g^{-1/2} F\right)^\sim \left(T_\theta(\bar{x}, \bar{\xi})\right) \omega(\theta) d\theta, \quad (192)$$

where the function $\Theta(x, \xi)$ must have the following properties:

1. $\Theta(x, \xi) = 1$ in the plane space case;
2. $\Theta$ is transformed as a scalar under transformations (291);
3. $\Theta(x, 0) = 1$;
4. $\Theta(x, \xi)$ is real and even with respect to the variable $\xi$ ($\Theta(x, \xi) = \Theta(x, -\xi)$).

Property 1. provides the first equality (191), which means that the quantization $\mathbf{Q}_{[\omega,\Theta]}(F)$ is a generalization of the quantization $Q_{[\omega]}(F)$. Property 2. is necessary to guarantee the covariance of $\mathbf{Q}_{[\omega,\Theta]}(F)$ under general coordinate transformations. Properties 3. and 4. provide the consistency (in particular, the correct classical limit (the correspondence principle)) of the quantization $\mathbf{Q}_{[\omega,\Theta]}(F)$.

Using Eq. (118), we can restore an operator $\mathbf{Q}_{[\omega,\Theta]}(F)$ which corresponds to matrix elements (192),

$$\mathbf{Q}_{[\omega,\Theta]}(F) = \int \varpi \langle y | \mathbf{Q}_{[\omega,\Theta]}(F) | z \rangle_\varpi \left(|y\rangle_\varpi \ \varpi \langle z|\right)$$
$$\times \sqrt{g(y) g(z)} dy dz. \quad (193)$$

In what follows, we call the quantization (193) the general covariant quantization.

One can see that the quantization $\mathbf{Q}_{[\omega]}(F)$ is a particular case of the general covariant quantization $\mathbf{Q}_{[\omega,\Theta]}(F)$ for $\Theta(x, \xi) = 1$,

$$\mathbf{Q}_{[\omega]}(F) = \mathbf{Q}_{[\omega,1]}(F). \quad (194)$$

We define Weyl versions of the general covariant quantization, setting $\omega(\theta) = \delta(\theta)$, similarly to Eq. (177), $\mathbf{Q}_{[\delta(\theta),\Theta]} = \mathbf{Q}^{\text{w}}_{[\Theta]}$,

$$_0 \langle y | \mathbf{Q}^{\text{w}}_{[\Theta]}(F) | z \rangle_0 = \frac{\Theta(\bar{x}, \bar{\xi})}{\mathscr{M}(\bar{x}, \bar{\xi})} \left(g^{-1/2} F\right)^\sim (\bar{x}, \bar{\xi})$$
$$= \frac{\Theta(\bar{x}, \bar{\xi})}{\mathscr{M}(\bar{x}, \bar{\xi})} \frac{\tilde{F}(\bar{x}, \bar{\xi})}{\sqrt{g(\bar{x})}}. \quad (195)$$

Thus, the general covariant quantization creates already a family (parametrized by the function $\Theta$) of the corresponding Weyl quantizations.

Similar to the definitions introduced before, we say that $F_{[\omega,\Theta]}(x, p)$ is the $[\omega, \Theta]$-symbol of the operator is $\hat{F}$, if $\hat{F} = \mathbf{Q}_{[\omega,\Theta]}(F_{[\omega,\Theta]})$, and $F^{\text{w}}_{[\Theta]}(x, p)$ is the $[\text{w}, \Theta]$-symbol of the operator is $\hat{F}$, if $\hat{F} = \mathbf{Q}^{\text{w}}_{[\Theta]}\left(F^{\text{w}}_{[\Theta]}\right)$.

Comparing Eqs. (192) and (195) with (175) and (177), we see that a relation between the symbols $F_{[\omega,\Theta]}$ and $F^{\text{w}}_{[\Theta]}$ is the same as a relation between the symbols $F_{[\omega]}$ and $F^{\text{w}}$, i.e.,

$$F^{\text{w}}_{[\Theta]}(x, p) = \Lambda_\varpi \left(x, -i\hbar \frac{\partial}{\partial p}\right) \sum_{n=0}^{\infty} \frac{\omega^n}{n!} (-\hbar \boldsymbol{\Pi})^n F_{[\omega,\Theta]}(x, p)$$
$$= \Lambda_\varpi \left(x, -i\hbar \frac{\partial}{\partial p}\right) \Omega(-\hbar \boldsymbol{\Pi}) F_{[\omega,\Theta]}(x, p). \quad (196)$$

Moreover, Eqs. (177) and (195) imply the following relations:

$$\widetilde{F^{\text{w}}}(x, \xi) = \Theta(x, \xi) \widetilde{F^{\text{w}}_{[\Theta]}}(x, \xi)$$
$$\implies F^{\text{w}}(x, p) = \Theta\left(x, -i\hbar \frac{\partial}{\partial p}\right) F^{\text{w}}_{[\Theta]}(x, p). \quad (197)$$

Thus, the symbol $F^{\text{w}}$ can be expressed via the symbol $F_{[\omega,\Theta]}$ as follows:

$$F^{\text{w}}(x, p) = \Theta\left(x, -i\hbar \frac{\partial}{\partial p}\right) \Lambda_\varpi \left(x, -i\hbar \frac{\partial}{\partial p}\right) \Omega(-\hbar \boldsymbol{\Pi})$$
$$\times F_{[\omega,\Theta]}(x, p). \quad (198)$$

Equation (190) allows us to express the symbol $F_{[\omega]}$ via the symbol $F_{[\omega,\Theta]}$,

$$F_{[\omega]}(x, p) = [\Omega(-\hbar \boldsymbol{\Pi})]^{-1} \Theta\left(x, -i\hbar \frac{\partial}{\partial p}\right)$$
$$\times \Omega(-\hbar \boldsymbol{\Pi}) F_{[\omega,\Theta]}(x, p). \quad (199)$$

According to Eq. (6), we have the equality $\mathbf{Q}_{[\omega,\Theta]}\left(F_{[\omega,\Theta]}\right) = \mathbf{Q}_{[\omega]}\left(F_{[\omega]}\right)$, which allows us to relate the quantizations $\mathbf{Q}_{[\omega,\Theta]}$ and $\mathbf{Q}_{[\omega]}$ as

$$\mathbf{Q}_{[\omega,\Theta]}(F) = \mathbf{Q}_{[\omega]}\left([\Omega(-\hbar\mathbf{\Pi})]^{-1}\Theta\left(x, -i\hbar\frac{\partial}{\partial p}\right)\right.$$
$$\left. \times \Omega(-\hbar\mathbf{\Pi})F\right). \quad (200)$$

We use this relation to verify the basic properties of the quantization $\mathbf{Q}_{[\omega,\Theta]}(F)$.

The operator $\Omega(-\hbar\mathbf{\Pi})$ is real, therefore $\mathbf{\Pi}$ is real as well. Then Eqs. (48) and (45) imply that the function $\Omega(k)$ is real. Since $\Theta(x, \xi)$ is even in the variable $\xi$, the operator $\Theta(x, -i\hbar\frac{\partial}{\partial p})$ is real as well. Moreover, if $F$ is real then the function $[\Omega(-\hbar\mathbf{\Pi})]^{-1}\Theta(x, -i\hbar\frac{\partial}{\partial p})\Omega(-\hbar\mathbf{\Pi})F$ is also real. Then it follows from Eq. (200) and from the hermiticity of the quantization $\mathbf{Q}_{[\omega]}$ that the operator $\mathbf{Q}_{[\omega,\Theta]}(F)$ is self-adjoint which means, in our terminology, that the quantization $\mathbf{Q}_{[\omega,\Theta]}(F)$ is hermitian.

Let us consider a function $F = F(x)$. Then $\mathbf{\Pi}F = 0$, according to definition (186). Taking into account that $\Omega(0) = 1$, see Eq. (46), and the property $\Theta(x, 0) = 1$, we have

$$[\Omega(-\hbar\mathbf{\Pi})]^{-1}\Theta\left(x, -i\hbar\frac{\partial}{\partial p}\right)\Omega(-\hbar\mathbf{\Pi})F$$
$$= [\Omega(0)]^{-1}\Theta(x, 0)\Omega(0)F = F. \quad (201)$$

Then, as follows from Eq. (200),

$$\mathbf{Q}_{[\omega,\Theta]}(F(x)) = \mathbf{Q}_{[\omega]}(F(x)) = F(\hat{x})$$
$$\Longrightarrow \mathbf{Q}_{[\omega,\Theta]}(1) = \hat{I}, \quad \mathbf{Q}_{[\omega,\Theta]}(x^\mu) = \hat{x}^\mu. \quad (202)$$

Let us consider a polynomial $T^\mu(x)p_\mu$. Then it follows from Eq. (186) that $\Omega(-\hbar\mathbf{\Pi})(T^\mu(x)p_\mu)$ is a polynomial of the same structure. Due to the properties of the function $\Theta(x, \xi)$, it has the form $\Theta(x, \xi) = 1 + O(\xi^2)$, which implies the relation

$$[\Omega(-\hbar\mathbf{\Pi})]^{-1}\Theta\left(x, -i\hbar\frac{\partial}{\partial p}\right)\Omega(-\hbar\mathbf{\Pi})T^\mu(x)p_\mu$$
$$= T^\mu(x)p_\mu.$$

Then it follows from Eq. (200) that

$$\mathbf{Q}_{[\omega,\Theta]}(T^\mu(x)p_\mu) = \mathbf{Q}_{[\omega]}(T^\mu(x)p_\mu). \quad (203)$$

In particular,

$$\mathbf{Q}_{[\omega,\Theta]}(p_\mu) = \hat{p}_\mu \Longrightarrow [\mathbf{Q}_{[\omega,\Theta]}(p_\mu)]_x$$
$$= -i\hbar\left[\partial_\mu + \left(\frac{1}{2} - i\varpi\right)\Gamma_\mu\right]. \quad (204)$$

### 4.2.2 Verification of the covariance

Taking into account definition (192) of the quantization $\mathbf{Q}_{[\omega,\Theta]}(F)$ and definition (265) of the geodesic flux, we can write

$$_\varpi\langle y|\mathbf{Q}_{[\omega,\Theta]}(F)|z\rangle_\varpi = \frac{\Theta(\bar{x}, \bar{\xi})}{\mathcal{M}(\bar{x}, \bar{\xi})}\int(g^{-1/2}F)^\sim$$
$$\times\left(\gamma(\bar{x}, \theta\bar{\xi}), \frac{1}{\theta}\beta(\bar{x}, \theta\bar{\xi})\right)\omega(\theta)\mathrm{d}\theta,$$

where $\bar{x} = \bar{x}(y, z)$ and $\bar{\xi} = \bar{\xi}(y, z)$ are a middle point and a tangent vector of the minimal geodesic line that connects $y$ and $z$. Using the transformation (114), we can write

$$_\varpi\langle y|\mathbf{Q}_{[\omega,\Theta]}|z\rangle_\varpi = \frac{\Theta(\bar{x}, \bar{\xi})}{\mathcal{M}(\bar{x}, \bar{\xi})}\int\left(g'^{-1/2}F'\right)^\sim$$
$$\times\left(\varphi(\gamma(\bar{x}, \theta\bar{\xi})), \frac{\partial x'}{\partial x}(\gamma(\bar{x}, \theta\bar{\xi}))\frac{1}{\theta}\beta(\bar{x}, \theta\bar{\xi})\right)\omega(\theta)\mathrm{d}\theta.$$

Then taking into account transformation properties (269) and (270) of the exponential function, and the fact that $\Theta$ is a scalar, we obtain

$$_\varpi\langle y|\mathbf{Q}_{[\omega,\Theta]}|z\rangle_\varpi = \frac{\Theta'(\bar{x}', \bar{\xi}')}{\mathcal{M}'(\bar{x}', \bar{\xi}')}\int\left(g'^{-1/2}F'\right)^\sim$$
$$\times\left(\gamma'(\bar{x}', \theta\bar{\xi}'), \frac{1}{\theta}\beta'(\bar{x}', \theta\bar{\xi}')\right)\omega(\theta)\mathrm{d}\theta$$
$$= \frac{\Theta'(\bar{x}', \bar{\xi}')}{\mathcal{M}'(\bar{x}', \bar{\xi}')}\int\left(g'^{-1/2}F'\right)^\sim$$
$$\times\left(T_\theta'(\bar{x}', \bar{\xi}')\right)\omega(\theta)\mathrm{d}\theta, \quad (205)$$

where $\bar{x}' = \varphi(\bar{x})$ and $\bar{\xi}' = \frac{\partial x'}{\partial x}(\bar{x})\xi$. According to (269),

$$\gamma'\left(\bar{x}', \frac{-1}{2}\bar{\xi}'\right) = \varphi\left(\gamma\left(\bar{x}, \frac{-1}{2}\bar{\xi}\right)\right) = \varphi(y) = y',$$
$$\gamma'\left(\bar{x}', \frac{1}{2}\bar{\xi}'\right) = \varphi\left(\gamma\left(\bar{x}, \frac{1}{2}\bar{\xi}\right)\right) = \varphi(z) = z',$$

and, therefore, $\bar{x}'$ and $\bar{\xi}'$ are a middle point and a tangent vector of the minimal geodesic line that connects $y'$ and $z'$ in the reference frame $K_{x'}$. One can see that the last term in Eq. (205) is the matrix element (192) in such a reference frame. Thus, we have the equality

$$_\varpi\langle y|\mathbf{Q}_{[\omega,\Theta]}(F)|z\rangle_\varpi = {}_\varpi\langle y'|'\mathbf{Q}'_{[\omega,\Theta]}(F')|z'\rangle'_\varpi.$$

Now we have to determine which form Eq. (18) takes for the general quantization $\mathbf{Q}_{[\omega,\Theta]}$. We recall that these equations define a linear transformation $U : \mathfrak{H} \to \mathfrak{H}'$. The property $\mathbf{Q}_{[\omega,\Theta]}(F(x)) = F(\hat{x})$, derived in the previous section, shows that the first equation of (18) has the same form as (157). It follows from Eq. (203) that (160) holds true for the second equation of (18). As in Sect. 4.2, this fact means that if vectors $|x\rangle \in \mathfrak{H}$ satisfy Eq. (118), then vectors $|x'\rangle' = |\varphi(x)\rangle' = U|x\rangle \in \mathfrak{H}'$ will satisfy similar equations in the reference frame $K_{x'}$.

Now we calculate the operator $U\mathbf{Q}_{[\omega,\Theta]}(F)U^\dagger$ using Eq. (193) and the relation $U|x\rangle = |\varphi(x)\rangle'$,

$$U\mathbf{Q}_{[\omega,\Theta]}(F)U^\dagger$$
$$= \int_\varpi \langle y| \mathbf{Q}_{[\omega,\Theta]}(F) |z\rangle_\varpi \, |\varphi(y)\rangle' \, \langle\varphi(z)|' \sqrt{g(y)g(z)} \mathrm{d}y\mathrm{d}z$$
$$= \int_\varpi \langle y'|' \mathbf{Q}'_{[\omega,\Theta]}(F') |z'\rangle'_\varpi \, |y'\rangle_\varpi \, \varpi \langle z'|' \sqrt{g'(y')g'(z')} \mathrm{d}y'\mathrm{d}z'$$
$$= \mathbf{Q}'_{[\omega,\Theta]}(F'). \tag{206}$$

In the course of this calculation, we have used the variable change $y' = \varphi(y)$ and $z' = \varphi(z')$ and the equality $\sqrt{g}\mathrm{d}x = \sqrt{g'}\mathrm{d}x'$. Equation (206) justifies the covariance of the general quantization $\mathbf{Q}_{[\omega,\Theta]}(F)$.

### 4.2.3 Correspondence principle

According to definition (195), a matrix element of an operator $\hat{F} = \mathbf{Q}_{[\mathrm{w},\mathscr{M}]}\left(F^{\mathrm{w}}_{[\mathscr{M}]}\right)$ is related to his $[\mathrm{w}, \mathscr{M}]$-symbol $F^{\mathrm{w}}_{[\mathscr{M}]}$ as

$$_0\langle y| \hat{F} |z\rangle_0 = \left[g^{-1/2} F\left(\bar{x}(y,z), \bar{\xi}(y,z)\right)\right]^\sim.$$

The $[\mathrm{w}, \mathscr{M}]$-star product satisfies the relation $\mathbf{Q}_{[\mathrm{w},\mathscr{M}]}\left(F^{\mathrm{w}}_{[\mathscr{M}]} * G^{\mathrm{w}}_{[\mathscr{M}]}\right) = \hat{F}\hat{G}$, which implies

$$_0\langle y| \hat{F}\hat{G} |z\rangle_0$$
$$= \left[g^{-1/2}\left(F^{\mathrm{w}}_{[\mathscr{M}]} * G^{\mathrm{w}}_{[\mathscr{M}]}\right)\right]^\sim\left(\bar{x}(y,z), \bar{\xi}(y,z)\right).$$

It follows from the completeness of the states $|x\rangle$, see (118), that

$$_0\langle y| \hat{F}\hat{G} |z\rangle_0 = \int _0\langle y| \hat{F} |w\rangle_0 \, _0\langle w| \hat{G} |z\rangle_0 \sqrt{g(w)} \mathrm{d}w.$$

Therefore, we can obtain the following relation:

$$\left[g^{-1/4}\left(F^{\mathrm{w}}_{[\Theta]} * G^{\mathrm{w}}_{[\Theta]}\right)\right]^\sim\left(\bar{x}(y,z), \bar{\xi}(y,z)\right)$$
$$= \int \left[g^{-1/2} F\left(\bar{x}(y,w), \bar{\xi}(y,w)\right)\right]^\sim$$
$$\times \left[g^{-1/2} G\left(\bar{x}(w,z), \bar{\xi}(w,z)\right)\right]^\sim \sqrt{g(w)} \mathrm{d}w. \tag{207}$$

Let us find an expansion of the $[\mathrm{w}, \Theta]$-star product in powers of $\hbar$. If $x = \bar{x}(y,z)$ and $\xi = \bar{\xi}(y,z)$, then $y = \gamma\left(x, \frac{-1}{2}\xi\right)$ and $z = \gamma\left(x, \frac{1}{2}\xi\right)$. According to expansion (278), we have

$$c^\mu = \frac{y^\mu + z^\mu}{2} = x^\mu - \frac{1}{8}\Gamma^\mu_{\mu_1\mu_2}(x)\xi^{\mu_1}\xi^{\mu_2} + O\left(\xi^4\right), \tag{208}$$

$$l^\mu = z^\mu - y^\mu = \xi^\mu + \frac{1}{48}\Gamma^\mu_{\mu_1\mu_2\mu_3}(x)\xi^{\mu_1}\xi^{\mu_2}\xi^{\mu_3} + O\left(\xi^5\right). \tag{209}$$

If $A(x, p)$ is an arbitrary function, then it follows from Eq. (208) that

$$\tilde{A}(x, \xi) = \tilde{A}(c, \xi) + \frac{1}{8}\Gamma^\mu_{\mu_1\mu_2}(x)\xi^{\mu_1}\xi^{\mu_2}\left[\frac{\partial A}{\partial x^\mu}\right]^\sim(c, \xi)$$
$$+ O(\xi^4)$$
$$= \tilde{A}(c, \xi) - \frac{\hbar^2}{8}\left[\Gamma^\mu_{\mu_1\mu_2}(x) \frac{\partial^3 A}{\partial p_{\mu_1}\partial p_{\mu_2}\partial x^\mu}\right]^\sim(c, \xi)$$
$$+ O(\hbar^4). \tag{210}$$

Equation (209) between $l^\mu$ and $\xi^\mu$ can be inverted,

$$\xi^\mu = l^\mu - \frac{1}{48}\Gamma^\mu_{\mu_1\mu_2\mu_3}(x)l^{\mu_1}l^{\mu_2}l^{\mu_3} + O(l^5),$$

and used in Eq. (210),

$$\tilde{A}(c, \xi) = \tilde{A}(c, l) - \frac{1}{48}\Gamma^\mu_{\mu_1\mu_2\mu_3}(x)l^{\mu_1}l^{\mu_2}l^{\mu_3}\left(\frac{\partial A}{\partial \xi^\mu}\right)^\sim(c, l)$$
$$+ O(l^5)$$
$$= \tilde{A}(c, l) - \frac{\hbar^2}{48}\left[\Gamma^\mu_{\mu_1\mu_2\mu_3}(x) \frac{\partial^3(p_\mu A)}{\partial p_{\mu_1}\partial p_{\mu_2}\partial p_{\mu_3}}\right]^\sim(c, l)$$
$$+ O(\hbar^4).$$

Thus, we obtain

$$\tilde{A}\left(\bar{x}(y,z), \bar{\xi}(y,z)\right) = \tilde{A}\left(\frac{y+z}{2}, z-y\right) + O(\hbar^2).$$

Taking into account definitions (208) and (209), we rewrite Eq. (207) as

$$\left[g^{-1/4}(F * G)\right]^\sim(x, \xi)$$
$$= \int \left[g^{-1/2} F\right]^\sim\left(x - \frac{z-w}{2}, w - y\right)\left[g^{-1/2} G\right]^\sim$$
$$\times \left(x + \frac{w-y}{2}, z - w\right)\sqrt{g(w)}\mathrm{d}w + O(\hbar^2). \tag{211}$$

In terms of new variables $\eta = w - y, \zeta = z - w$, the integrand in (211) can be written as

$$\left(g^{-1/2}F\right)^\sim\left(x - \frac{\zeta}{2}, \eta\right)\left(g^{-1/2}G\right)^\sim\left(x + \frac{\eta}{2}, \zeta\right)$$
$$= \left[\left(g^{-1/2}F\right)^\sim(x, \eta) - \frac{\zeta^\mu}{2}\partial_\mu\left(g^{-1/2}F\right)^\sim(x, \eta) + O(\zeta^2)\right]$$
$$\times \left[\left(g^{-1/2}G\right)^\sim(x, \zeta) + \frac{\eta^\mu}{2}\partial_\mu\left(g^{-1/2}G\right)^\sim(x, \zeta) + O(\eta^2)\right]. \tag{212}$$

Then using the equalities

$$\zeta^\mu \left(g^{-1/2}G\right)^\sim(x, \zeta) = -i\hbar\left(g^{-1/2}\frac{\partial G}{\partial p_\mu}\right)^\sim(x, \zeta),$$

$$\eta^\mu \left(g^{-1/2}F\right)^\sim(x, \eta) = -i\hbar\left(g^{-1/2}\frac{\partial F}{\partial p_\mu}\right)^\sim(x, \eta),$$

and the equation $\partial_\mu(g^{-1/2}F) = g^{-1/2}(\partial_\mu F - \Gamma_\mu F)$, which can be derived from Eq. (252), we transform the l.h.s. of Eq. (212) to the following form:

$$\left(g^{-1/2}F\right)^{\sim}\left(x-\frac{\zeta}{2},\eta\right)\left(g^{-1/2}G\right)^{\sim}\left(x+\frac{\eta}{2},\zeta\right)$$
$$= g^{-1}\widetilde{F}(x,\eta)\widetilde{G}(x,\zeta)$$
$$+ \frac{i\hbar g^{-1}}{2}\left\{\frac{\widetilde{\partial F}}{\partial x^{\mu}}(x,\eta)\frac{\widetilde{\partial G}}{\partial p_{\mu}}(x,\zeta) - \frac{\widetilde{\partial F}}{\partial p_{\mu}}(x,\eta)\frac{\widetilde{\partial G}}{\partial x^{\mu}}(x,\zeta)\right.$$
$$+ \Gamma_{\mu}(x)\left[\frac{\widetilde{\partial F}}{\partial p_{\mu}}(x,\eta)\tilde{G}(x,\zeta) - \tilde{F}(x,\eta)\frac{\widetilde{\partial G}}{\partial p_{\mu}}(x,\zeta)\right]\right\}$$
$$+ O(\hbar^{2}).$$

Thus, integral (211) can be represented as a sum of three integrals,

$$\left[g^{-1/2}(F*G)\right]^{\sim}(x,\xi) = I_1 + I_2 + I_3 + O(\hbar^2),$$
$$I_1 = g^{-1}\int \widetilde{F}(x,\eta)\widetilde{G}(x,\zeta)\sqrt{g(w)}dw,$$
$$I_2 = \frac{i\hbar g^{-1}}{2}\int\left(\frac{\widetilde{\partial F}}{\partial x^{\mu}}(x,\eta)\frac{\widetilde{\partial G}}{\partial p_{\mu}}(x,\zeta)\right.$$
$$\left. - \frac{\widetilde{\partial F}}{\partial p_{\mu}}(x,\eta)\frac{\widetilde{\partial G}}{\partial x^{\mu}}(x,\zeta)\right)\sqrt{g(w)}dw,$$
$$I_3 = \frac{i\hbar g^{-1}}{2}\Gamma_{\mu}(x)\int\left[\frac{\widetilde{\partial F}}{\partial p_{\mu}}(x,\eta)\tilde{G}(x,\zeta)\right.$$
$$\left. - \tilde{F}(x,\eta)\frac{\widetilde{\partial G}}{\partial p_{\mu}}(x,\zeta)\right]\sqrt{g(w)}dw. \qquad (213)$$

One can see that all integrals in (213) have the same structure,

$$\int \widetilde{F}(x,\eta)\widetilde{G}(x,\zeta)\sqrt{g(w)}dw$$
$$= \int \widetilde{F}(x,w-y)\widetilde{G}(x,z-w)\sqrt{g(w)}dw. \qquad (214)$$

Now we calculate the first terms of the expansion of the typical integral (214) in powers of $\hbar$. For simplicity, let us suppose that we deal with polynomial functions in their arguments. Then one can write

$$\widetilde{F}(x,w-y) = F\left(x,i\hbar\frac{\partial}{\partial k}\right)\delta(w-y+k)\bigg|_{k=0},$$

such that

$$\int \widetilde{F}(x,\eta)\widetilde{G}(x,\zeta)\sqrt{g(w)}dw$$
$$= F\left(x,i\hbar\frac{\partial}{\partial k_1}\right)G\left(x,i\hbar\frac{\partial}{\partial k_2}\right)\delta(\xi+k_1+k_2)$$
$$\times \sqrt{g\left(x+\frac{k_2-k_1}{2}\right)}\bigg|_{k_1=k_2=0}. \qquad (215)$$

From Eq. (252) one can derive the following relations:

$$\frac{\partial}{\partial k_s^{\mu}}\sqrt{g\left(x+\frac{k_2-k_1}{2}\right)} = \sqrt{g\left(x+\frac{k_2-k_1}{2}\right)}$$
$$\times \left(\frac{\partial}{\partial k_s^{\mu}} + \frac{(-1)^s}{2}\Gamma_{\mu}\left(x+\frac{k_2-k_1}{2}\right)\right), \quad s=1,2,$$

which, being used in Eq. (215), result in the representation

$$\int \widetilde{F}(x,\eta)\widetilde{G}(x,\zeta)\sqrt{g(w)}dw$$
$$= g^{1/2}F\left(x,i\hbar\frac{\partial}{\partial k_1} - \frac{i\hbar}{2}\Gamma_{\mu}\left(x+\frac{k_2-k_1}{2}\right)\right)$$
$$\times G\left(x,i\hbar\frac{\partial}{\partial k_2} + \frac{i\hbar}{2}\Gamma_{\mu}\left(x+\frac{k_2-k_1}{2}\right)\right)\delta(\xi+k_1+k_2)\bigg|_{k_1=k_2=0}. \qquad (216)$$

Decompositions of operators in the r.h.s. of (216) in a power series with respect to $\hbar$ have the form

$$F\left(x,i\hbar\frac{\partial}{\partial k_1} - \frac{i\hbar}{2}\Gamma_{\mu}\left(x+\frac{k_2-k_1}{2}\right)\right)$$
$$= F\left(x,i\hbar\frac{\partial}{\partial k_1}\right) - \frac{i\hbar}{2}\Gamma_{\mu}\left(x+\frac{k_2-k_1}{2}\right)$$
$$\times \frac{\partial F}{\partial p_{\mu}}\left(x,i\hbar\frac{\partial}{\partial k_1}\right) + O(\hbar^2);$$
$$G\left(x,i\hbar\frac{\partial}{\partial k_2} + \frac{i\hbar}{2}\Gamma_{\mu}\left(x+\frac{k_2-k_1}{2}\right)\right)$$
$$= G\left(x,i\hbar\frac{\partial}{\partial k_2}\right) + \frac{i\hbar}{2}\Gamma_{\mu}\left(x+\frac{k_2-k_1}{2}\right)$$
$$\times \frac{\partial G}{\partial p_{\mu}}\left(x,i\hbar\frac{\partial}{\partial k_2}\right) + O(\hbar^2). \qquad (217)$$

Then, taking into account that

$$F\left(x,i\hbar\frac{\partial}{\partial k_1}\right)\left[\frac{i\hbar}{2}\Gamma_{\mu}\left(x+\frac{k_2-k_1}{2}\right)\right]$$
$$= \left[\frac{i\hbar}{2}\Gamma_{\mu}\left(x+\frac{k_2-k_1}{2}\right)\right]F\left(x,i\hbar\frac{\partial}{\partial k_1}\right) + O(\hbar^2),$$

we obtain

$$\int \widetilde{F}(x,\eta)\widetilde{G}(x,\zeta)\sqrt{g(w)}dw = g^{1/2}(FG)^{\sim}(x,\xi)$$
$$+ g^{1/2}\frac{i\hbar}{2}\Gamma_{\mu}(x)\left(F\frac{\partial G}{\partial p_{\mu}} - \frac{\partial F}{\partial p_{\mu}}G\right)^{\sim}(x,\xi) + O(\hbar^2). \qquad (218)$$

Using this result, we obtain decompositions of integrals (213) in powers of $\hbar$ up to second power terms,

$$I_1 = g^{-1/2}(FG)^{\sim}(x,\xi) + \frac{i\hbar g^{-1/2}}{2}\Gamma_{\mu}(x)$$
$$\times \left(F\frac{\partial G}{\partial p_{\mu}} - \frac{\partial F}{\partial p_{\mu}}G\right)^{\sim}(x,\xi) + O(\hbar^2),$$
$$I_2 = \frac{i\hbar g^{-1/2}}{2}(\{F,G\})^{\sim}(x,\xi) + O(\hbar^2), \qquad (219)$$
$$I_3 = \frac{i\hbar g^{-1/2}}{2}\Gamma_{\mu}(x)\left(\frac{\partial F}{\partial p_{\mu}}G - F\frac{\partial G}{\partial p_{\mu}}\right)^{\sim}(x,\xi) + O(\hbar^2).$$

Finally, we get

$$g^{-1/2}(F*G)^\sim(x,\xi) = g^{-1/2}(FG)^\sim(x,\xi)$$
$$+ \frac{i\hbar g^{-1/2}}{2}(\{F,G\})^\sim(x,\xi) + O(\hbar^2)$$
$$\Longrightarrow (F*G)(x,p) = (FG)(x,p)$$
$$+ \frac{i\hbar}{2}\{F,G\}(x,p) + O(\hbar^2). \tag{220}$$

Equation (220) implies that the $[w,\Theta]$-star product satisfies two conditions,

$$(F*G)(x,p) = (FG)(x,p) + O(\hbar),$$
$$(F*G - G*F)(x,p) = i\hbar\{F,G\}(x,p) + O(\hbar^2), \tag{221}$$

which means that for the quantization $\mathbf{Q}_{[w,\Theta]}$ the correspondence principle holds true; see Eq. (7).

It follows from Eqs. (197) that

$$F^w(x,\xi) = \Theta\left(x, -i\hbar\frac{\partial}{\partial p}\right) F^w_{[\Theta]}(x,\xi)$$
$$= \mathcal{M}\left(x, -i\hbar\frac{\partial}{\partial p}\right) F^w_{[\mathcal{M}]}(x,\xi).$$

Therefore,

$$F^w(x,\xi) = \mathcal{G}\left(x, -i\hbar\frac{\partial}{\partial p}\right) F^w_{[\mathcal{M}]}(x,\xi),$$
$$\mathcal{G}(x,\xi) = \frac{\mathcal{M}(x,\xi)}{\Theta(x,\xi)} = 1 + \frac{1}{2}\mathcal{G}_{\alpha\beta}(x)\xi^\alpha\xi^\beta + O(\xi^4), \tag{222}$$

with $\mathcal{G}_{\alpha\beta}(x) = \mathcal{G}_{\beta\alpha}(x)$. The form of the decomposition of the function $\mathcal{G}(x,\xi)$ stems from the fact that $\mathcal{M}$ and $\Theta$ have the same structure. The latter relation is useful to transform the $[w,\Theta]$-star products.

Let $F^w_{[\Theta]}$, $F^w_{[\mathcal{M}]}$, $G^w_{[\Theta]}$, and $G^w_{[\mathcal{M}]}$ be the corresponding symbols of $\hat{F}$ and $\hat{G}$ operators, then

$$F^w_{[\Theta]} * G^w_{[\Theta]} = \mathcal{G}\left(x, -i\hbar\frac{\partial}{\partial p}\right)\left(F^w_{[\mathcal{M}]} * G^w_{[\mathcal{M}]}\right).$$

We have the inverse relations

$$F^w_{[\mathcal{M}]} = \mathcal{G}^{-1}\left(x, -i\hbar\frac{\partial}{\partial p}\right) F^w_{[\Theta]}$$

allowing us to find

$$F^w_{[\Theta]} * G^w_{[\Theta]} = \mathcal{G}\left(x, -i\hbar\frac{\partial}{\partial p}\right)$$
$$\times \left[\left(\mathcal{G}^{-1}F\right)^w_{[\mathcal{M}]} * \left(\mathcal{G}^{-1}G\right)^w_{[\mathcal{M}]}\right].$$

Using expansion (222) for $\mathcal{G}(x,\xi)$, we obtain

$$F^w_{[\Theta]} * G^w_{[\Theta]} = F^w_{[\mathcal{M}]} * G^w_{[\mathcal{M}]}$$
$$- \frac{\hbar^2}{2}\left[\mathcal{G}_{\alpha\beta}\frac{\partial^2}{\partial p_\alpha \partial p_\beta}\left(F^w_{[\mathcal{M}]} * G^w_{[\mathcal{M}]}\right)\right.$$
$$- \left(\mathcal{G}_{\alpha\beta}\frac{\partial^2 F}{\partial p_\alpha \partial p_\beta}\right)^w_{[\mathcal{M}]} * G^w_{[\mathcal{M}]}$$
$$\left. - F^w_{[\mathcal{M}]} * \left(\mathcal{G}_{\alpha\beta}\frac{\partial^2 G}{\partial p_\alpha \partial p_\beta}\right)^w_{[\mathcal{M}]}\right] + O(\hbar^4).$$

Substituting approximation (220) in this equation, we obtain

$$F^w_{[\Theta]} * G^w_{[\Theta]} = F^w_{[\mathcal{M}]} * G^w_{[\mathcal{M}]}$$
$$-\hbar^2 \mathcal{G}_{\alpha\beta}\frac{\partial F}{\partial p_\alpha}\frac{\partial G}{\partial p_\beta} + O(\hbar^3) = F^w_{[\mathcal{M}]} * G^w_{[\mathcal{M}]} + O(\hbar^2).$$

Therefore,

$$F^w_{[\Theta]} * G^w_{[\Theta]} = F^w_{[\Theta]}G^w_{[\Theta]} + \frac{i\hbar}{2}\{F^w_{[\Theta]}, G^w_{[\Theta]}\} + O(\hbar^2).$$

Thus, the $[w,\Theta]$-star product satisfies properties (221), which means that the correspondence principle holds true for the quantization $\mathbf{Q}_{[w,\Theta]}(F)$.

Equation (196) can be written as

$$F^w_{[\Theta]} = \mathcal{O} F_{[\omega,\Theta]}, \quad \mathcal{O} = \Lambda_\varpi\left(x, -i\hbar\frac{\partial}{\partial p}\right)\Omega(-\hbar\mathbf{\Pi}).$$

As follows from Eq. (188), the function $\Omega(k)$ has the expansion $\Omega(k) = 1 + \varpi k + O(k^2)$. Then Eqs. (186) and (181) imply

$$\mathcal{O} = 1 - \hbar\varpi\left(\frac{\partial^2}{\partial x^\mu \partial p_\mu} + p_\mu \Gamma^\mu_{\alpha\beta}(x)\frac{\partial^2}{\partial p_\alpha \partial p_\beta}\right) + O(\hbar^2),$$
$$\mathcal{O}^{-1} = 1 + \hbar\varpi\left(\frac{\partial^2}{\partial x^\mu \partial p_\mu} + p_\mu \Gamma^\mu_{\alpha\beta}(x)\frac{\partial^2}{\partial p_\alpha \partial p_\beta}\right) + O(\hbar^2). \tag{223}$$

Now we can relate $[w,\Theta]$- and $[\omega,\Theta]$-star product in the same manner as the related $w$-star product and $[w,\Theta]$-star product,

$$F_{[\omega,\Theta]} * G_{[\omega,\Theta]} = \mathcal{O}^{-1}\left[(\mathcal{O}F)^w_{[\Theta]} * (\mathcal{O}G)^w_{[\Theta]}\right]. \tag{224}$$

With the help of representations (223), we rewrite (224) as

$$F_{[\omega,\Theta]} * G_{[\omega,\Theta]} = F^w_{[\Theta]} * G^w_{[\Theta]}$$
$$+ \hbar\varpi\left[\left(\frac{\partial^2}{\partial x^\mu \partial p_\mu} + p_\mu \Gamma^\mu_{\alpha\beta}\frac{\partial^2}{\partial p_\alpha \partial p_\beta}\right)\right.$$
$$\times \left(F^w_{[\Theta]} * G^w_{[\Theta]}\right) - \left(\frac{\partial^2 F}{\partial x^\mu \partial p_\mu} + p_\mu \Gamma^\mu_{\alpha\beta}\frac{\partial^2 F}{\partial p_\alpha \partial p_\beta}\right) * G^w_{[\Theta]}$$
$$\left. + F^w_{[\Theta]} * \left(\frac{\partial^2 G}{\partial x^\mu \partial p_\mu} + p_\mu \Gamma^\mu_{\alpha\beta}\frac{\partial^2 G}{\partial p_\alpha \partial p_\beta}\right)\right] + O(\hbar^2).$$

In this formula, we use the approximation

$$F^w_{[\Theta]} * G^w_{[\Theta]} = F^w_{[\Theta]} G^w_{[\Theta]} + O(\hbar)$$

to obtain

$$F_{[\omega,\Theta]} * G_{[\omega,\Theta]} = F_{[\omega,\Theta]} * G_{[\omega,\Theta]} + \frac{i\hbar}{2}\{F_{[\omega,\Theta]}, G_{[\omega,\Theta]}\}$$
$$+ \hbar\varpi\left[\frac{\partial F}{\partial x^\mu}\frac{\partial G}{\partial p_\mu} + \frac{\partial F}{\partial p_\mu}\frac{\partial G}{\partial x^\mu}\right.$$
$$\left.+ p_\mu \Gamma^\mu_{\alpha\beta}\left(\frac{\partial F}{\partial p_\alpha}\frac{\partial G}{\partial p_\beta} + \frac{\partial F}{\partial p_\beta}\frac{\partial G}{\partial p_\alpha}\right)\right] + O(\hbar^2). \quad (225)$$

The quantity in the square brackets in the r.h.s. of Eq. (225) is symmetric with respect to the permutation of $F$ e $G$. Therefore $[\omega, \Theta]$-star products of $[\omega, \Theta]$-symbols satisfy conditions (7), which proves the correspondence principle for all the family of quantizations $\mathbf{Q}_{[\omega,\Theta]}(F)$, in particular, for the quantizations $\mathbf{Q}_{[\omega]}(F) = \mathbf{Q}_{[\omega,1]}(F)$.

## 5 Some physical applications

### 5.1 Quantization of a particle in a spherically symmetric field

Let us start with the classical Hamiltonian of a particle moving in a spherically symmetric field, written from the beginning in polar coordinates $(x'^1, x'^2, x'^3) = (r, \theta, \varphi)$,

$$H^P(r,\theta,\varphi) = \frac{1}{2m}g^{\mu'\nu'}(r,\theta,\varphi)p_{\mu'}p_{\nu'} + U(r)$$
$$= \frac{1}{2m}[p_r^2 + r^{-2}(p_\theta^2 + p_\varphi^2 \sin^{-2}\theta)] + U(r),$$
$$g^{\mu'\nu'} = \text{diag}(1, r^{-2}, r^{-2}\sin^{-2}\theta). \quad (226)$$

According to Eq. (136), the quantization $Q_{[\omega]}$ of the homogeneous polynomial $H^P(r,\theta,\varphi)$ does not contain any ambiguity and reads

$$Q_{[\omega]}(H^P(r,\theta,\varphi)) = \hat{H}^P$$
$$= -\frac{\hbar^2}{2m}g^{-1/2}\partial_\mu g^{1/2} g^{\mu\nu}\partial_\nu + U(r). \quad (227)$$

This is just Hamiltonian (13). As was already said, its kinetic part is proportional to the Laplace–Beltrami operator, $g^{-1/2}\partial_\mu g^{1/2} g^{\mu\nu}\partial_\nu$.

### 5.2 Quantization of a non-relativistic particle in curved space

Consider here a quantization $\mathbf{Q}_{[\omega,\Theta]}(H)$ of the non-relativistic particle Hamiltonian $H$

$$H = \frac{1}{2m}g^{\mu\nu}(x)p_\mu p_\nu \quad (228)$$

in a curved space with a metric tensor $g^{\mu\nu}(x)$ – see the discussion in Sect. 1.

This quantization can be done according to Eq. (200). We now present the corresponding calculations.

It turns out that the operator $\boldsymbol{\Pi}$ (186) annihilates the classical function $H$,

$$\boldsymbol{\Pi} H = \frac{1}{2m}\left(\frac{\partial^2}{\partial x^\mu \partial p_\mu} + \Gamma^\mu_\alpha \frac{\partial}{\partial p_\alpha} + p_\mu \Gamma^\mu_{\alpha\beta}\frac{\partial^2}{\partial p_\alpha \partial p_\beta}\right)g^{\rho\sigma}p_\rho p_\sigma$$
$$= \frac{p_\rho}{m}\left(\partial_\mu g^{\rho\mu} + g^{\rho\sigma}\Gamma_\sigma + \Gamma^\rho_{\alpha\beta}g^{\alpha\beta}\right) = 0.$$

In these calculations, we have used Eq. (252) for the Christoffel symbols. Thus, taking into account initial condition (27), we obtain

$$\Omega(-\hbar\boldsymbol{\Pi})H = \Omega(0)H = H. \quad (229)$$

According to properties assumed for the function $\Theta(x,\xi)$ (see Sect. 4.2), its expansion in powers of $\xi$ has the form

$$\Theta(x,\xi) = 1 + \frac{1}{2}\Theta_{\alpha\beta}(x)\xi^\alpha\xi^\beta + O(\xi^4),$$
$$\Theta_{\alpha\beta}(x) = \left.\frac{\partial^2 \Theta}{\partial \xi^\alpha \partial \xi^\beta}\right|_{\xi=0}. \quad (230)$$

This expansion implies the following exact result:

$$\Theta\left(x, -i\hbar\frac{\partial}{\partial p}\right)H$$
$$= \frac{1}{2m}\left(1 - \frac{\hbar^2}{2}\Theta_{\alpha\beta}(x)\frac{\partial^2}{\partial p_\alpha \partial p_\beta}\right)g^{\mu\nu}(x)p_\mu p_\nu$$
$$= \frac{1}{2m}g^{\mu\nu}(x)p_\mu p_\nu - \frac{\hbar^2}{2m}\Theta_{\alpha\beta}(x)g^{\alpha\beta}(x)$$
$$= \frac{1}{2m}g^{\mu\nu}(x)p_\mu p_\nu + \frac{\hbar^2}{6m}R(x), \quad (231)$$

where $R(x)$ is the scalar curvature.

Taking into account that $\boldsymbol{\Pi}\left(\Theta_{\alpha\beta}(x)g^{\alpha\beta}(x)\right) = 0$, since $\Theta_{\alpha\beta}(x)g^{\alpha\beta}(x)$ does not depend on $p$, we finally calculate $\mathbf{Q}_{[\omega,\Theta]}(H)$ via $\mathbf{Q}_{[\omega]}(H)$ using definition (200):

$$\mathbf{Q}_{[\omega,\Theta]}(H) = \mathbf{Q}_{[\omega]}\left(\Omega^{-1}(-\hbar\boldsymbol{\Pi})\Theta\left(x, -i\hbar\frac{\partial}{\partial p}\right)\Omega(-\hbar\boldsymbol{\Pi})H\right)$$
$$= \mathbf{Q}_{[\omega]}\left(\Omega^{-1}(-\hbar\boldsymbol{\Pi})\Theta\left(x, -i\hbar\frac{\partial}{\partial p}\right)H\right)$$
$$= \mathbf{Q}_{[\omega]}\left(H - \frac{\hbar^2}{2m}\Theta_{\alpha\beta}(x)g^{\alpha\beta}(x)\right). \quad (232)$$

Since $\mathbf{Q}_{[\omega]}(F(x)) = F(\hat{x})$, we finally obtain

$$\left[\mathbf{Q}_{[\omega,\Theta]}(H)\right]_x = -\frac{\hbar^2}{2m}g^{-1/2}\partial_\mu\sqrt{g}g^{\mu\nu}\partial_\nu$$
$$- \frac{\hbar^2}{2m}\Theta_{\alpha\beta}(x)g^{\alpha\beta}(x). \quad (233)$$

Thus, applying the general covariant quantization (200) to the non-relativistic particle Hamiltonian, we obtain the Laplace–

Beltrami operator as the kinetic part of the quantum Hamiltonian and the quantum potential term

$$U(x) = -\frac{\hbar^2}{2m} \Theta_{\alpha\beta}(x) g^{\alpha\beta}(x), \tag{234}$$

which is as arbitrary as the tensor $\Theta_{\alpha\beta}(x)$. Equation (234) represents possible ambiguities in the quantization of the non-relativistic particle Hamiltonian.

If we choose $\Theta = \mathscr{M}$, where the function $\mathscr{M}$ is given by its expansion (304), we obtain the quantum potential in one of its particular forms

$$U(x) = \frac{\hbar^2}{6m} R(x),$$

where $R(x)$ is the scalar curvature.

One can demonstrate that for any quantum potential of the form

$$U(x) = \frac{\hbar^2}{m} f(R(x)), \tag{235}$$

where the function $f(R)$ is almost arbitrary, there exists a covariant quantization $\mathbf{Q}_{[\omega,\Theta]}(H)$, which produces this potential, i.e.,

$$\left[\mathbf{Q}_{[\omega,\Theta]}(H)\right]_x = -\frac{\hbar^2}{2m} g^{-1/2} \partial_\mu g^{1/2} g^{\mu\nu} \partial_\nu + \frac{\hbar^2}{m} f(R(x)). \tag{236}$$

Indeed, the result (236) of the quantization corresponds to any $\Theta$ of the form

$$\Theta(x, \xi) = 1 + \frac{f(R(x))}{R(x)} R_{\alpha\beta}(x) + \cdots. \tag{237}$$

The only natural restriction of the function $f(R)$ reads

$$\lim_{R \to 0} f(R)/R < \infty. \tag{238}$$

We interpret the existence of the above ambiguity in the following way: The mathematical analysis shows that in principle there exist infinite possibilities to quantize a classical theory without violating the correspondence principle. The choice of a concrete quantization has to be made by a physicist on the basis of experimental observations.

## 6 Summary

1. We have constructed a class $Q^c_{[\omega]}(F)$ of quantizations in flat spaces and Cartesian coordinates parametrized by a function $\omega(\theta)$, $\theta \in \mathbb{R}$. We have proved their consistence (which includes, in particular, verification of the correct classical limit-correspondence principle) and a covariance under coordinate transformations from the group $O(r, s)$. We have demonstrated that commonly discussed quantizations, such that $px$ and $xp$ quantizations, the Weyl quantization and the Born–Jordan quantization, are particular cases of the constructed class for a specific choice of the function $\omega(\theta)$.

2. Then we have generalized quantizations $Q^c_{[\omega]}(F)$ formulated in flat spaces in Cartesian coordinate systems to quantizations $Q_{[\omega]}(F)$ formulated also in flat spaces but already in arbitrary coordinate systems. We have proved that the latter quantizations are consistent and covariant under arbitrary coordinate transformations. Applying this quantization to the old problem of constructing quantum Hamiltonian in polar coordinates, we directly obtain a correct result.

3. We have derived a coordinate representation for the quantization $Q_{[\omega]}(F)$. Noteworthy are the following applications of these quantizations:

   (i) Quantization of the coordinate $x^\mu$:
   $$\left[Q_{[\omega]}(x^\mu)\right]_x = \left(\hat{x}^\mu\right)_x = x^\mu. \tag{239}$$

   (ii) Quantization of the momentum $p_\mu$:
   $$\left[Q_{[\omega]}(p_\mu)\right]_x = \left(\hat{p}_\mu\right)_x = -i\hbar \left[\partial_\mu + \phi \Gamma_\mu\right],$$
   $$\phi = \int_{-\infty}^{+\infty} (\theta + 1/2) \omega(\theta) d\theta. \tag{240}$$

   (iii) Quantization of the free non-relativistic particle Hamiltonian $H = \frac{1}{2m} g^{\mu\nu}(x) p_\mu p_\nu$ in a flat space:
   $$\left[Q_{[\omega]}(H)\right]_x = -\frac{\hbar^2}{2m} g^{-1/2} \partial_\mu g^{1/2} g^{\mu\nu} \partial_\nu. \tag{241}$$

   The latter result, (241), is independent of the function $\omega(\theta)$ and is therefore unique in any flat space.

4. We have constructed a class of quantizations $\mathbf{Q}_{[\omega]}(F)$, which is, in fact, a direct generalization of the quantizations $Q_{[\omega]}(F)$ to the curved space case. Such a generalization can be called minimal and is parametrized by the same function $\omega(\theta)$. We have proved its consistency and covariance under general coordinate transformations.

5. Then we have constructed an extended class $\mathbf{Q}_{[\omega,\Theta]}(F)$ of quantizations in a curved space, which is already parametrized by the function $\omega(\theta)$ and a new function $\Theta(x, \xi)$. This class is a generalization of the quantizations $\mathbf{Q}_{[\omega]}(F)$ and includes the latter quantizations as a particular case at $\Theta = 1$. We have presented a rather nontrivial proof of their consistency and their covariance under general coordinate transformations.

6. In particular, it was interesting to apply the quantizations $\mathbf{Q}_{[\omega,\Theta]}(F)$ to the coordinate $x^\mu$, to the momentum $p_\mu$, and to a free non-relativistic particle Hamiltonian $H = \frac{1}{2m} g^{\mu\nu}(x) p_\mu p_\nu$ in a curved space in order to find a difference with the quantizations of the same physical quantities in a flat space; see item 3. It turns out that the quantizations $\mathbf{Q}_{[\omega,\Theta]}(x^\mu)_x$ and $\mathbf{Q}_{[\omega,\Theta]}(p_\mu)_x$ lead to the same results (239) and (240),

$$\left[\mathbf{Q}_{[\omega,\Theta]}\left(x^{\mu}\right)\right]_{x}=\left(\hat{x}^{\mu}\right)_{x}=x^{\mu},$$
$$\left[\mathbf{Q}_{[\omega,\Theta]}\left(p_{\mu}\right)\right]_{x}=-i\hbar\left[\partial_{\mu}+\phi\Gamma_{\mu}\right],$$
$$\phi=\int_{-\infty}^{+\infty}(\theta+1/2)\,\omega(\theta)\mathrm{d}\theta,$$

whereas the course of the quantization of the free non-relativistic particle Hamiltonian $H=\frac{1}{2m}g^{\mu\nu}(x)p_{\mu}p_{\nu}$ in a curved space contains an ambiguity. A quantum potential $U$ appears in a sum with the Laplace–Beltrami operator,

$$\left[\mathbf{Q}_{[\omega,\Theta]}(H)\right]_{x}=-\frac{\hbar^{2}}{2m}g^{-1/2}\partial_{\mu}\sqrt{g}g^{\mu\nu}\partial_{\nu}+U(x),$$
$$U(x)=-\frac{\hbar^{2}}{2m}\Theta_{\alpha\beta}(x)g^{\alpha\beta}(x),\ \Theta_{\alpha\beta}(x)=\left.\frac{\partial^{2}\Theta}{\partial\xi^{\alpha}\partial\xi^{\beta}}\right|_{\xi=0}.$$

The ambiguity of $U$ is due to the ambiguity of the tensor $\Theta_{\alpha\beta}(x)$. One can demonstrate that, for any quantum potential of the form

$$U(x)=\frac{\hbar^{2}}{m}f\left(R(x)\right),$$

where the function $f(R)$ is almost arbitrary, there exists a covariant quantization $\mathbf{Q}_{[\omega,\Theta]}(H)$, which produces this potential.

7. In subsequent publications, we plan to generalize the obtained results to quantize classical theories with constraints and consider a covariant path-integral quantization.

**Acknowledgements** D.M.G. is grateful to the Brazilian foundation FAPESP, grant 2016/03319-6, Sã o Paulo Research Foundation (FAPESP), and to the CNPq for their permanent support. His work is also partially supported by Tomsk State University Competitiveness Improvement Program and by RFBR, research project No. 15-02-00293a. Both authors are grateful to Dr. Geraldo for friendly support during their work.

## A Appendix

### A.1 The operator $e^{i(\eta\hat{x}+\hat{p}\xi)}$ and its matrix elements

I. Using the Baker–Campbell–Hausdorff formula $e^{\hat{A}+\hat{B}}=e^{-\frac{1}{2}\left[\hat{A},\hat{B}\right]}e^{\hat{A}}e^{\hat{B}}$, we can obtain

$$e^{i(\eta\hat{x}+\hat{p}\xi)}=e^{-\frac{i\hbar}{2}\eta\xi}e^{i\hat{p}\xi}e^{i\eta\hat{x}}. \quad (242)$$

II. Representation (242) allows us to calculate commutation relations between $e^{i(\eta\hat{x}+\hat{p}\xi)}$ and canonical operators $\hat{x}^{\mu}$ and $\hat{p}_{\mu}$,

$$\begin{aligned}\left[\hat{x}^{\mu},e^{i(\eta\hat{x}+\hat{p}\xi)}\right]&=e^{-\frac{i\hbar}{2}\eta\xi}\left[\hat{x}^{\mu},e^{i\hat{p}\xi}\right]e^{i\eta\hat{x}}\\&=-\hbar\xi^{\mu}e^{i(\eta\hat{x}+\hat{p}\xi)},\\\left[\hat{p}_{\mu},e^{i(\eta\hat{x}+\hat{p}\xi)}\right]&=e^{-\frac{i\hbar}{2}\eta\xi}e^{i\hat{p}\xi}\left[\hat{p}_{\mu},e^{i\eta\hat{x}}\right]\\&=\hbar\eta_{\mu}e^{i(\eta\hat{x}+\hat{p}\xi)}.\end{aligned} \quad (243)$$

III. Using successively the Baker–Campbell–Hausdorff formula, we obtain a generalization of Eq. (242)

$$\begin{aligned}e^{i(\eta\hat{x}+\hat{p}\xi)}&=e^{i\left[(\eta\hat{x}+\alpha\hat{p}\xi)+(1-\alpha)\hat{p}\xi\right]}\\&=e^{i\hbar\frac{1-\alpha}{2}\eta\xi}e^{i(\eta\hat{x}+\alpha\hat{p}\xi)}e^{i(1-\alpha)\hat{p}\xi}\\&=e^{-i\hbar\left(\alpha-\frac{1}{2}\right)\eta\xi}e^{i\alpha\hat{p}\xi}e^{i\eta\hat{x}}e^{i(1-\alpha)\hat{p}\xi}.\end{aligned} \quad (244)$$

IV. Using representation (242) and a complete set of generalized eigenvectors $\{|x\rangle\}$ of the operator $\hat{x}$, with properties (86), we calculate the matrix element $\langle x|e^{i(\eta\hat{x}+\hat{p}\xi)}|y\rangle$ as follows:

$$\begin{aligned}\langle x|e^{i(\eta\hat{x}+\hat{p}\xi)}|y\rangle&=e^{i\eta\left(y-\frac{\hbar}{2}\xi\right)}\langle x|e^{i\hat{p}\xi}|y\rangle\\&=e^{i\eta\left(y-\frac{\hbar}{2}\xi\right)}\delta(x+\hbar\xi-y)\\&=e^{i\eta\frac{x+y}{2}}\delta(x+\hbar\xi-y)\\&=\hbar^{-D}e^{i\eta\frac{x+y}{2}}\delta\left(\xi-\frac{y-x}{\hbar}\right).\end{aligned} \quad (245)$$

V. Any operator $\hat{F}$ defined in the Hilbert space $\mathfrak{H}$ can be represented as an integral over operators $e^{i(\eta\hat{x}+\hat{p}\xi)}$ as follows:

$$\begin{aligned}\hat{F}&=\int c(\eta,\xi)e^{i(\eta\hat{x}+\hat{p}\xi)}\mathrm{d}\eta\mathrm{d}\xi,\\c(\eta,\xi)&=\left(\frac{\hbar}{2\pi}\right)^{D}\int\left\langle x-\frac{\hbar}{2}\xi\right|\hat{A}\left|x+\frac{\hbar}{2}\xi\right\rangle e^{-i\eta x}\mathrm{d}x.\end{aligned} \quad (246)$$

The validity of Eq. (246) can be justified as follows:

$$\begin{aligned}&\langle x|\int c(\eta,\xi)e^{i(\eta\hat{x}+\hat{p}\xi)}\mathrm{d}\eta\mathrm{d}\xi|y\rangle\\&=\hbar^{-D}\int c(\eta,\xi)e^{i\eta\frac{x+y}{2}}\delta\left(\xi-\frac{y-x}{\hbar}\right)\mathrm{d}\eta\mathrm{d}\xi\\&=\left(\frac{1}{2\pi}\right)^{D}\int\left\langle z-\frac{1}{2}(y-x)\right|\hat{F}\left|z+\frac{1}{2}(y-x)\right\rangle e^{i\eta\left(\frac{x+y}{2}-z\right)}\mathrm{d}z\mathrm{d}\eta\\&=\int\left\langle z-\frac{1}{2}(y-x)\right|\hat{F}\left|z+\frac{1}{2}(y-x)\right\rangle\\&\quad\times\delta\left(\frac{x+y}{2}-z\right)\mathrm{d}z=\langle x|\hat{F}|y\rangle.\end{aligned}$$

### A.2 A generalization of the Leibnitz formula

Let $f(x)$ and $g(x)$ be two $x$-dependent functions ($x = (x^1, x^2, \ldots x^D)$) and $T^{\mu_1\cdots\mu_n}(x)$ are $x$-dependent coefficients completely antisymmetric in the indices $\mu_1\cdots\mu_n$. One can prove that the following generalization of the well-known Leibnitz formula[12] holds true:

$$T^{\mu_1\cdots\mu_n}\partial_{\mu_1}\cdots\partial_{\mu_n}fg = T^{\mu_1\cdots\mu_n}\sum_{k=0}^{n}\binom{n}{k}$$
$$\times \left(\partial_{\mu_1}\cdots\partial_{\mu_k}f\right)\left(\partial_{\mu_{k+1}}\cdots\partial_{\mu_n}g\right). \quad (247)$$

The above statement can be proved by induction. It is obviously true for $n = 1$. Under the supposition that (247) holds true for $n$, we then will prove that it holds true for $n+1$ as well.

Let us consider the left hand side of (247) at $n+1$,

$$T^{\mu_1\cdots\mu_n\mu_{n+1}}\partial_{\mu_1}\cdots\partial_{\mu_n}\partial_{\mu_{n+1}}fg = T^{\mu_1\cdots\mu_n\mu_{n+1}}$$
$$\times \left\{\partial_{\mu_1}\cdots\partial_{\mu_n}\left[\left(\partial_{\mu_{n+1}}f\right)g + g\left(\partial_{\mu_{n+1}}f\right)\right]\right\}. \quad (248)$$

Since $T^{\mu_1\cdots\mu_n\mu_{n+1}}$ is symmetric in all the indices, it is also symmetric in the subgroup $\mu_1\cdots\mu_n$ of all the indices. Then we can use Eq. (247) to calculate the terms in the curled brackets. Thus, we obtain

$$T^{\mu_1\cdots\mu_n\mu_{n+1}}\left[\sum_{k=0}^{n}\binom{n}{k}\left(\partial_{\mu_1}\cdots\partial_{\mu_k}\partial_{\mu_{n+1}}f\right)\left(\partial_{\mu_{k+1}}\cdots\partial_{\mu_n}g\right)\right.$$
$$\left.+ \sum_{k=0}^{n}\binom{n}{k}\left(\partial_{\mu_1}\cdots\partial_{\mu_k}f\right)\left(\partial_{\mu_{k+1}}\cdots\partial_{\mu_n}\partial_{\mu_{n+1}}g\right)\right]. \quad (249)$$

Due to the complete symmetry of $T^{\mu_1\cdots\mu_n\mu_{n+1}}$, the first term in the RHS of Eq. (249) can be written as

$$T^{\mu_1\cdots\mu_n\mu_{n+1}}\sum_{k=0}^{n}\binom{n}{k}\left(\partial_{\mu_1}\cdots\partial_{\mu_k}\partial_{\mu_{n+1}}f\right)\left(\partial_{\mu_{k+1}}\cdots\partial_{\mu_n}g\right)$$
$$= T^{\mu_1\cdots\mu_n\mu_{n+1}}\sum_{k=1}^{n+1}\binom{n}{k-1}\left(\partial_{\mu_1}\cdots\partial_{\mu_k}f\right)\left(\partial_{\mu_{k+1}}\cdots\partial_{\mu_{n+1}}g\right).$$

Thus,

$$T^{\mu_1\cdots\mu_n\mu_{n+1}}\partial_{\mu_1}\cdots\partial_{\mu_n}\partial_{\mu_{n+1}}fg = T^{\mu_1\cdots\mu_n\mu_{n+1}}$$
$$\times \left\{\sum_{k=1}^{n}\left[\binom{n}{k} + \binom{n}{k-1}\right]\right.$$

---

[12] The many-dimensional Leibnitz formula has the following form:

$$\frac{d^n}{dx^n}(fg) = \sum_{k=0}^{n}\binom{n}{k}\frac{d^k f}{dx^k}\frac{d^{n-k}g}{dx^{n-k}}, \quad \binom{n}{k} = \frac{n!}{k!(n-k)!}.$$

$$\times \left(\partial_{\mu_1}\cdots\partial_{\mu_k}f\right)\left(\partial_{\mu_{k+1}}\cdots\partial_{\mu_{n+1}}g\right)$$
$$+\binom{n}{0}\left(\partial_{\mu_1}\cdots\partial_{\mu_{n+1}}f\right)g + \binom{n}{n}f\left(\partial_{\mu_1}\cdots\partial_{\mu_{n+1}}g\right)\right\}. \quad (250)$$

Then, using properties of the binomial coefficients,

$$\binom{n}{k} + \binom{n}{k-1} = \binom{n+1}{k},$$
$$\binom{n+1}{0} = \binom{n}{0} = \binom{n+1}{n+1} = \binom{n}{n} = 1, \quad (251)$$

we justify Eq. (247) for $n+1$.

### A.3 Exponential function

In constructions of the present article, an exponential function (exponential map) in a Riemannian space is used. Now we recall its definition and some properties [42].

Let $M$ be a Riemannian space and $x \in M$. An affine connection on $M$ allows one to define the notion of a geodesic through the point $x$. In the case considered in the present article, we have Christoffel symbols $\Gamma^{\mu}_{\alpha\beta}(x)$ as the connection, they and their contraction $\Gamma_{\alpha}(x) = \Gamma^{\mu}_{\alpha\mu}(x)$ are expressed via the metric tensor,

$$\Gamma^{\mu}_{\alpha\beta}(x) = \frac{1}{2}g^{\mu\rho}\left(\frac{\partial g_{\rho\alpha}}{\partial x^{\beta}} + \frac{\partial g_{\rho\beta}}{\partial x^{\alpha}} - \frac{\partial g_{\alpha\beta}}{\partial x^{\rho}}\right),$$
$$\Gamma_{\alpha}(x) = \left(\sqrt{g}\right)^{-1}\partial_{\alpha}\sqrt{g} = \partial_{\alpha}\ln\left(\sqrt{g}\right), \quad g = \left|\det g_{\mu\nu}\right|. \quad (252)$$

Christoffel symbols define covariant derivatives $\nabla_{\alpha}$ of a tensor $T^{\mu_1\cdots\mu_s}_{\nu_1\cdots\nu_r}$ as follows:

$$\nabla_{\alpha}\left(T^{\mu_1\cdots\mu_s}_{\nu_1\cdots\nu_r}\right) = \partial_{\alpha}\left(T^{\mu_1\cdots\mu_s}_{\nu_1\cdots\nu_r}\right)$$
$$+ \Gamma^{\mu_1}_{\varkappa\alpha}T^{\varkappa\cdots\mu_s}_{\nu_1\cdots\nu_r} + \cdots + \Gamma^{\mu_s}_{\varkappa\alpha}T^{\mu_1\cdots\varkappa}_{\nu_1\cdots\nu_r}$$
$$- \Gamma^{\varkappa}_{\nu_1\alpha}T^{\mu_1\cdots\mu_s}_{\varkappa\cdots\nu_r} - \cdots - \Gamma^{\varkappa}_{\nu_r\alpha}T^{\mu_1\cdots\mu_s}_{\nu_1\cdots\varkappa}. \quad (253)$$

Then curved lines $x^{\mu}(\tau) = C^{\mu}(\tau)$ (parametrized by the proper time $\tau$), satisfying the equations

$$\ddot{C}^{\mu} = -\Gamma^{\mu}_{\alpha\beta}(C)\dot{C}^{\alpha}\dot{C}^{\beta}, \quad g_{\mu\nu}(x)\dot{C}^{\mu}\dot{C}^{\nu} = \dot{C}^2 = 1, \quad (254)$$

are geodesic lines. The geodesic lines provide the minimum (extremum) of the action

$$S[C] = \int \sqrt{g_{\mu\nu}(x)\dot{C}^{\mu}\dot{C}^{\nu}}\,d\tau \quad (255)$$

under the gauge condition

$$\frac{d}{d\tau}\left[\sqrt{g_{\mu\nu}(C(\tau))\dot{C}^{\mu}(\tau)\dot{C}^{\nu}(\tau)}\right] = 0, \quad (256)$$

which fixes the nature of the parameter $\tau$.

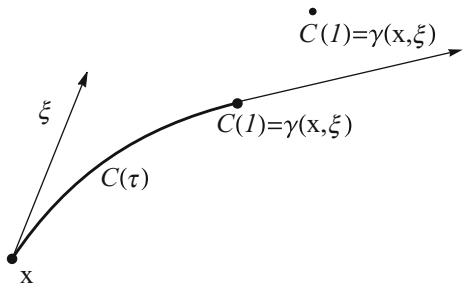

**Fig. 2** The exponential function

There exists a unique geodesic $C_\xi(\tau)$ satisfying the initial conditions

$$C_\xi(0) = x, \quad \dot{C}_\xi(0) = \xi, \tag{257}$$

where $\xi$ is the tangent vector to the geodesic at $\tau = 0$, $\xi \in T_x M$.

Using the introduced geodesic $C_\xi(\tau)$, we can define two vector functions

$$\gamma^\mu(x, \xi) = C^\mu_\xi(1) \quad \text{and} \quad \beta^\mu(x, \xi) = \dot{C}^\mu_\xi(1). \tag{258}$$

The corresponding exponential map $\xi \longrightarrow M$ is defined by the vector function $\gamma^\mu(x, \xi)$, which is called the exponential function, as follows: $\exp_x(\xi) = \gamma(x, \xi)$.

The point $\gamma^\mu(x, \xi)$ is a translation of the point $x$ along a geodesic from $\tau = 0$ to $\tau = 1$ with an initial velocity $\xi$; and $\beta(x, \xi)$ is a result of the parallel translation of $\xi$ from the initial point $x$ to the final point $\gamma(x, \xi)$ (from $\tau = 0$ to $\tau = 1$) (see Fig. 2).

The exponential function satisfies the condition

$$\gamma^\mu(x, 0) = x^\mu, \tag{259}$$

which follows from the fact that the constant function $C(\tau) = x$ is a solution of Eq. (254) with the initial conditions

$$C(0) = x, \quad \dot{C}(0) = 0. \tag{260}$$

The second equation of (254) implies that vectors $\dot{C}(1) = \beta(x, \xi)$ and $\dot{C}(0) = \xi$ have the same length,

$$g_{\mu\nu}(x)\xi^\mu \xi^\nu = g_{\mu\nu}(\gamma(x, \xi))\beta^\mu(x, \xi)\beta^\nu(x, \xi). \tag{261}$$

Let $C_\xi(\tau)$ be a solution of Eq. (254) with initial conditions (257), then one can easily verify that $C_\xi(\lambda\tau) = C_{\lambda\xi}(\tau)$. Thus, the solution $C_\xi(\tau)$ can be expressed via the function $\gamma(x, \xi)$. Indeed,

$$C_\xi(\lambda) = C_{\lambda\xi}(1) = \gamma(x, \lambda\xi) \Longrightarrow C_\xi(\tau) = \gamma(x, \tau\xi). \tag{262}$$

Then

$$\beta^\mu(x, \xi) = \frac{d}{d\tau}\gamma^\mu(x, \tau\xi)\Big|_{\tau=1} = \frac{\partial \gamma^\mu}{\partial \xi^\alpha}(x, \xi)\xi^\alpha. \tag{263}$$

Let us consider the differential equation (254) in a non-canonical phase space $(x, \xi) \in \mathbb{R}^{2D}$. It describes a propagation of initial data (257) by the geodesic flux $T_\tau(x, \xi) = (C(\tau), \dot{C}(\tau))$. Coordinates of the geodesic flux $T_\tau(x, \xi) = (x_\tau(x, \xi), \xi_\tau(x, \xi))$ satisfy first order equations

$$\dot{x}^\mu_\tau = \xi^\mu_\tau, \quad \dot{\xi}^\mu_\tau = -\Gamma^\mu_{\alpha\beta}(x_\tau)\xi^\alpha_\tau \xi^\beta_\tau. \tag{264}$$

Equation (262) represents an explicit form for $x^\mu_\tau$, namely, $x_\tau(x, \xi) = \gamma(x, \tau\xi)$. It follows from Eqs. (264), (262) and (263) that $\xi_\tau(x, \xi) = \frac{1}{\tau}\beta(x, \tau\xi)$. Thus, the geodesic flux $T_\tau : \mathbb{R}^{2D} \to \mathbb{R}^{2D}$ is

$$T_\tau(x, \xi) = \left(\gamma(x, \tau\xi), \frac{1}{\tau}\beta(x, \tau\xi)\right). \tag{265}$$

The geodesic flux as a flux of any differential equation represents a diffeomorphism, which has the commutative group property

$$T_\tau(T_\lambda(x, \xi)) = T_{\tau+\lambda}(x, \xi), \quad T^{-1}_\tau = T_{-\tau}, \tag{266}$$

where $T^{-1}_\tau(x, \xi)$ is an inverse function to the one $T_\tau(x, \xi)$. Setting $\lambda = 1$ in the latter equations and considering the first component only, we find the composition law

$$\gamma(x, (1+\tau)\xi) = \gamma(\gamma(x, \xi), \tau\beta(x, \xi)). \tag{267}$$

The set of Eqs. (255) and (256) is equivalent to Eq. (254) and is invariant under a general coordinate transformation $x' = \varphi(x)$, where $\varphi$ is an arbitrary diffeomorphism. Therefore, if $C(\tau)$ is a solution of Eq. (254) in a coordinate system $K_x$ with coordinates $x$ with the initial conditions (260), then $C'(\tau) = \varphi(C(\tau))$ is a solution of Eq. (254) in a new coordinate system $K_{x'}$ with coordinates $x'$ (with new connections $\Gamma'^\mu_{\alpha\beta}$) and with the initial conditions

$$C'^\mu(0) = x'^\mu = \varphi^\mu(x), \quad \dot{C}'^\mu(0) = \xi'^\mu = \frac{\partial x'^\mu}{\partial x^\alpha}(x)\xi^\alpha. \tag{268}$$

In a coordinate system $K_{x'}$ the exponential function $\gamma'(x', \xi')$ is defined by the connections $\Gamma'^\mu_{\alpha\beta}$. One can find relations between $\gamma'(x', \xi')$ and $\gamma(x, \xi)$ as follows:

$$\gamma'(x', \xi') = C'(1) = \varphi(C(1)) = \varphi(\gamma(x, \xi))$$
$$\Longrightarrow \varphi(\gamma(x, \xi)) = \gamma'\left(\varphi(x), \frac{\partial x'}{\partial x}(x)\xi\right). \tag{269}$$

Velocities related to a geodesic line in coordinate systems $K_{x'}$ and $K_x$ are related as $\dot{C}'^\mu = \partial_\alpha \varphi^\mu(C) \dot{C}^\alpha$. At $\tau = 1$, we obtain a transformation law for $\beta$,

$$\beta'^\mu(x', \xi') = \beta'^\mu\left(\varphi(x), \frac{\partial x'}{\partial x}(x)\xi\right)$$
$$= \frac{\partial x'^\mu}{\partial x^\alpha}(\gamma(x, \xi))\beta^\alpha(x, \xi). \tag{270}$$

It follows that $\beta^\alpha(x,\xi)$ is not a vector field, since the transformation matrix $\frac{\partial x'^\mu}{\partial x^\alpha}$ is taken in the point $\gamma(x,\xi)$ but not in the point $x$. With the help of Eqs. (269) and (270), we find a transformation law of the geodesic flux $T_\tau(x,\xi) = (x_\tau(x,\xi), \xi_\tau(x,\xi))$,

$$\left(x'_\tau, \xi'_\tau\right) = \left(\varphi(x_\tau), \frac{\partial x'}{\partial x}(x_\tau)\xi_\tau\right). \tag{271}$$

We note that in the case of a flat space with Cartesian coordinates all the connections are zero, such that Eq. (254) reads $\ddot{C}^\mu = 0$. Its solutions with initial conditions (257) are $C(\tau) = x + \tau\xi$ such that $\gamma(x,\xi) = x + \xi$ and $\beta(x,\xi) = \xi$. Then the exponential function in an arbitrary coordinate system $K_{x'}$ is represented by Cartesian coordinates $x$ as follows:

$$\gamma'(x',\xi') = \gamma'\left(\varphi(x), \frac{\partial x'}{\partial x}(x)\xi\right) = \varphi(x+\xi). \tag{272}$$

With the help of the exponential function, one can write a covariant Taylor formula,

$$F(\gamma(x,\xi)) = \sum_{n=0}^\infty \frac{\xi^{\mu_1}\cdots\xi^{\mu_n}}{n!}\nabla_{(\mu_1}\ldots\nabla_{\mu_n)}F(x), \tag{273}$$

where indices inside the brackets are totally symmetrized. Equation (273) is a generalization of the well-known formula $F(x+\xi) = \exp\left(\xi^\mu\partial_\mu\right)\cdot F(x)$ in a flat space.

To prove Eq. (273), we calculate successively derivatives of $F(C)$, reducing all the higher derivatives to the first one, using Eq. (254). By induction, we can prove that

$$\dot{F}(C) = \dot{C}^\mu\partial_\mu F(C),\ \ddot{F}(C) = \dot{C}^\mu\dot{C}^\nu\nabla_\mu\nabla_\nu F(C), \ldots,$$
$$\frac{d^n}{d\tau^n}F(C) = \dot{C}^{\mu_1}\ldots\dot{C}^{\mu_n}\nabla_{\mu_1}\ldots\nabla_{\mu_n}F(C),\ldots. \tag{274}$$

Using the initial condition (257), we obtain

$$\left.\frac{d^n}{d\tau^n}F(C(\tau))\right|_{\tau=0} = \xi^{\mu_1}\ldots\xi^{\mu_n}\nabla_{(\mu_1}\ldots\nabla_{\mu_n)}F(x). \tag{275}$$

This result allows us to decompose a function $F(C(\tau))$ in a power series with respect to $\tau$. Setting $\tau = 1$ in such a series, we obtain Eq. (273).

All the derivatives in $\xi$ have a simple representation,

$$\left.\frac{\partial}{\partial\xi^{\mu_1}}\cdots\frac{\partial}{\partial\xi^{\mu_n}}F(\gamma(x,\xi))\right|_{\xi=0} = \nabla_{(\mu_1}\ldots\nabla_{\mu_n)}F(x). \tag{276}$$

We list three first nonsymmetrized products of the operators $\nabla_\mu$:

$$\nabla_{\mu_1} = \partial_{\mu_1},$$
$$\nabla_{\mu_1}\nabla_{\mu_2} = \left(\partial_{\mu_1}\delta^{\nu_2}_{\mu_2} - \Gamma^{\nu_2}_{\mu_1\mu_2}\right)\partial_{\nu_2},$$
$$\nabla_{\mu_1}\nabla_{\mu_2}\nabla_{\mu_3} = \left(\partial_{\mu_1}\delta^{\nu_2}_{\mu_2}\delta^{\nu_3}_{\mu_3} - \Gamma^{\nu_2}_{\mu_1\mu_2}\delta^{\nu_3}_{\mu_3} - \Gamma^{\nu_3}_{\mu_1\mu_3}\delta^{\nu_2}_{\mu_2}\right)$$
$$\times \left(\partial_{\nu_2}\delta^{\rho_3}_{\nu_3} - \Gamma^{\rho_3}_{\nu_2\nu_3}\right)\partial_{\rho_3}. \tag{277}$$

Setting $F(x) = x^\mu$ in Eq. (273), we obtain a decomposition of the exponential function $\gamma^\mu$ in powers of $\xi$,

$$\gamma^\mu(x,\xi) = x^\mu + \xi^\mu - \frac{1}{2}\Gamma^\mu_{\mu_1\mu_2}(x)\xi^{\mu_1}\xi^{\mu_2}$$
$$+ \frac{1}{6}\Gamma^\mu_{\mu_1\mu_2\mu_3}(x)\xi^{\mu_1}\xi^{\mu_2}\xi^{\mu_3} + \cdots,$$
$$\Gamma^\mu_{\mu_1\mu_2\mu_3} = 2\Gamma^\nu_{(\mu_1\mu_2}\Gamma^\mu_{\mu_3)\nu} - \Gamma^\mu_{(\mu_1\mu_2,\mu_3)}. \tag{278}$$

Using Eq. (263), we derive a similar decomposition for $\beta^\mu(x,\xi)$,

$$\beta^\mu(x,\xi) = \xi^\mu - \Gamma^\mu_{\mu_1\mu_2}(x)\xi^{\mu_1}\xi^{\mu_2}$$
$$+ \frac{1}{2}\Gamma^\mu_{\mu_1\mu_2\mu_3}(x)\xi^{\mu_1}\xi^{\mu_2}\xi^{\mu_3} + \cdots. \tag{279}$$

A.4 Normal coordinates and minimal geodesic line

Equation (278) implies

$$\left.\frac{\partial\gamma^\mu}{\partial\xi^\nu}\right|_{\xi=0} = \delta^\mu_\nu. \tag{280}$$

This matrix is invertible; therefore, the function $\xi \longmapsto \gamma(x_0,\xi)$ for any $x_0$ represents a diffeomorphism in the vicinity of the point $\xi = 0$. New coordinates $x'$ given by the equation $x = \varphi^{-1}(x') = \gamma(x_0, x')$ are called normal coordinates around the point $x_0$. It follows from Eq. (259) that the point $x = x_0$ corresponds to the point $x' = 0$. Property (280) implies that $\frac{\partial x^\mu}{\partial x'^\nu}(0) = \delta^\mu_\nu$, and, therefore, a transition to normal coordinates does not change tensors in the point $x = x_0 \Longleftrightarrow x' = 0$.

The curves $C'^\mu(\tau) = \tau\xi^\mu$ in normal coordinates $x'$ correspond to the curves $C^\mu(\tau) = \gamma^\mu(x_0, \tau\xi)$ in the coordinates $x$. The latter curves are geodesic lines which pass through the point $x_0$, according to Eq. (262). This means that in the reference frame $K_{x'}$ geodesic lines which pass through the point $x' = 0$ are straight lines $C'^\mu(\tau) = \tau\xi^\mu$. Thus, as follows from the same Eq. (262), the exponent function in the reference frame $K_{x'}$ has the property

$$\gamma'(0,\xi) = \xi. \tag{281}$$

As follows from the quadratic term of expansion (278), we have

$$\Gamma'^\mu_{\mu_1\mu_2}(0) = 0 \tag{282}$$

in normal coordinates. From the cubic term of expansion (278), we have

$$\Gamma'^\mu_{(\mu_1\mu_2,\mu_3)}(0) = \frac{1}{6}\left(\Gamma'^\mu_{\mu_1\mu_2,\mu_3} + \Gamma'^\mu_{\mu_1\mu_3,\mu_2} + \Gamma'^\mu_{\mu_2\mu_1,\mu_3}\right.$$
$$\left. + \Gamma'^\mu_{\mu_2\mu_3,\mu_1} + \Gamma'^\mu_{\mu_3\mu_1,\mu_2} + \Gamma'^\mu_{\mu_3\mu_2,\mu_1}\right)_{x'=0}$$
$$= 0.$$

Contracting indices $\mu$ and $\mu_3$ and using a symmetry of connections in lower indices, we obtain the identity

$$\Gamma'^\mu_{\mu_1\mu_2,\mu}(0) + \Gamma'_{\mu_1,\mu_2}(0) + \Gamma'_{\mu_2,\mu_1}(0) = 0.$$

Using Eq. (252), we obtain $\Gamma'_{\mu_1,\mu_2} = \Gamma'_{\mu_2,\mu_1}$, which implies an identity that holds true in the normal coordinates,

$$\Gamma'^\mu_{\mu_1\mu_2,\mu}(0) + 2\Gamma'_{\mu_1,\mu_2}(0) = 0. \tag{283}$$

With the help of Eqs. (282) and (283), we can express the Ricci tensor as follows:

$$R'_{\mu\nu}(0) = \Gamma'^\rho_{\mu\nu,\rho}(0) - \Gamma'_{\mu,\nu}(0) = -3\Gamma'_{\mu,\nu}(0)$$
$$= \frac{3}{2}\Gamma'^\rho_{\mu\nu,\rho}(0). \tag{284}$$

If a point $x$ is in the region where $\gamma(x_0, \xi)$ represents a diffeomorphism, then points $x_0$ and $x$, can be related by a geodesic line $C(\tau)$. In this case, $C(0) = x_0$ and $C(1) = x$ and $C(\tau) = \gamma(x_0, \tau\xi)$; see Eq. (262). In the general case, there exist more than one geodesic line which relates two given points. Now, we define the so-called minimal geodesic line.

Let us consider a vicinity of the point $x_0$ where the correspondence $\xi \longmapsto \gamma(x_0, \xi)$ is a diffeomorphism. In addition, we define a diffeomorphism $\xi = \varphi(x)$, where $\varphi^{-1}(\xi) = \gamma(x_0, \xi)$ in this vicinity. Let us consider a $D$-dimensional sphere $E' = \{\xi \in \mathbb{R}^D | \xi^2 = g_{\mu\nu}(x_0)\xi^\mu\xi^\nu < r^2\}$ which is completely embedded in the above mentioned vicinity (such a sphere exists with $r > 0$ because the vicinity is an open manifold). Then we define the correspondence $E = \varphi^{-1}(E')$, which is also open since $\varphi$ is a diffeomorphism. The function $\xi \longmapsto \gamma(x_0, \xi)$ is a diffeomorphism in $E'$, if for any point $x \in E$ there exists one unique point $\xi \in E'$ such that $x = \gamma(x_0, \xi)$. Consider a geodesic line $C_{x_0x}(\tau) = \gamma(x_0, \tau\xi)$ which relates $x_0$ with $x$. This geodesic line is called a minimal geodesic line only if $x \in E$, as was constructed above. Its length is $S[C_{x_0,x}] = \sqrt{g_{\mu\nu}(x_0)\xi^\mu\xi^\nu} < r$. Let us suppose that there exists another geodesic line $K(\tau)$ (different from $C_{x_0x}$), which relates $x_0$ with $x$, i.e., $K(0) = x_0$ and $K(1) = x$. According to the definition of the exponential line $x = \gamma(x_0, \xi_K)$ with $\xi_K = \dot{K}(0)$. Since the function $\xi \longmapsto \gamma(x_0, \xi)$ is a diffeomorphism, which represents a bijection in $E'$, we see that $\xi_K \notin E'$ as $S[K] \geq r$. Thus, a minimal geodesic line is a geodesic line of a minimal length which relates $x_0$ with $x$. The argument

$$\tau_f = r\left(g_{\mu\nu}(x_0)\xi^\mu_K\xi^\nu_K\right)^{-1/2}$$

is restricted, $0 < \tau_f \leq 1$, because $K(\tau_f) = \gamma(x_0, \tau_f\xi_K)$ is a point on the geodesic line. We have

$$\sqrt{g_{\mu\nu}(x_0)(\tau_f\xi_K)^\mu(\tau_f\xi_K)^\nu} = r,$$

if $\tau_f\xi_K \notin E'$ and $\tau_f\xi_K \in \overline{E'}$, where $\bar{A}$ is a closure (in the topological sense) of the set $A$. Since $\xi \longmapsto \gamma(x_0, \xi)$ is a diffeomorphism, $K(\tau_f) = \gamma(x_0, \tau_f\xi_K) \notin E$ and $K(\tau_f) \in \overline{E}$. Thus, a minimal geodesic line is a unique geodesic line which is entirely contained in $E$. Because $\varphi$ is a diffeomorphism, $\lim_{x \to x_0} \xi = 0$ and, therefore,

$$\lim_{x \to x_0} S\left[C_{x_0x}\right] = 0. \tag{285}$$

Therefore, a minimal geodesic line is a unique geodesic line between $x_0$ and $x$, which has the property (285).

### A.5 Jacobian of the geodesic flux

The differential equation (254) is the Euler–Lagrange equation for the following nonsingular Lagrangian:

$$L(x, \dot{x}) = \frac{1}{2}g_{\mu\nu}(x)\dot{x}^\mu\dot{x}^\nu.$$

Let us define a function $G : \mathbb{R}^{2D} \to \mathbb{R}^{2D}$,

$$G(x, \xi) = \left((x^\mu), (g_{\mu\nu}(x)\xi^\nu)\right), \tag{286}$$

which transforms a point $(x, \xi)$ in a noncanonical phase space into a point $(x, p = \partial L/\partial \dot{x}^\mu = g_{\mu\nu}(x)\dot{x}^\nu)$ of a canonical phase space. The inverse function has the form

$$G^{-1}(x, p) = \left((x^\mu), (g^{\mu\nu}(x)p_\nu)\right). \tag{287}$$

Using the introduced function, we can transform a trajectory $(C(\tau), P(\tau))$, given in the canonical phase space, into a trajectory $(C(\tau), P(\tau)) = G(C(\tau), \dot{C}(\tau))$ in the noncanonical phase space. With the help of geodesic flux (264), we can write $(C(\tau), P(\tau)) = G(T_\tau(x, \xi))$, where $x = C(0)$ and $\xi = \dot{C}(0)$ are initial conditions in the non-canonical phase space. These initial conditions can be, in turn, reformulated as initial conditions in the canonical phase space with the help of the $(x, p) = G^{-1}(x, \xi)$, and choosing $p = \dot{P}(0)$ as an initial condition for the momentum. Thus, we have the relation $(C(\tau), P(\tau)) = G(T_\tau(G^{-1}(x, p)))$, which defines the canonical geodesic flux

$$T^C_\tau(x, p) = G(T_\tau(G^{-1}(x, p))). \tag{288}$$

According to the Liouville theorem [43] the Jacobian of the canonical geodesic flux is equal to 1. This fact implies

$$1 = \frac{\partial(T^C_\tau)}{\partial(x, p)} = \frac{\partial(G)}{\partial(x, \xi)}(T_\tau(x, \xi)) \frac{\partial(T_\tau)}{\partial(x, \xi)}(x, \xi)$$
$$\times \frac{\partial(G^{-1})}{\partial(x, p)}(x, p).$$

The Jacobians involved can easily be calculated,

$$\frac{\partial(G)}{\partial(x,\xi)} = \det\begin{pmatrix} \frac{\partial x^\alpha}{\partial x^\beta} = \delta^\alpha_\beta & \frac{\partial x^\alpha}{\partial \xi^\beta} = 0 \\ \frac{\partial g_{\alpha\nu}}{\partial x^\beta}\xi^\nu & g_{\alpha\nu}\frac{\partial \xi^\nu}{\partial \xi^\beta} = g_{\alpha\beta} \end{pmatrix} = g(x),$$

$$\frac{\partial(G^{-1})}{\partial(x,p)} = \det\begin{pmatrix} \frac{\partial x^\alpha}{\partial x^\beta} = \delta^\alpha_\beta & \frac{\partial x^\alpha}{\partial p_\beta} = 0 \\ \frac{\partial g^{\alpha\nu}}{\partial x^\beta}p_\nu & g^{\alpha\nu}\frac{\partial p_\nu}{\partial p_\beta} = g^{\alpha\beta} \end{pmatrix} = g^{-1}(x).$$

Finally we obtain

$$\frac{\partial(T_\tau)}{\partial(x,\xi)}(x,\xi) = \frac{g(x)}{g(\gamma(x,\tau\xi))}. \tag{289}$$

A.6 Properties of the function $\mathscr{M}$

Above we have expressed matrix elements of the operators $\mathbf{Q}_{[\omega]}(F)$ via the function $\mathscr{M}(x,\xi)$ defined by Eqs. (171) and (167). Here we study this function in more detail. First, we represent it as follows:

$$\mathscr{M}(x,\xi) = g^{-1}(x)\sqrt{g(\gamma_{-1/2}(x,\xi))\,g(\gamma_{1/2}(x,\xi))}f(x,\xi),$$

$$f(x,\xi) = \left|\det\left(\frac{\partial\Phi}{\partial(x,\xi)}\right)\right| = \left|\det\begin{pmatrix} \frac{\partial\gamma_{-1/2}}{\partial x} & \frac{\partial\gamma_{-1/2}}{\partial \xi} \\ \frac{\partial\gamma_{1/2}}{\partial x} & \frac{\partial\gamma_{1/2}}{\partial \xi} \end{pmatrix}\right|, \tag{290}$$

where

$$\gamma_a^\mu(x,\xi) = \gamma^\mu(x,a\xi), \quad \frac{\partial\gamma_a}{\partial x} = \left(\frac{\partial\gamma_a^\mu}{\partial x^\nu}\right), \frac{\partial\gamma_a}{\partial \xi} = \left(\frac{\partial\gamma_a^\mu}{\partial \xi^\nu}\right).$$

From this representation, we see that $\mathscr{M}(x,\xi)$ is a function even in $\xi$ since $\sqrt{g(\gamma_{-1/2}(x,\xi))g(\gamma_{1/2}(x,\xi))}$ and $f(x,\xi)$ are even in $\xi$,

$$\mathscr{M}(x,\xi) = \mathscr{M}(x,-\xi).$$

In a new coordinate system $K_{x'}$, we have

$$x'^\mu = \varphi(x), \quad \xi'^\mu = \frac{\partial\varphi^\mu}{\partial x^\nu}(x)\xi^\nu, \quad x^\mu = (\varphi^{-1})^\mu(x'),$$

$$\xi^\mu = \frac{\partial x^\mu}{\partial x'^\nu}\xi'^\nu, \tag{291}$$

such that

$$\mathscr{M}'(x',\xi') = g^{-1}(x)\sqrt{g'(\gamma'_{-1/2})\,g(\gamma'_{1/2})}f'(x',\xi'),$$

$$f'(x',\xi') = \left|\det\begin{pmatrix} \frac{\partial\gamma'_{-1/2}}{\partial x'} & \frac{\partial\gamma'_{-1/2}}{\partial \xi'} \\ \frac{\partial\gamma'_{1/2}}{\partial x'} & \frac{\partial\gamma'_{1/2}}{\partial \xi'} \end{pmatrix}\right|, \tag{292}$$

where

$$\gamma_a'^\mu(x',\xi') = \gamma'^\mu(x',a\xi') = \varphi^\mu(\gamma(x,a\xi))$$
$$= \varphi^\mu(\gamma_a(x,\xi)),$$

according to transformation law (269). In addition we have

$$\frac{\partial\gamma_a'^\mu}{\partial x'^\nu} = \frac{\partial\varphi^\mu}{\partial x^\alpha}(\gamma_a)\left(\frac{\partial\gamma_a^\alpha}{\partial x^\beta}\frac{\partial x^\beta}{\partial x'^\nu} + \frac{\partial\gamma_a^\alpha}{\partial \xi^\beta}\frac{\partial \xi^\beta}{\partial x'^\nu}\right),$$

$$\frac{\partial\gamma_a'^\mu}{\partial \xi'^\nu} = \frac{\partial\varphi^\mu}{\partial x^\alpha}(\gamma_a)\frac{\partial\gamma_a^\alpha}{\partial \xi^\beta}\frac{\partial \xi^\beta}{\partial \xi'^\nu} = \frac{\partial\varphi^\mu}{\partial x^\alpha}(\gamma_a)\frac{\partial\gamma_a^\alpha}{\partial \xi^\beta}\frac{\partial x^\beta}{\partial x'^\nu}.$$

Defining matrices

$$A^\mu_{a\,\nu} = \frac{\partial\varphi^\mu}{\partial x^\nu}(\gamma_a), \quad B^\mu_{\;\nu} = \frac{\partial x^\mu}{\partial x'^\nu}, \quad C^\mu_{\;\nu} = \frac{\partial \xi^\mu}{\partial x'^\nu},$$

we express $f'$ (292) as a modulus of the determinant of the following matrix:

$$\begin{pmatrix} A_{-\frac{1}{2}}\frac{\partial\gamma_{-1/2}}{\partial x}B + A_{-\frac{1}{2}}\frac{\partial\gamma_{-1/2}}{\partial \xi}C & A_{-\frac{1}{2}}\frac{\partial\gamma_{-1/2}}{\partial \xi}B \\ A_{\frac{1}{2}}\frac{\partial\gamma_{1/2}}{\partial x}B + A_{\frac{1}{2}}\frac{\partial\gamma_{1/2}}{\partial \xi}C & A_{\frac{1}{2}}\frac{\partial\gamma_{1/2}}{\partial \xi}B \end{pmatrix}$$

$$= \begin{pmatrix} A_{-\frac{1}{2}} & 0 \\ 0 & A_{\frac{1}{2}} \end{pmatrix}$$

$$\times \begin{pmatrix} \frac{\partial\gamma_{-1/2}}{\partial x} & \frac{\partial\gamma_{-1/2}}{\partial \xi} \\ \frac{\partial\gamma_{1/2}}{\partial x} & \frac{\partial\gamma_{1/2}}{\partial \xi} \end{pmatrix}\begin{pmatrix} B & 0 \\ 0 & B \end{pmatrix}\begin{pmatrix} I & 0 \\ B^{-1}C & I \end{pmatrix}. \tag{293}$$

Taking into account the transformation law $\sqrt{g'} = \sqrt{g}\left|\det(\partial x/\partial x')\right|$, we calculate some of the determinants from (293),

$$\left|\det\begin{pmatrix} A_{-\frac{1}{2}} & 0 \\ 0 & A_{\frac{1}{2}} \end{pmatrix}\right| = \left|\det A_{-\frac{1}{2}}\det A_{\frac{1}{2}}\right|$$

$$= \sqrt{\frac{g(\gamma_{-1/2})}{g'(\gamma'_{-1/2})}}\sqrt{\frac{g(\gamma_{1/2})}{g'(\gamma'_{1/2})}},$$

$$\left|\det\begin{pmatrix} B & 0 \\ 0 & B \end{pmatrix}\right| = |\det B|^2 = \frac{g'(x')}{g(x)}, \quad \left|\det\begin{pmatrix} I & 0 \\ B^{-1}C & I \end{pmatrix}\right| = 1.$$

Thus, the quantity $f'$ from Eq. (292) is related to $f$ from Eq. (290) as

$$f' = \sqrt{\frac{g(\gamma_{-1/2})}{g'(\gamma'_{-1/2})}}\sqrt{\frac{g(\gamma_{1/2})}{g'(\gamma'_{1/2})}}\frac{g'(x')}{g(x)}f. \tag{294}$$

From Eqs. (290) and (292) one can see that $\mathscr{M}$ is a scalar under transformations (291),

$$\mathscr{M}(x,\xi) = \mathscr{M}'(x',\xi'). \tag{295}$$

If the space is a plane, there exists a reference $K_x$ where $g_{\mu\nu}$ has the form (19) and $\Gamma^\mu_{\alpha\beta} \equiv 0$. In such a reference frame, $g(x) = 1$ and $\gamma(x,\xi) = x + \xi$, such that

$$\mathscr{M}(x,\xi) = f(x,\xi) = \left|\det\begin{pmatrix} I & -I/2 \\ I & I/2 \end{pmatrix}\right| = 1.$$

According to Eq. (295) $\mathscr{M}(x,\xi) = 1$ in any coordinate system.

Let us consider the decompositions

$$\mathscr{M}(x,\xi) = \sum_{n=0}^{\infty} m_{\mu_1\cdots\mu_n}(x)\xi^{\mu_1}\ldots\xi^{\mu_n},$$

$$\mathscr{M}'(x',\xi') = \sum_{n=0}^{\infty} m'_{\mu_1\cdots\mu_n}(x')\xi'^{\mu_1}\ldots\xi'^{\mu_n}. \quad (296)$$

Comparing them, taking into account the transformation laws (291) and (295), we obtain

$$m_{\mu_1\cdots\mu_n}(x) = m'_{\nu_1\cdots\nu_n}(x') \frac{\partial x'^{\nu_1}}{\partial x^{\mu_1}}\cdots\frac{\partial x'^{\nu_n}}{\partial x^{\mu_n}}, \quad (297)$$

which means that coefficients $m_{\mu_1\cdots\mu_n}(x)$ are tensors. This allows one to calculate the first terms of decompositions (296) in a most simple coordinate system. To this aim, it is most convenient to use the normal coordinates discussed in Appendix A.4. Let $x$ are normal coordinates. Then it follows from Eq. (281) that

$$\gamma(0,\xi) = \xi \Leftrightarrow \gamma_a(0,\xi) = a\xi. \quad (298)$$

In turn, this relation implies

$$\frac{\partial \gamma_a^\mu}{\partial \xi^\nu}(0,\xi) = a\delta_\nu^\mu, \quad \frac{\partial \gamma_a^\mu}{\partial x^\nu}(0,\xi) \neq \delta_\nu^\mu.$$

One can see that, in the normal coordinates, we have the following result:

$$f(0,\xi) = \left|\det\begin{pmatrix} \frac{\partial \gamma_{-1/2}}{\partial x} & \frac{\partial \gamma_{-1/2}}{\partial \xi} \\ \frac{\partial \gamma_{1/2}}{\partial x} & \frac{\partial \gamma_{1/2}}{\partial \xi} \end{pmatrix}\right|$$

$$= \left|\det\left[\frac{1}{2}\frac{\partial}{\partial x}\left[\gamma\left(x,-\frac{1}{2}\xi\right) + \gamma\left(x,\frac{1}{2}\xi\right)\right]\right]\right|_{x=0}. \quad (299)$$

Using expansion (278), we obtain

$$\frac{1}{2}\left[\gamma\left(x,-\frac{1}{2}\xi\right) + \gamma\left(x,\frac{1}{2}\xi\right)\right]$$

$$= x^\mu - \frac{1}{8}\Gamma_{\alpha\beta}^\mu(x)\xi^\alpha\xi^\beta + O(\xi^4)$$

$$\implies \frac{1}{2}\partial_\nu\left[\gamma^\mu\left(x,-\frac{1}{2}\xi\right) + \gamma^\mu\left(x,\frac{1}{2}\xi\right)\right]$$

$$= \delta_\nu^\mu - \frac{1}{8}\partial_\nu\Gamma_{\alpha\beta}^\mu(x)\xi^\alpha\xi^\beta + O(\xi^4). \quad (300)$$

Then with the help of equation $\det(I + A) = 1 + \mathrm{tr}A + O(A^2)$, we obtain from (299) and (300)

$$f(0,\xi) = 1 - \frac{1}{8}\partial_\mu\Gamma_{\alpha\beta}^\mu(0)\xi^\alpha\xi^\beta + O(\xi^4). \quad (301)$$

It follows from Eq. (298) that

$$\sqrt{g(\gamma_a(0,\xi))/g(0)} = \sqrt{g(a\xi)/g(x)}$$

$$= 1 + a\xi^\alpha \left. g^{-1/2}\partial_\alpha g^{1/2}\right|_{x=0}$$

$$+ \frac{a^2}{2}\xi^\alpha\xi^\beta \left. g^{-1/2}\partial_\alpha\partial_\beta g^{1/2}\right|_{x=0} + O(\xi^3).$$

Taking unto account Eq. (252) and the fact that the Levi-Civita connections are zero in normal coordinates, we obtain

$$\sqrt{g(\gamma_a(0,\xi))/g(x)} = 1 + \frac{a^2}{2}\xi^\alpha\xi^\beta\partial_\alpha\Gamma_\beta(0) + O(\xi^3).$$

Thus, the following expansion holds true:

$$\sqrt{g(\gamma_{-1/2}(0,\xi))g(\gamma_{1/2}(0,\xi))/g^2(x)}$$

$$= 1 + \frac{1}{4}\xi^\alpha\xi^\beta\partial_\alpha\Gamma_\beta(0) + O(\xi^4). \quad (302)$$

With the help of results (301) and (302), we find

$$\mathscr{M}(0,\xi) = 1 - \frac{1}{4}\left[\frac{1}{2}\partial_\mu\Gamma_{\alpha\beta}^\mu(0) - \partial_\alpha\Gamma_\beta(0)\right]\xi^\alpha\xi^\beta + O(\xi^4). \quad (303)$$

Equation (284) allows one to simplify the latter expression,

$$\mathscr{M}(0,\xi) = 1 - \frac{1}{6}R_{\alpha\beta}(0)\xi^\alpha\xi^\beta + O(\xi^4).$$

Since coefficients $m_{\mu_1\cdots\mu_n}(x)$ are tensors, see Eq. (297), the above equation holds true in any reference frame. The point 0, in normal coordinates, corresponds to any point of the space. That is why the formula below is valid in any reference frame and at any point $x$,

$$\mathscr{M}(x,\xi) = 1 - \frac{1}{6}R_{\alpha\beta}(x)\xi^\alpha\xi^\beta + O(\xi^4). \quad (304)$$

**Appendix B: Notation**

- By $\mathbb{R} = (-\infty, \infty)$ the set of all real numbers is denoted.
- Derivatives in $x^\mu$ of a function $A(x)$ are commonly denoted as

  $\partial A/\partial x^\mu = \partial_\mu A$, $\partial/\partial\theta = \partial_\theta$, and so on.

- Greek and Latin vector and tensor indices take on the values $1,\ldots,D$ unless otherwise specified; the convention about summation over repeated indices is adopted unless otherwise specified.
- By $\mathfrak{H}$ a Hilbert space is denoted.
- By $\hat{I}$ an identical operator is denoted.
- By $K_x$ a reference frame that generates coordinates $x = 1,\ldots,D$ is denoted; by $K_x^C$ a Cartesian reference frame

in a flat space that generates coordinates $x = 1, \ldots, D$ is denoted
- By $\tilde{F}(\eta, \xi) = [F(x, p)]^{\approx}(\eta, \xi)$ a double Fourier transform of a function $F(x, p)$ is denoted; see Eq. (22).
- By $\tilde{F}(x, \xi) = [F(x, p)]^{\sim}(x, \xi)$ a partial Fourier transform of a function $F(x, p)$ with respect to momenta is denoted; see Eq. (83).
- By $Q_{\Omega}(F(x, p)) = Q_{[\omega]}(F(x, p))$ we denote a family of covariant quantizations in a flat space, which are parametrized by a weight function $\Omega(k)$ or by its Fourier transform $\omega(\theta) = \frac{1}{2\pi} \int_{-\infty}^{+\infty} \Omega(k) e^{-i\theta k} dk$.
- By $Q_{(\theta)}(F(x, p))$ we denote the so-called basic quantization, which is the quantization $Q_{[\omega]}(F)$ with $\omega(\theta') = \delta(\theta' - \theta)$ or the quantization $Q_{\Omega}(F(x, p))$ with $\Omega(k) = \Omega_{(\theta)}(k) = e^{i\theta k}$, $k = \eta\xi$; see Eq. (49).
- By $\mathbf{Q}_{[\omega, \Theta]}(F(x, p))$ we denote an a family of covariant quantizations in a curved space parametrized by two functions $\omega(\theta)$ and $\Theta(x, \xi)$. The corresponding $[\omega, \Theta]$-symbols are denoted by $F_{[\omega, \Theta]}$.
- By $\mathbf{Q}_{[\omega]}(F(x, p)) = \mathbf{Q}_{[\omega, 1]}(F(x, p))$ we denote a family of covariant quantizations in a curved space parametrized by a function $\omega(\theta)$. The corresponding $[\omega]$-symbols are denoted by $F_{[\omega]}$.
- By $\mathbf{Q}^{\mathrm{w}}_{[\Theta]}$ we denote a family of Weyl quantizations in a curved space, parametrized by a function $\Theta(x, \xi)$. The corresponding $[\mathrm{w}, \Theta]$-symbols are denoted by $F^{\mathrm{w}}_{[\Theta]}$.
- By $\mathbf{Q}^{\mathrm{w}}(F) = \mathbf{Q}^{\mathrm{w}}_{[1]}$ we denote a Weyl quantizations in a curved space. The corresponding w-symbols are denoted by $F^{\mathrm{w}}$.
- By $\Gamma^{\mu}_{\alpha\beta}(x)$ Christoffel symbols are denoted.
- By a vector function $\gamma^{\mu}(x, \xi)$, $x \in M$, $\xi \in T_x M$ an exponential map $\xi \longrightarrow M$ (exponential function) is denoted. Here $M$ is a Riemannian space, and $T_x M$ is a tangent space to $M$.